\documentclass[fleqn,usenatbib]{mnras}

\usepackage{newtxtext,newtxmath}
\usepackage[T1]{fontenc}
\usepackage{ae,aecompl}

\usepackage{graphicx}	
\usepackage{amsmath}	
\usepackage{amssymb}	
\usepackage[utf8]{inputenc}
\usepackage{multirow}
\usepackage{array}
\usepackage{caption}
\usepackage[export]{adjustbox}
\usepackage{xcolor}
\newcolumntype{P}[1]{>{\centering\arraybackslash}p{#1}}
\newcolumntype{C}[1]{>{\centering\arraybackslash}m{#1}}
\newcommand{\HL}[1]{\textcolor{red}{#1}}

\title[Global Site Selection]{Global Site Selection for Astronomy}
\author[N. Aksaker et al.]{%
N. Aksaker$^{1,2}$\thanks{E-mail:naksaker@cu.edu.tr},
S.K. Yerli$^{3}$,
M.A. Erdoğan$^{1}$,
Z. Kurt$^{4,2}$,
K. Kaba$^{5}$,
M. Bayazit$^{4}$, \newauthor
C. Yesilyaprak$^{5,6}$
\\
$^1$Adana Organised Industrial Zones Vocational School of Technical Science, University of Çukurova, 01410, Adana, Turkey\\
$^2$Space Science and Solar Energy Research and Application Center (UZAYMER), University of Çukurova, 01330, Adana, Turkey\\
$^3$Department of Physics, Ortadoğu Teknik Üniversitesi, 06800, Ankara, Turkey\\
$^4$Remote Sensing and Geographic Information System, University of Cukurova, 01330, Adana, Turkey\\
$^5$Astrophysics Research and Application Center (ATASAM), Atatürk University, 25240, Erzurum, Turkey \\
$^6$Department of Astronomy and Astrophysics, Science Faculty, Atatürk University, 25240, Erzurum, Turkey
}

\date{Accepted 2020 January 20. Received 2020 January 20; in original form 2019 December 05}
\pubyear{2020}

\begin{document}
\label{firstpage}
\pagerange{\pageref{firstpage}--\pageref{lastpage}}
\maketitle

\begin{abstract}
A global site selection for astronomy was performed with 1 km spatial resolution ($\sim$ 1 Giga pixel in size) using long term and up-to-date datasets to classify the entire terrestrial surface on the Earth.
Satellite instruments are used to get the following datasets of Geographical Information System (GIS) layers:
Cloud Coverage, Digital Elevation Model, Artificial Light, Precipitable Water Vapor, Aerosol Optical Depth, Wind Speed and Land Use -- Land Cover.
A Multi Criteria Decision Analysis (MCDA) technique is applied to these datasets creating four different series where each layer will have a specific weight.
We introduce for the first time a ``Suitability Index for Astronomical Sites'' namely, SIAS.
This index can be used to find suitable locations and to compare different sites or observatories.
Mid-western Andes in South America and Tibetan Plateau in west China were found to be the best in all SIAS Series.
Considering all the series, less than 3 \% of all terrestrial surfaces are found to be the best regions to establish an astronomical observatory.
In addition to this, only approximately 10 \% of all current observatories are located in good locations in all SIAS series.
Amateurs, institutions or countries aiming to construct an observatory could create a short-list of potential site locations using layout of SIAS values for each country without spending time and budget.
The outcomes and datasets of this study has been made available through a web site, namely ``Astro GIS Database'' on \texttt{\url{www.astrogis.org}}.
\end{abstract}


\section{Introduction}
\label{sec:Introduction}
Constructing an astronomical observatory hosting a telescope is an engineering application with cost-efficiency in mind.
However, the main reason to have a protected telescope is to collect and focus as much photons as possible without being disturbed by the atmosphere or light sources not originated from space.
Therefore, astronomers have been trying to find the best locations so that their investments on ``telescope time'' could become cost--efficient when number of clear nights goes above 90 \% annually, corresponding almost eleven months.
At the moment, the best observatory sites are highly telescope populated.
However, they are not many in number.
So, while Thirty Meter Telescope (TMT -- 30 m) and European Extremely Large Telescope (E-ELT -- 39 m) are being built on these kind of sites, the future of ground-based astronomy is based on finding more of the best of the best sites \citep{sarazin2006, TMT2009}.
On other hand, demand for small-- to medium--sized dedicated telescopes have always been on the rise; simply because the first concern is to collect photons for each of the question asked by the astronomers.
Therefore, demand to find \textit{suitable} sites for each telescope and observatory pair will continue to be an important topic for both astronomers and engineers.

Astronomers have been walking through the path of \textit{observation} for many decades while science and technology working hand to hand to support them.
Therefore, this path has to be simplified for both observers and engineers by finding an answer to the following question:
How one chooses and/or builds an observatory with all the required constraints involved in the question to be solved?

In the last decade, an empirical relation between the cost ($C$) and size ($D$) of the telescope has been widely used:
\begin{equation}
C \sim D^{\alpha}
\end{equation}
where $D$ is the aperture of the telescope in meters.
Due to improvements in technology within the past decade, the value $\alpha$ has decreased from 2.7 to 2.0 \citep{vanbelle2004}.
Therefore, it is safe to assume that almost every institution or organization could find required budget to have their questions answered with their own instruments under an observatory constructed in a \textit{\textbf{suitable}} observatory site.

Even though there is no official list of observatories with all the mandatory properties listed, unofficial list with simple details contains more than two thousands observatories (recent number is $\sim$2122; shown in Fig. \ref{F:layers}).
In addition to this, due to the nature of astronomy, observers usually do not register their observatories or their instruments unless they are used in scientific campaigns.
Therefore, it is easy to assume that in future we would expect to have more and more observatories being built hosting telescopes with varying sizes.
However, even though all is good for the technologies used in observatories, statistically and practically it is not the same story for selecting \textbf{observatory sites}. In the past;
\begin{itemize}
\item Observers as well as engineers, didn't pay much attention to the importance of the site selection (see Section \ref{S:conc}). Note that even in the early years of now mid--sized telescope constructions site testings were not very detailed.
\item Atmospheric conditions weren't measured, recorded or cataloged as much as needed.
\item Due to technological limitations small-sized telescopes were preferred over costly large telescopes and they were allocated mostly close to the institution.
\end{itemize}

Therefore, when site selection is involved in today's ecology of astronomy and engineering, one needs to not spend budget and time over and over again to find just a single \textbf{suitable} site for all the constraints involved.

This work attempts to solve the overhead of ``site selection'' once for all globally by using the whole land surface of Earth; for amateurs, astronomers, institutions, governments, in general for anybody who wants to have an idea of the site before starting up a project.

\subsection{Site Selection}
Site selection procedure is complex due to its nature:
Data from different sources are combined and analyzed together.
Tools and techniques of Geographic Information System (GIS) have already provided many solutions to these kind of problems where it offers cost and time efficient solutions to decision makers.
Therefore, GIS techniques in site selection is widely used in many fields: 
hospitals, solar and wind farmsteads, urban solid waste plants, as well as observatories \citep{nas2009,soltani2011,uyan2013,noorollahi2016}.

In the case of an astronomical observatory site, another important parameter would be the major wavelengths aimed in the scientific rationale of the observatory.
Thus, when observations on longer wavelengths, especially in infrared, are aimed then observers expect to have lower water vapor content in the atmosphere. 
This can be achieved if PWV measurements are involved in the selection process \HL{\citep{2019PASP..131d5001O,kucuk2012}.}
For radio astronomy, RFI (Radio Frequency Interference) should also be included as a layer as well as PWV \citep{umar2014}.
Note also that a recent site testing study by \citet{2019PASP..131d5001O} included PWV, Temperature and Wind data from radiosonde measurements for a limited time span and for a localized region centered mainly on observatory sites. They summarize that PWV and Temperature are very important in site selection for radio astronomy however, wind vertical profile has to be obtained for the final decision.

Another recent site selection exercise was carried out for Iran’s 3.4m optical telescope \citep{nasiri2019}.
They have used available long--term meteorological and geographical parameters.
Similarly, in Pakistan \citet{daniyal2019} conducted a study to identify potential optical telescope sites by using ``Analytical Hierarchy Process (AHP) with a Multi Criteria Decision Analysis (MCDA)'' technique. 

The most comprehensive work on the global scale site selection was carried out for the 39 meter ``{\it European Extremely Large Telescope} (E-ELT)'': FriOWL \citep{sarazin2006}.
They have used satellite data including various climatological parameters.
FriOWL used GIS techniques by including several astro--climatological layers together: monthly mean temperature, outgoing longwave radiation, water vapor, cloud coverage, wind, aerosol index \citep{graham2005}.
It's spatial resolution was around $2.5\times2.5\degr$ (approximately $300\times300$ km$^{2}$).
It contains several climatic layers with a time coverage of approximately 15 years.
Since many parameters with many different data types (satellite data, processed ground--based data etc.) were involved together, it was found to be useful even though it has a limited spatial resolution with limited temporal resolution up to year 2002.
However, at the end, it was successfully used for finding the best site for the E-ELT.

Another recent study by \citet{hellemeier2019} which uses again FriOWL on site testing compiled astro-climatological data from 15 cites distributed all around the world for a period of 45 years.
They conclude that having a long-term dataset combined with seasonal variation could elevate the importance of the astronomical observatory sites.

Continuous improvements on meteorological models result in lower spatial resolution of the atmospheric layers.
One of the latest studies in these kind of modeling is done by \citet{2018MNRAS.480.1278O}.
They applied ``the European Centre for Medium range Weather Forecasts (ECMWF)'' model successfully in astronomical site testing.

In Table \ref{T:EarlyWork}, a summary of earlier astronomical site selection studies by different groups are given along with GIS layers used.
In all these works, two layers found to be common:
Cloud Coverage and Altitude.
We also include Artificial Light (AL) as an important layer because the light pollution in operational observatories becomes more and more invasive.
\section{Datasets}
\label{sec:Datasets}
When constructing a data collection for GIS analysis it is always preferred the layer to cover a long time series. This can easily be achieved when polar orbiting satellites monitor the Earth from space to surface. Since their lifetimes are long enough, they will naturally produce long, continuous and consistent databases. Typical date range for such the Earth observing satellites are around 5 years.

In this study, we have collected the following datasets from different missions and instruments on board these satellites. Links, websites, resources or repositories of dataset we downloaded are listed in Table \ref{T:source}. Downloading the layers are automated using an in-house coded python script.

Note that the Antarctica continent has been excluded from the analysis because the continent does not have datasets for all layers; it has only CC, DEM and PWV. However, due to the continent's special location, these datasets have to be studied separately. In this continent, CC is expected to be very small on average.

In all datasets (except AOD) not all time coverage was used, only twilight and nighttime were selected. The reason to ignore the daytime is simply because we aim to find ``clear skies'' during the night (optical telescopes work during nighttime). However, we have to also include twilight not to make discrete time span in the dataset.

\subsection{Cloud Coverage - CC}
Cloud coverage or cloudiness is defined as the percentage of sky covered with clouds at the desired geographical coordinate \citep{glickman2000}. Note that astronomical observations rely on clear skies having almost no clouds in the sky. Therefore, CC is the first and the most important layer of astronomical site selection; see discussions in \cite{graham2005, sarazin2006, varela2008, aksaker2015}. This fact can also be deduced from Table \ref{T:EarlyWork}. CC can be recorded with many different methods: by synoptic naked-eye observations \citep{lau1997, breon1999, norris1999}, by using lidar and radar methods \citep{Intrieri2002}, by using an “All Sky Camera” \citep{ackerman1981, pfister2003, long2006} or using satellite monitoring \citep{ventkappa2019}.

Since satellite data have good temporal and spatial resolution in determining cloud coverage we preferred it over the other methods. MODIS (Moderate Resolution Imaging Spectroradiometer) instrument has MOD35\_L2 \& MYD35\_L2 cloud coverage products that contain “Cloud Mask” layer. Global astronomical site selection should also take into account geographical conditions, climatic conditions and orography as well as the cloud coverage. Therefore atmospheric datasets should be sensitive to certain extreme conditions such as desert, snow, ice and altitude etc. \citep{ackerman1998,frey2008}.

The MODIS instrument provides 288 HDF4 files per day except for AOD. MODIS data is in HDF4 format (5-minute granule size for swath 2330 x 1354 km). An open source python library was used to convert them into a GIS suitable format. Conversion into GEOTIFF format according to WGS84 DATUM criteria was handled by two python functions (warp \& translate) of GDAL (Geospatial Data Abstraction Library Virtual Format) library.

\begin{figure*}
    \centering
    \begin{tabular}{@{}c@{}c@{}}
    \includegraphics[height=5.5cm]{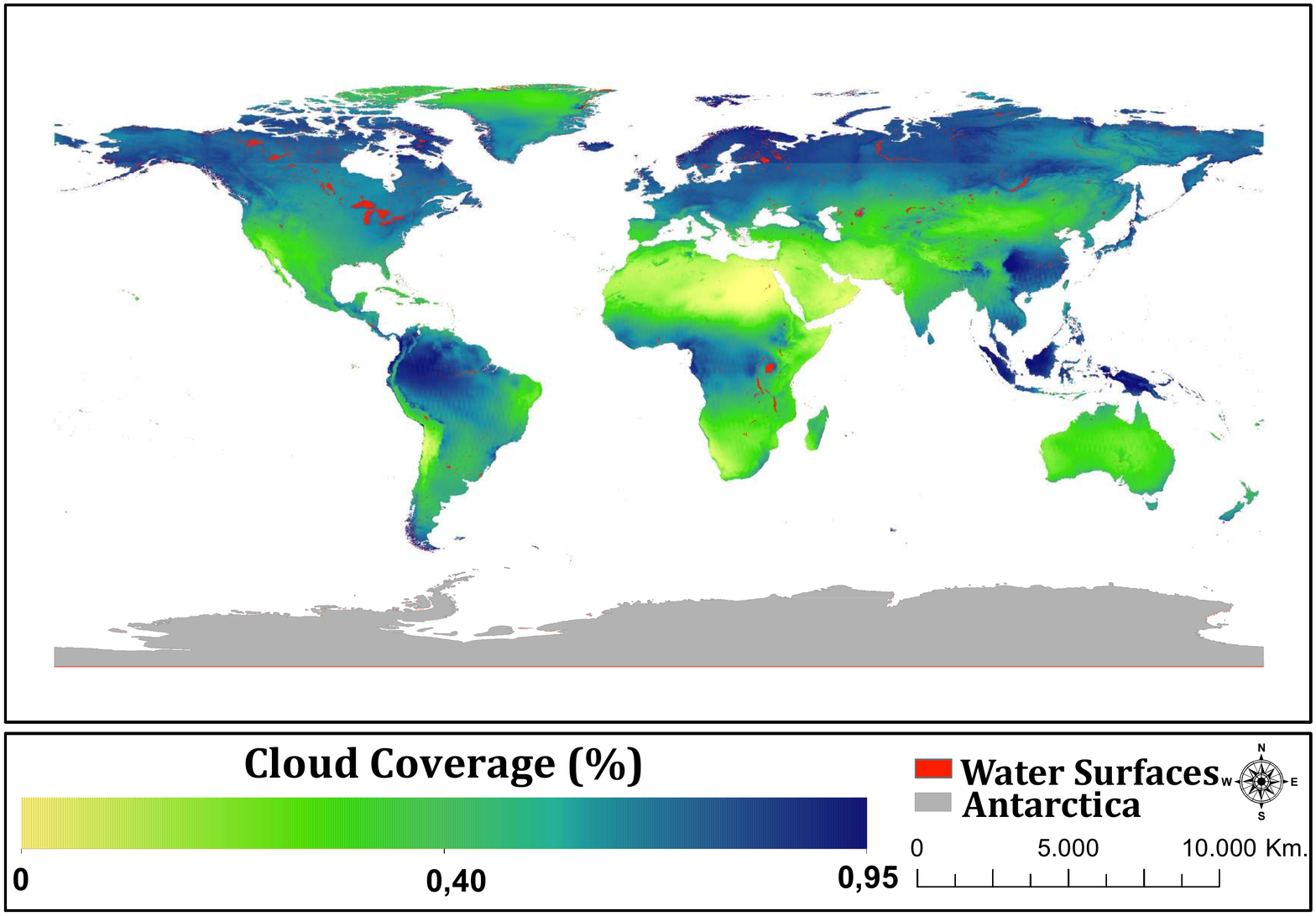}       &
    \includegraphics[height=5.5cm]{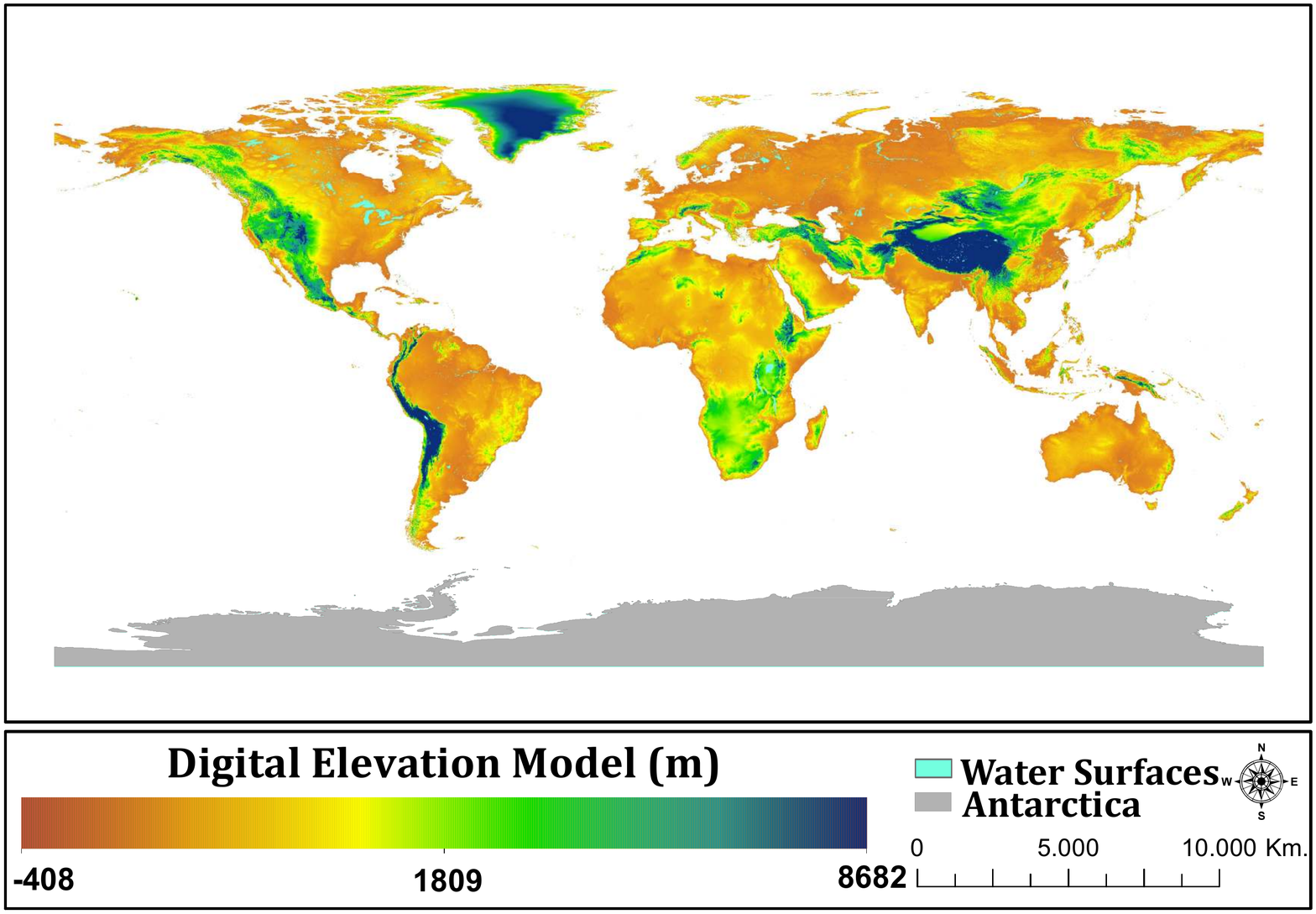}      \\
    \includegraphics[height=5.5cm]{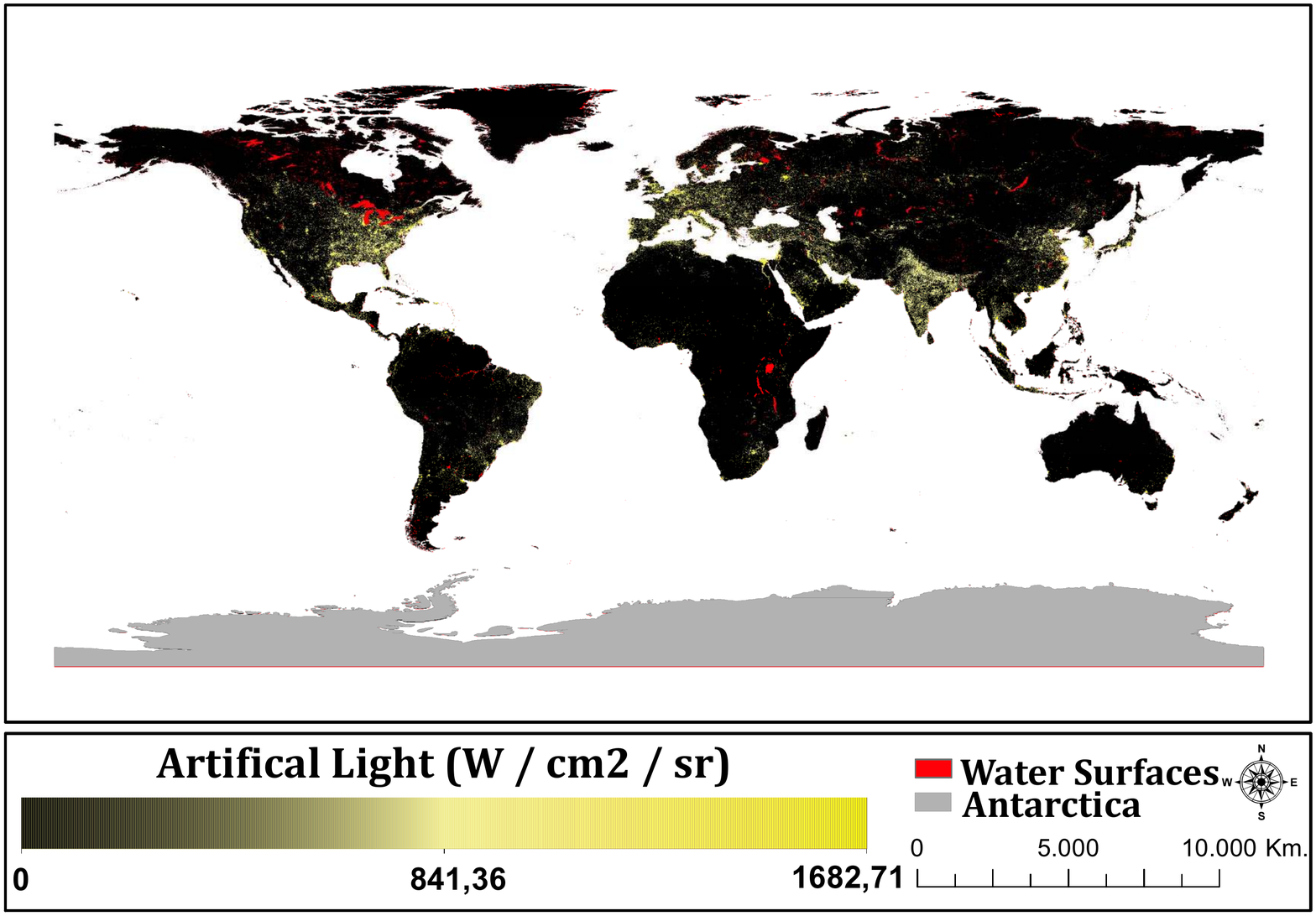}    &
    \includegraphics[height=5.5cm]{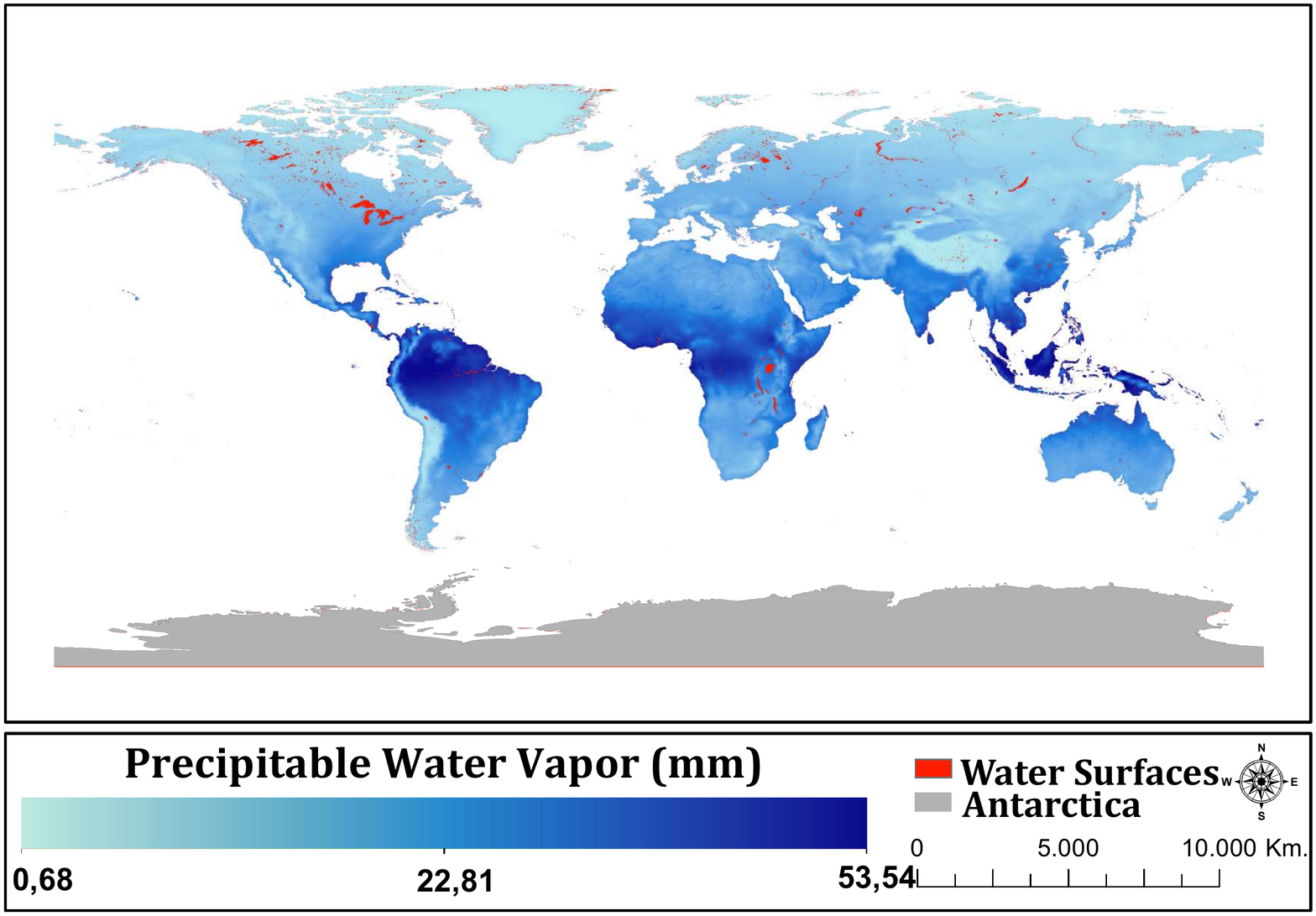}      \\
    \includegraphics[height=5.5cm]{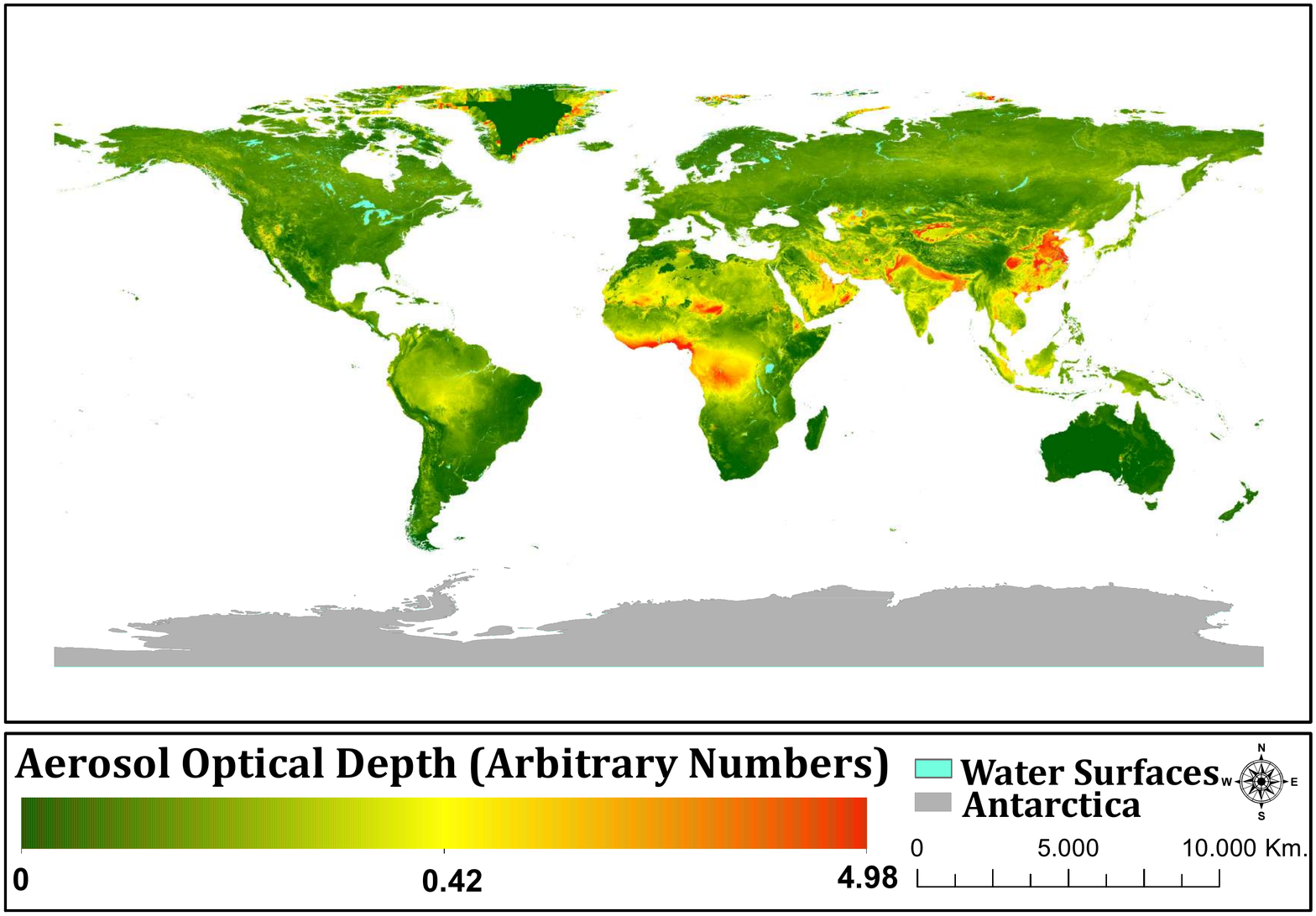}      &
    \includegraphics[height=5.5cm]{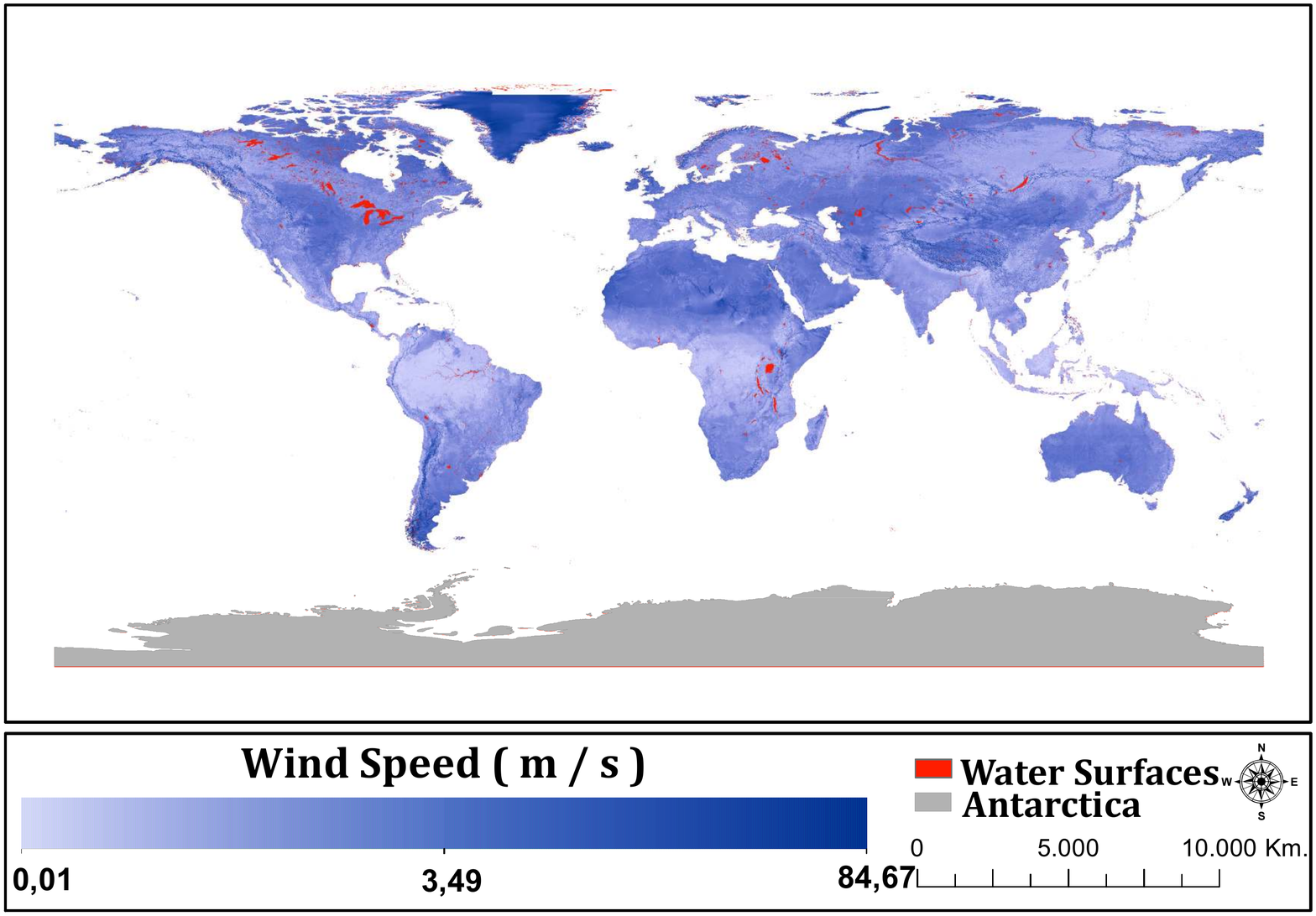}     \\
    \includegraphics[height=5.5cm]{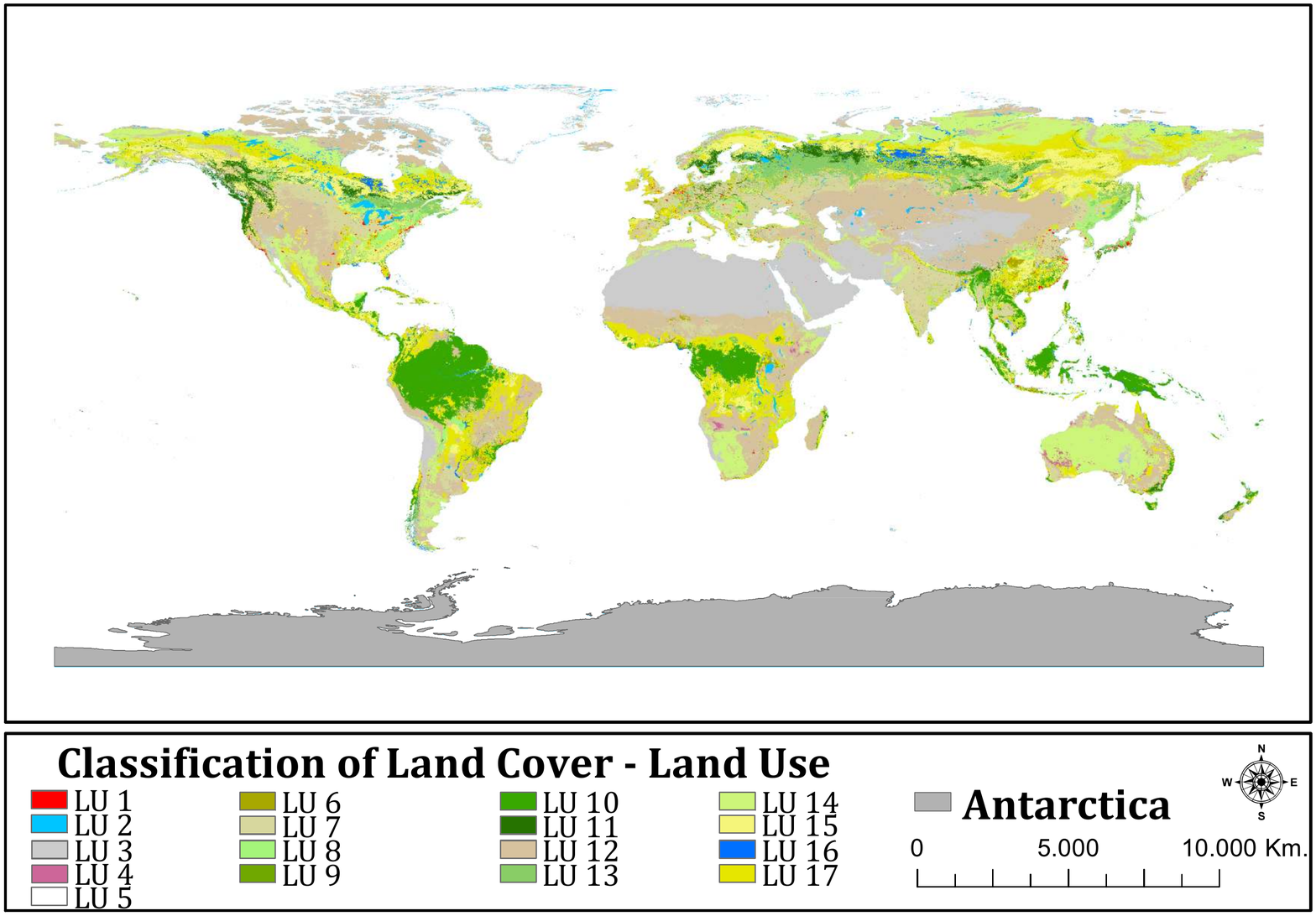}     &
    \includegraphics[height=5.5cm]{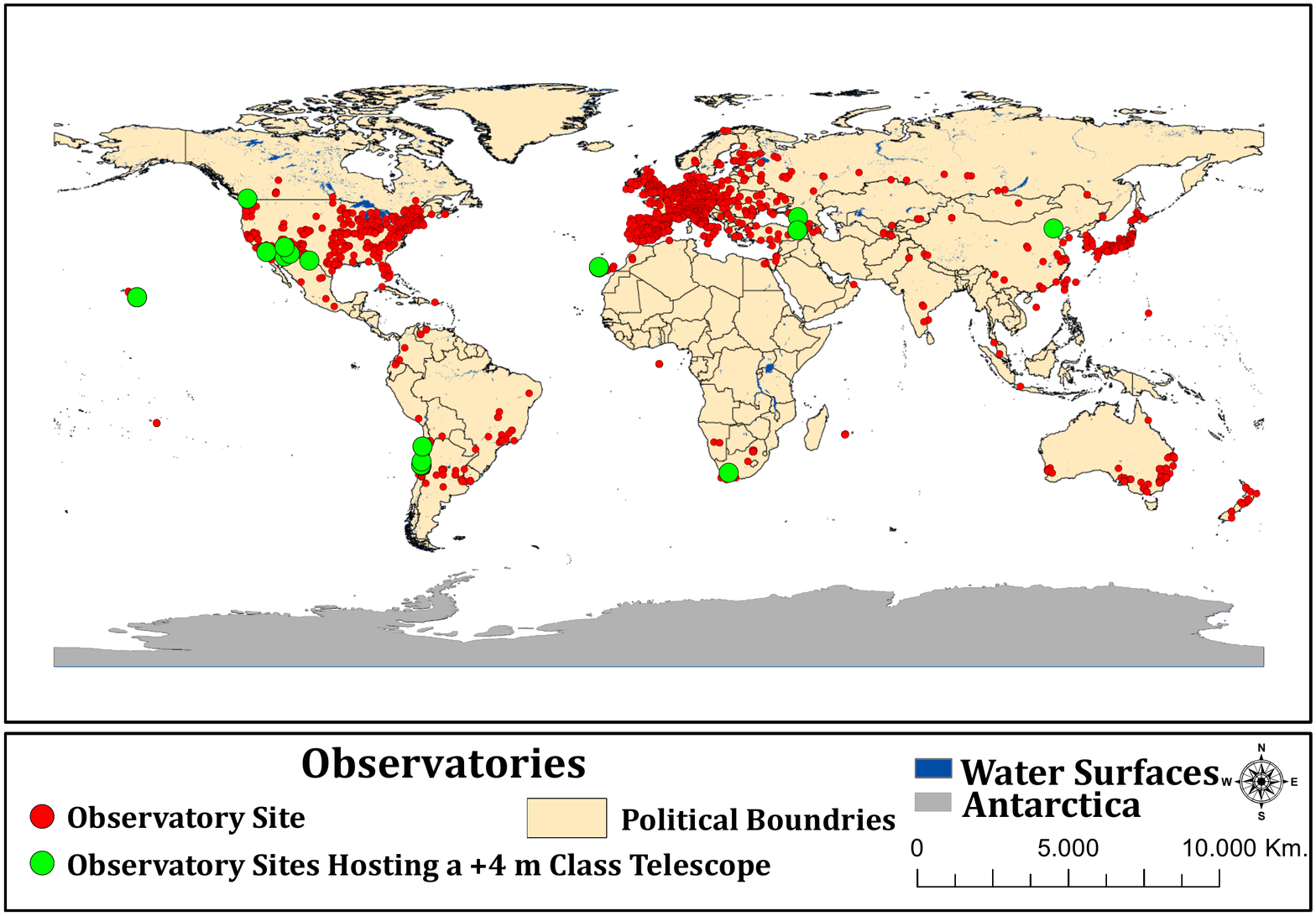}       \\
    \end{tabular}
\caption{%
    The GIS layers used in this study (from top to bottom):
    1st panel: CC (left), DEM (right);
    2nd panel: AL (left), PWV (right);
    3rd panel: AOD (left), WIND (right);
    4th panel: LULC (left), Observatory location (right).
    Abbreviations in LULC layer are enumerated from table 3 of \citet{sullamenashe}.
    See Section \ref{sec:Datasets} for details.%
}
\label{F:layers}
\end{figure*}

The “Cloud Mask” data of MODIS consist of 48-bits per pixel information. Only the first eight bits are extracted and used in classifying the clear sky and the cloud detection \citep{platnick2003}. The last bit of this extracted bit block determines whether cloud detection algorithm was used or not. The sixth and the seventh bit of the block are used to introduce an index for CC. Since bit-wise operation requires a lot of CPU power, parallel programming has been implemented in some parts of the code. Using this index a daily average has been calculated at each pixel. The long-term daily mean CC average obtained by combining datasets of both Aqua and Terra satellites. The resultant CC map is shown in Fig. \ref{F:layers}.

Cloud dynamics show instantaneous variations that are mainly due to temperature, wind speed, wind direction and orography. However, on the global scale cloud circulations are due to Earth’s shape, Earth’s rotation speed and direction, and dependence on the latitude. This global variation can also be marked as northeasterly and southeasterly trade winds on Fig. \ref{F:layers}.

Examination of CC reveals the following points on the global scale: a) Clear night sky (i.e having less cloud coverage during night) counts are higher for west of South and North America, north and northwest Africa and Australia than the rest of the world; b) Some of localized counts are due to climatological and geographical conditions; c) For west of South America, South Africa and Australia, clear night sky counts are due to the latitude effect; d) Similarly, for northern hemisphere latitude effect is generally less pronounced.

\subsection{Digital Elevation Model - DEM}
Elevation (Altitude) is also another important criterion in astronomical site selection.
As the altitude of the site gets higher the site becomes less affected from atmospheric events such as clouds and aerosols,
and therefore have potentially good seeing values \citep{aksaker2015}. Thus the best place will be above the atmosphere. DEM is a 3D model for visualizing coordinate components of any topography and the differences of the altitudes. The DEM datasets are stored as raster format digital images. The reference surface for any altitude point is generally taken as the average sea level.

There exists many different resolutions for DEM, ranging from 30 m to 1 km. GTOPO30 with 1 km resolution DEM is used in this work. It has a horizontal grid spacing of 30$\arcsec$ (corresponding approximately to 1 km) and vertical units represent altitude in meters above mean sea level. It covers latitudes from 90$\degr$ north to 90$\degr$ south, and longitudes between 180$\degr$ west to 180$\degr$ east. The DEM data set has 33 smaller pieces, or tiles in GEOTIFF format. The layer is given in Fig. \ref{F:layers}.

\subsection{Artificial Light - AL}
Humanity is perturbed the world surface that can be seen from space at night. The light created artificially increases the sky brightness, which is now defined as “light pollution”. In a light polluted region on the surface, the background of the sky increases and therefore number of stars observed decreases exponentially. At the moment, one-third of humanity cannot see the Milky Way. Moreover 80 \% of the world live in light polluted regions \citep{falchi2016}.

Astronomers look for sites having skies that are not polluted with artificial light. Thus, nighttime image of Earth’s surface will reveal “dark” regions as locations producing potentially less amount of artificial light. Therefore, collected nighttime images will be the AL dataset.

The nighttime data are from the Visible Infrared Imaging Radiometer Suite (VIIRS)-Day Night Band (DNB) where DNB is in visible band. The radiometric resolution of dataset is up to 14 bits in visible to red wavelengths ($0.5-0.9 \mu$m) with radiation counts around $2 \times 10^{-9}$ W cm$^{-2}$ sr$^{-1}$ \citep{nurbandi2016}.

The dataset covers the world from 75 N to 65 S latitudes with 15$\arcsec$ grid size in a 6--cell set GEOTIFF format with a spatial resolution of 742 m per pixel. Nighttime images are produced daily. However, \citet{elvidge2017} filtered out non-artificial light sources (eg. lightning, fishing boats, clouds) and they created a database of images (monthly and yearly) on the website of the dataset. The dataset is ready to be used in GIS. Two different types are used: 1) extension ``vcm-orm-ntl'' (version in which VIIRS Cloud Mask, Outlier Removed and Nighttime Lights); 2) extension “Avg\_rade9” (brightness normalized with $10^{-9}$). The layer is given in Fig. \ref{F:layers}.

\subsection{Precipitable Water Vapor - PWV}
PWV products of MODIS contain two different algorithms:
a) near infrared (daytime with $1\time 1$ km spatial resolution);
b) infrared (both day and nighttime with $5 \times 5$ km$^{2}$ spatial resolution).
Since astronomical site selection concerns nighttime sky conditions, infrared algorithms are preferred.
Note that MODIS atmospheric profile product (MOD07\_L2 \& MYD07\_L2) consists also other atmospheric datasets such as ozone, temperature profile, atmospheric stability and moisture profile.

The dataset contains 5 minutes swath images at 6.1 version (C6.1) in HDF4 format with 16--bit integer data type. The storage size is 288 files per day. Some days are missing in the dataset, which amounts to a maximum of 3 days per year. PWV dataset has three subsets for the whole atmosphere: low, high, total. In this study, we only use the last one. Values of dataset are in cm unit and spatial resolution is 5 km.

As with the other MODIS products, PWV is converted to WGS84 datum first, and then a mosaic is created later. They are stored in GEOTIFF format for further analyses in GIS. Using both satellites (Aqua and Terra), a mean PWV image was created for only nighttime. Note that MODIS generates PWV values only for unclouded pixels. The layer is given in Fig. \ref{F:layers}.

\subsection{Aerosol Optical Depth - AOD}
AOD is defined as the total optical extinction in a column of atmosphere normally interpolated to 550 nm \citep{sayer2014}. AOD is directly related to the transparency of the sky. Two major entities, clouds and aerosols, degrades the transparency \citep{varela2008}. Therefore, a lower annual AOD level will be good for an astronomical site.

In this study, MODIS instrument on both Aqua and Terra satellites were used. AOD products are produced using different algorithms for different contents, namely absence of cloud, snow and ice cover over land (using Deep Blue and Dark Target algorithms) and ocean (using Dark Target algorithm). The spatial resolution of AOD datasets are $10\times 10$ km and $3\times 3$ km for MOD04\_L2 and MOD04\_3K \& MYD04\_3K, respectively. Both datasets were used to have a complete coverage.

MODIS delivers 288 HDF files per day. However, since AOD information is not produced at night, only $\sim$ 144 files are used. This is also applied to the AOD dataset with ``Cloud Coverage'' where AOD information is obtained from the ``Land and Ocean'' Science Data Set (SDS). All data were merged according to WGS84 DATUM criteria and then converted to GEOTIFF format. The layer is given in Fig. \ref{F:layers}.

\subsection{Wind Speed - WS}
High wind speed increases kinematics of atmospheric particles, decreasing the stability of the atmosphere.
It also indirectly affects construction cost of the observatory dome.
Therefore, annual lower wind speed is good for an astronomical site
However, WS is a meteorological parameter that changes very rapidly from point to point.
Note also that since wind causes atmospheric turbulence it is directly correlated with astronomical seeing \citep{Liu2010}.  

WS dataset used in this study acquired from the World Bank supported project called ``Global Wind Atlas''. WS was produced for a height of 10 m above the surface together with DEM and then reproduced with a resolution of 225 m. The dataset were combined with the mosaic method. The layer is given in Fig. \ref{F:layers}.

\subsection{Land Use and Land Cover - LULC}
In an optimal astronomical site selection, selection criteria should also include the following surface details: the land use, land cover, water bodies, forest, agriculture land and barren lands.

The MODIS Land Cover Type Product (MCD12Q1) provides a series of science data sets (SDSs) that map global land cover at 5 km spatial resolution at annual time step for six land cover legends \citep{sullamenashe}. The MCD12Q1 product is created using supervised classification of MODIS reflectance data \citep{friedl2002, friedl2010}.
International Geosphere--Biosphere Programme (IGBL) legend and classification has also been adopted for this layer (table 3 of \citealp{sullamenashe}).

\subsection{Other supplementary datasets}
The country border data%
\footnote{\url{http://thematicmapping.org/downloads/world_borders.php}}
is a vector format layer containing 247 regions as countries.

In total 2115 observatory locations%
\footnote{\url{https://www.projectpluto.com/obsc.htm}}
has been added.
However, whenever available, they have been improved using observatory's official website.
\section{Geographic Information System (GIS) based Multi-Criteria Decision Analysis (MCDA)}
\label{sec:mcda}
In selecting an astronomical site, the analyst aims to determine an optimum location that would satisfy the selection criteria.
Deciding on an appropriate astronomical site is based on numerous datasets (aka. layers) collected according to criteria specific to astronomical sites.
In deciding for a site to accommodate an observatory, the selection typically involves the evaluation of multiple criteria according to several, often conflicting, objectives \citep{rikalovic2014}.
This can be dealt with a key framework of GIS based MCDA which is considered to be an important spatial analysis method in the decision-making process that allows information derived from different sources to be combined \citep{feizizadeh2014a}.

GIS is designed as a system to capture, store, analyze, model the spatially referenced data to produce solutions to complex problems in many different fields \citep{garcia2013}.
The MCDA technique along with GIS can assist to categorize, examine and appropriately organize the collected information for spatial based selection of astronomical sites.

In this study an MCDA method named as Simple Additive Weighting (SAW) by \citet{churchman1954} was used to assess astronomical site selection process.
Since SAW uses weighted sums, it is accepted as one of the most applicable methods to deal with MCDA, especially under GIS environment \citep{tzeng2011}.
Therefore, we have also used weighted criteria all along the process.
Then, the weight of each criteria layer was multiplied with the pixel based importance score of the criteria.
Therefore, the Suitability Index for an Astronomical Site (SIAS) can then be calculated using the following equation:
\begin{equation}
SIAS_{i} = \sum_{j} W_{j} X_{ij}
\end{equation}
SIAS$_{i}$ is the index for pixel $i$; $W_{j}$ is the relative weight for $j^{\text{th}}$ criteria; $X_{ij}$ is the importance score of pixel $i$ for the $j^{\text{th}}$ criteria.

MCDA framework of this study consists of three main stages:
\begin{enumerate}
\item Spatial regulations:
All criteria layers were transformed into target spatial resolution of 1 km.
CC and AL layers were processed using ``nearest neighbor resample method'' to decrease their original spatial resolutions of 3 km and 742 m to 1 km, respectively.

\item Standardization:
Multiple criteria of SIAS cannot be compared or weighted without transforming each layer into a standardized, common metric.
Fuzzy sets have been applied using sigmoidal function in order to standardize criterion layers by assigning a degree of membership to each pixel of the criteria using an asymptotic scale from 0 to 1 \citep{feizizadeh2014b,gorsevski2010,jiang2000}.

\item Aggregation:
In selecting suitable sites for an astronomical observatory, different types of datasets are combined with different types of criteria where at the end suitable locations can be subsetted from the global SIAS.

Overlay analysis is a spatial analysis technique used to merge and sum-up all layers into a single layer by rescaling each layer to a common 0--1 scale.
The overlay analysis within MCDA is called ``Weighted Averaging Overlay'' which is the sum of the standardized layers divided by the total weight.
Therefore, overlay analysis is applied in our SIAS study.
\end{enumerate}

\begin{figure*}
    \centering
	\includegraphics[width=\textwidth]{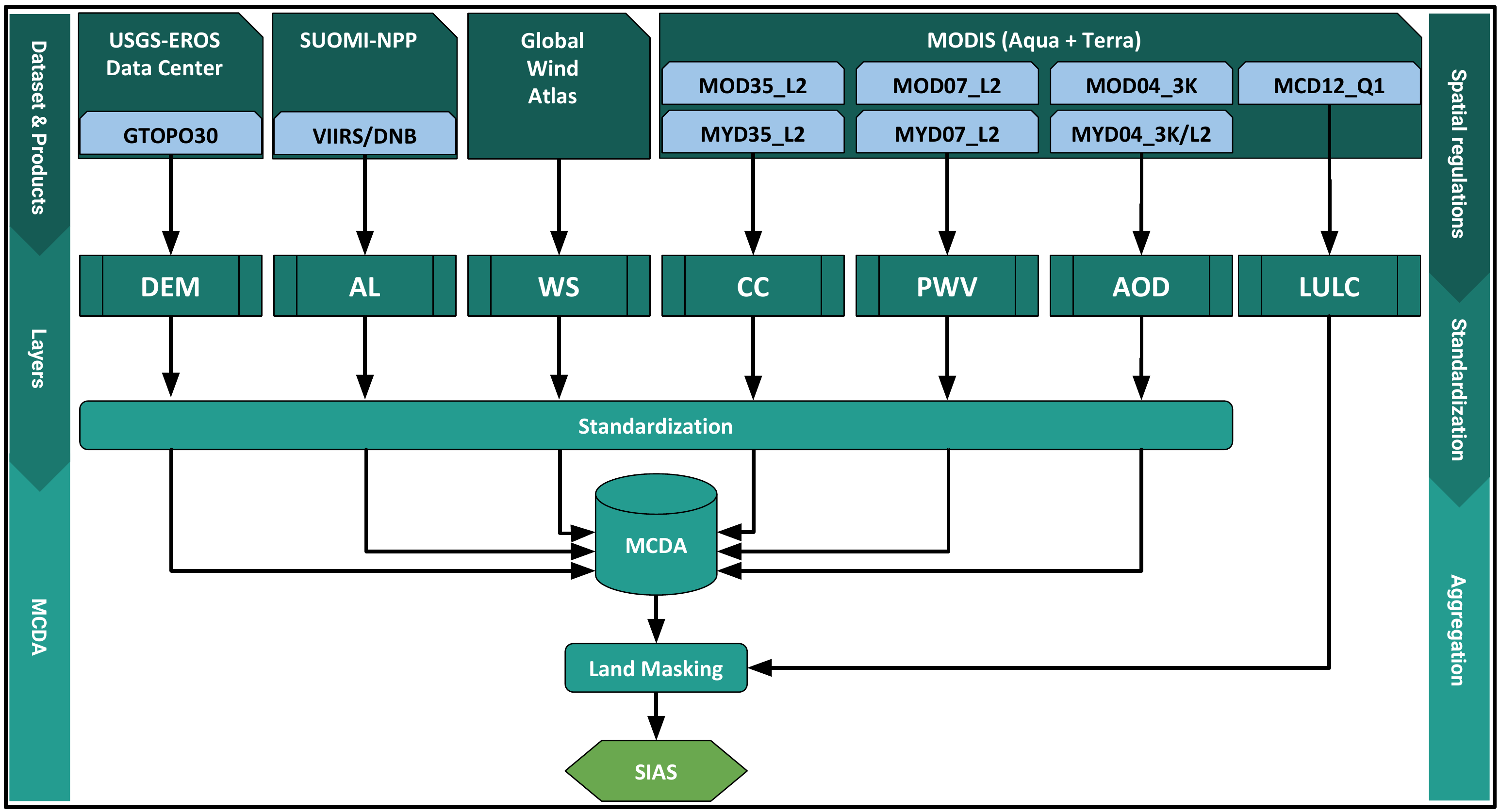}
	\caption{%
	The pipeline of the site selection process.
	See Section \ref{sec:mcda} for details.%
	}
	\label{F:pipeline}
\end{figure*}

These processing steps were applied to all layers, except LULC.
The pipeline of this analysis is given in Fig. \ref{F:pipeline}.
A summary is given in Table \ref{T:summary} where some additional notes are listed below:
\begin{itemize}
\item Values of DEM dataset for global land areas were between -4,941 and 8,685 m.
Values less than zero in terrestrial area indicate errors due to the production process of the data.
These incorrect values were rearranged according to the height of the Dead Sea, Israel (-408 m. It is the deepest in the terrestrial area of the world).

\item Values of AL dataset ranges (0--1,683) W cm$^{-2}$ sr$^{-1}$.
The highest value of AL was recorded as 452 W cm$^{-2}$ sr$^{-1}$ in Las Vegas (USA).
The dataset was skewed and it is corrected by using a logarithmic scale.
Note that since illumination varies by the inverse square law, logarithm of AL is used before it is fed into SFL membership function.
On this new scale, the average value of dataset became -1.24.

\item Ocean and sea area were excluded from all dataset leaving only terrestrial surface area. 

\item Values of AOD dataset for Sahara and Central Asian deserts were not covered in 3 km resolution.
Therefore, these regions are patched from 10 km resolution by resampling each pixel to 3 km.
\end{itemize}

\subsection{SIAS Series}
A simple methodology is aimed in applying the MCDA analysis.
SIAS series created and layer weights are given in Table \ref{T:layers}.
In all series created, oceans are masked out therefore, only terrestrial surface were included.
Moreover, the Antarctica continent has been excluded (see Section \ref{sec:mcda}).
In addition to masking, using LULC’s terrestrial identification, water surfaces within lands (e.g lakes, rivers etc.) were also excluded in SIAS Series.
\begin{description}
\item[\textbf{Series A:}]
A control series where weights of all layers are equal.
All six layers discussed above were included:
CC, DEM, AL, PWV, AOD and WS.
By doing so site’s meteorological, geographical and anthropogenic properties are considered to be equal.

\item[\textbf{Series B:}]
In this series, some of the correlated layers are excluded:
\begin{itemize}
\item PWV is highly correlated with altitude \citep{aksaker2015}.

\item Astronomical observations usually work during the night.
However, the AOD dataset has not been measured in the nighttime \citep{remer2009} and it decreases as altitude increases i.e. correlated.

\item A limited number of global WS datasets exist:
(a) The ERA5 model dataset%
\footnote{\url{https://confluence.ecmwf.int/display/CKB/ERA5}}
has a 31 km resolution therefore it cannot be used in this study.
(b) World Bank’s wind data was prepared together with DEM and reproduced with a resolution of 225 m.
Although the resolution fits very well as a dataset it is excluded as layer in this series to avoid DEM to be used twice.
\end{itemize}
Therefore, as it is done in \citet{aksaker2015}, only CC, DEM and AL are taken in this series.

\item[\textbf{Series C:}]
A variation of Series B: We increased importance of CC twice as of DEM and AL.
Having less amount of CC would mean that a site would have more observing time.
Therefore, a location having a high SIAS Series C value would be chosen as an observatory site, for example, running in queue mode or performing long--term surveys.
This Series will then be named as ``\textbf{observing time preferred}''.

\item[\textbf{Series D:}]
Another variation of Series B: This time, the importance of CC and DEM are interchanged. 
Note that, the seeing is highly correlated with the altitude \citep{Racine2005} (as the altitude increases the seeing values get better).
Therefore, a location having a high SIAS Series D value would be chosen for an observatory site, for example, working in infrared band or expecting higher resolutions in both imaging and spectroscopy.
This Series will then be named as ``\textbf{seeing preferred}''.
\end{description}

The world layout of each series along with SIAS histogram are given in Fig. \ref{F:sias}.
A full country based data access will be available as a supplement through the publisher’s repository.
\section{Results, Conclusion, Discussion and Outcomes}
\label{sec:Results}
Using a GIS/MCDA analysis of 7 different layers with an up-to-date coverage having a fixed 1 km spatial resolution for all the terrestrial area, an index named as ``Suitability Index of Astronomical Sites -- SIAS'' is introduced for the first time.
Datasets having 1 km spatial resolutions correspond to approximately 1 Giga pixels in size.
Note that since PWV, AOD and WS are correlated with CC, DEM and AL, the SIAS Series A shows higher statistical values than the Series B, C and D.
Therefore, it has to be taken as a control series.
In total, four different SIAS series are created for different site selection purposes.

Since the datasets covers several different wavelength bands, outcome of this work will also be useful for other types of observatories working for example in radio astronomy.
Statistical importance of SIAS Series can easily be gathered when SIAS values are analyzed on a Normal Distribution (3$\sigma$ rule).
For each Series the followings are given in Table \ref{T:results}:
(a) minimum, mean, maximum and $\sigma$ values;
(b) 1$\sigma$, 2$\sigma$ and 3$\sigma$ values;
(c) corresponding land surface area to these three $\sigma$ levels and their percentages;
(d) number of observatories falling into these three $\sigma$ levels and their percentages.

\subsection{Conclusions}
\label{S:conc}
\begin{figure*}
    \centering
    \begin{tabular}{@{}c@{}c@{}}
    \includegraphics[height=5.45cm]{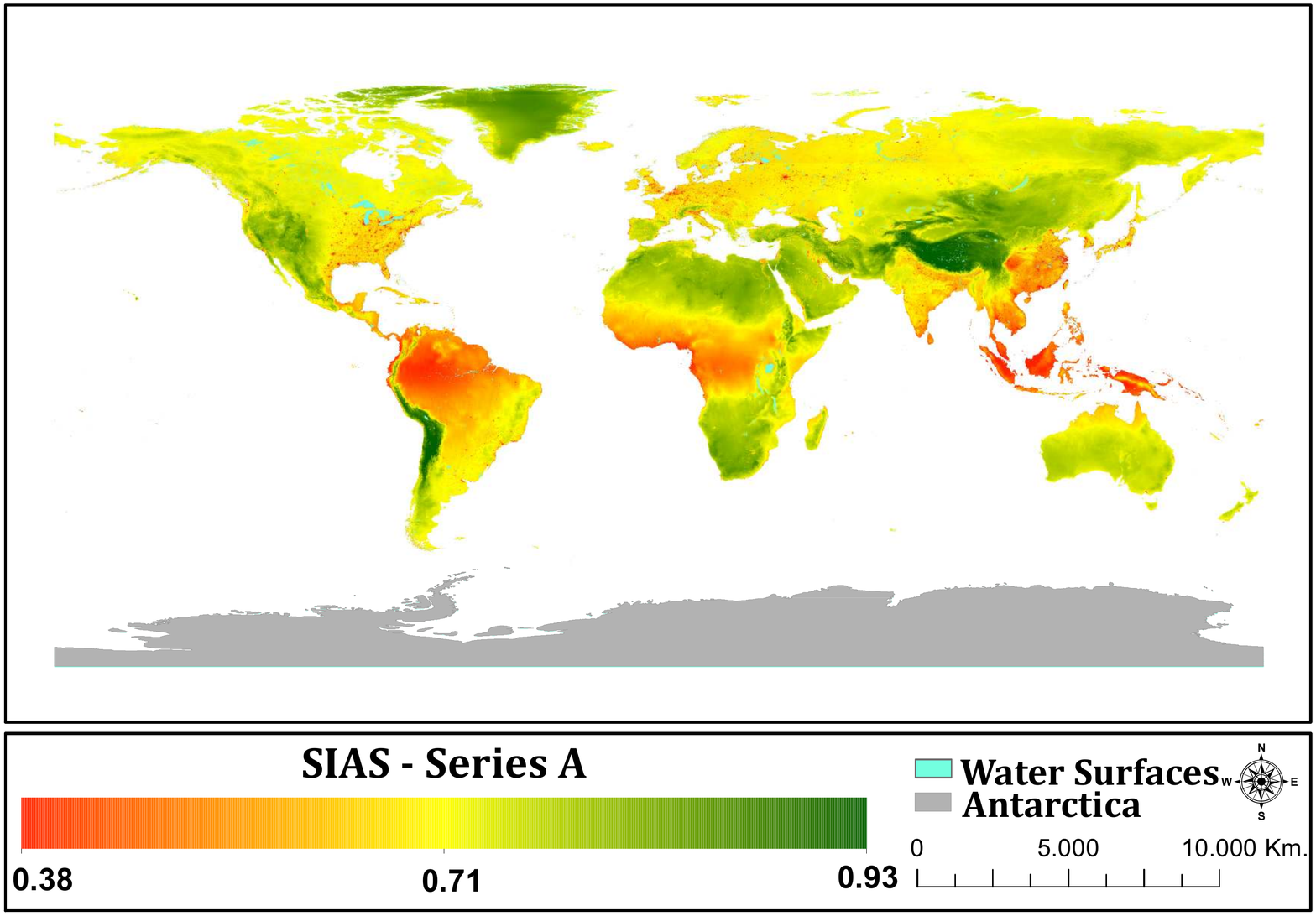}  &
    \includegraphics[height=5.45cm]{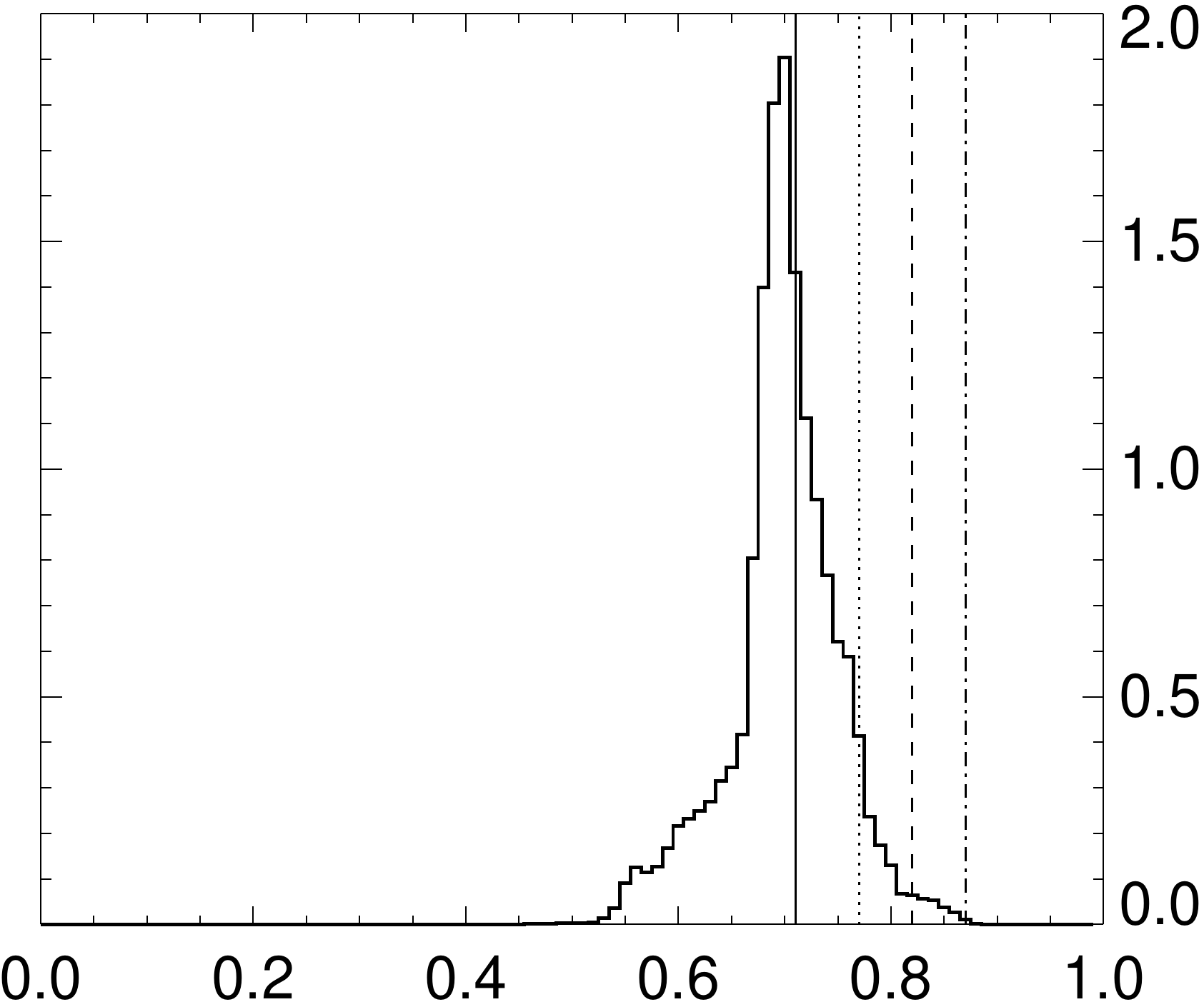}    \\
    \includegraphics[height=5.45cm]{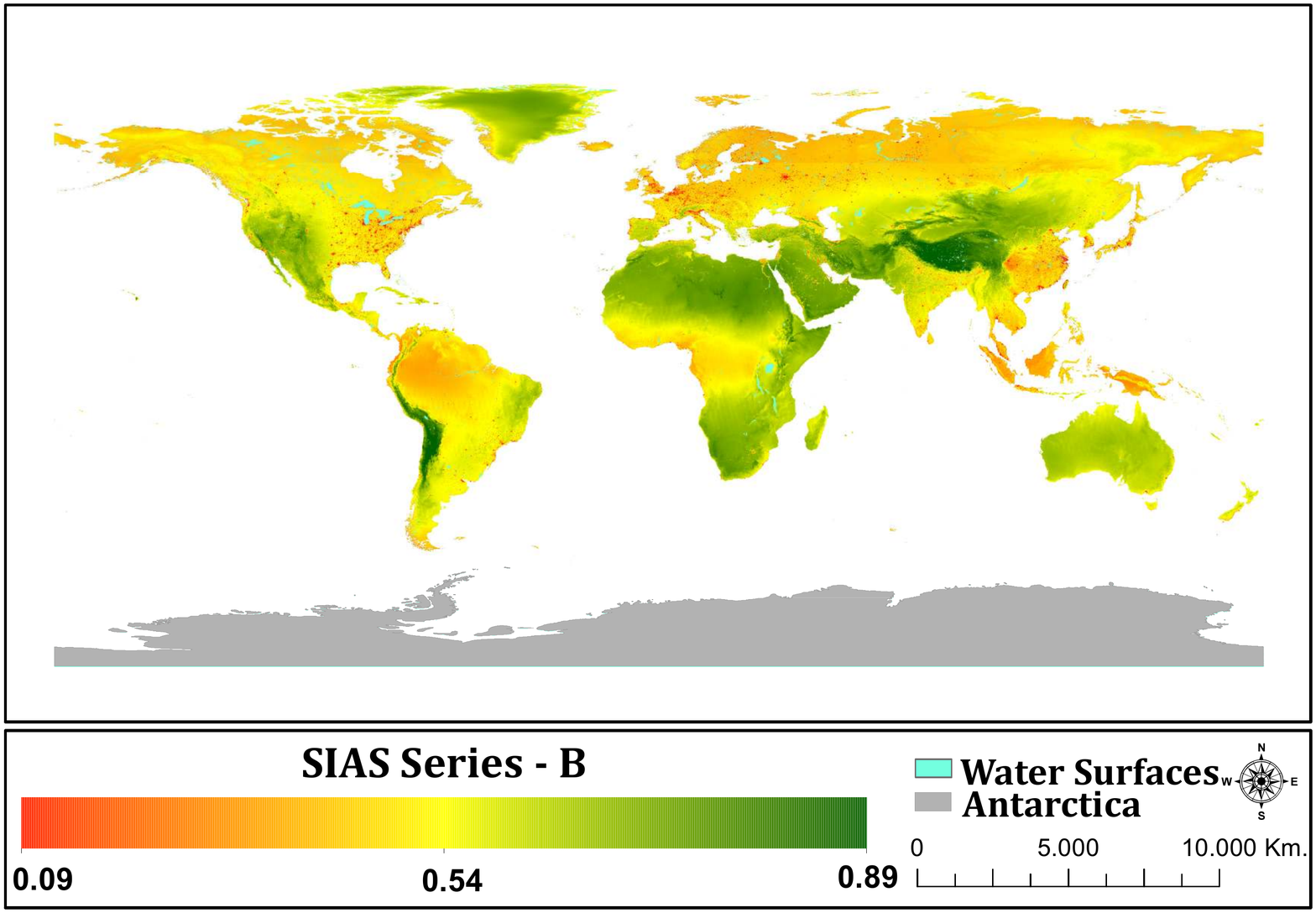}  &
    \includegraphics[height=5.45cm]{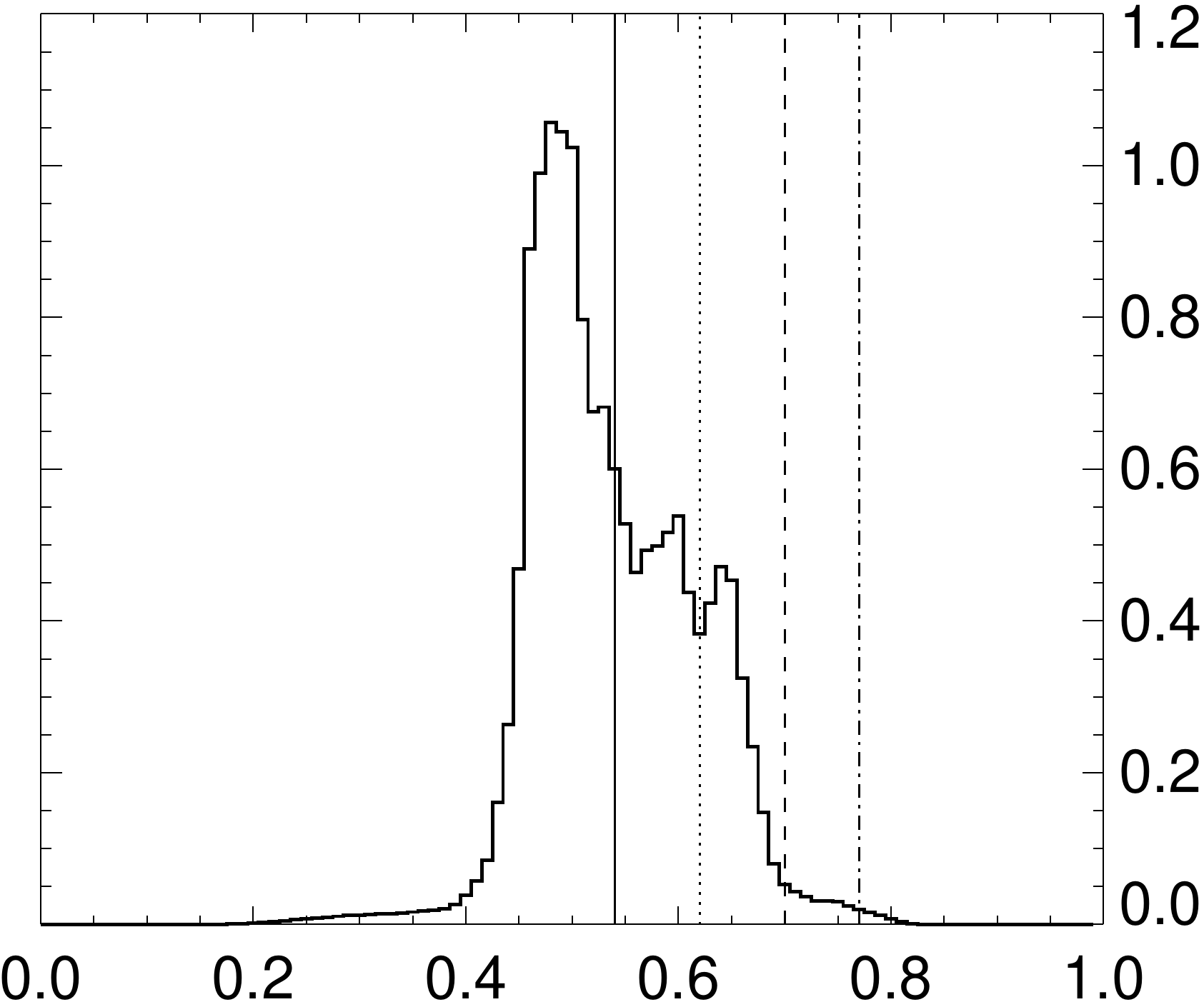}    \\
    \includegraphics[height=5.45cm]{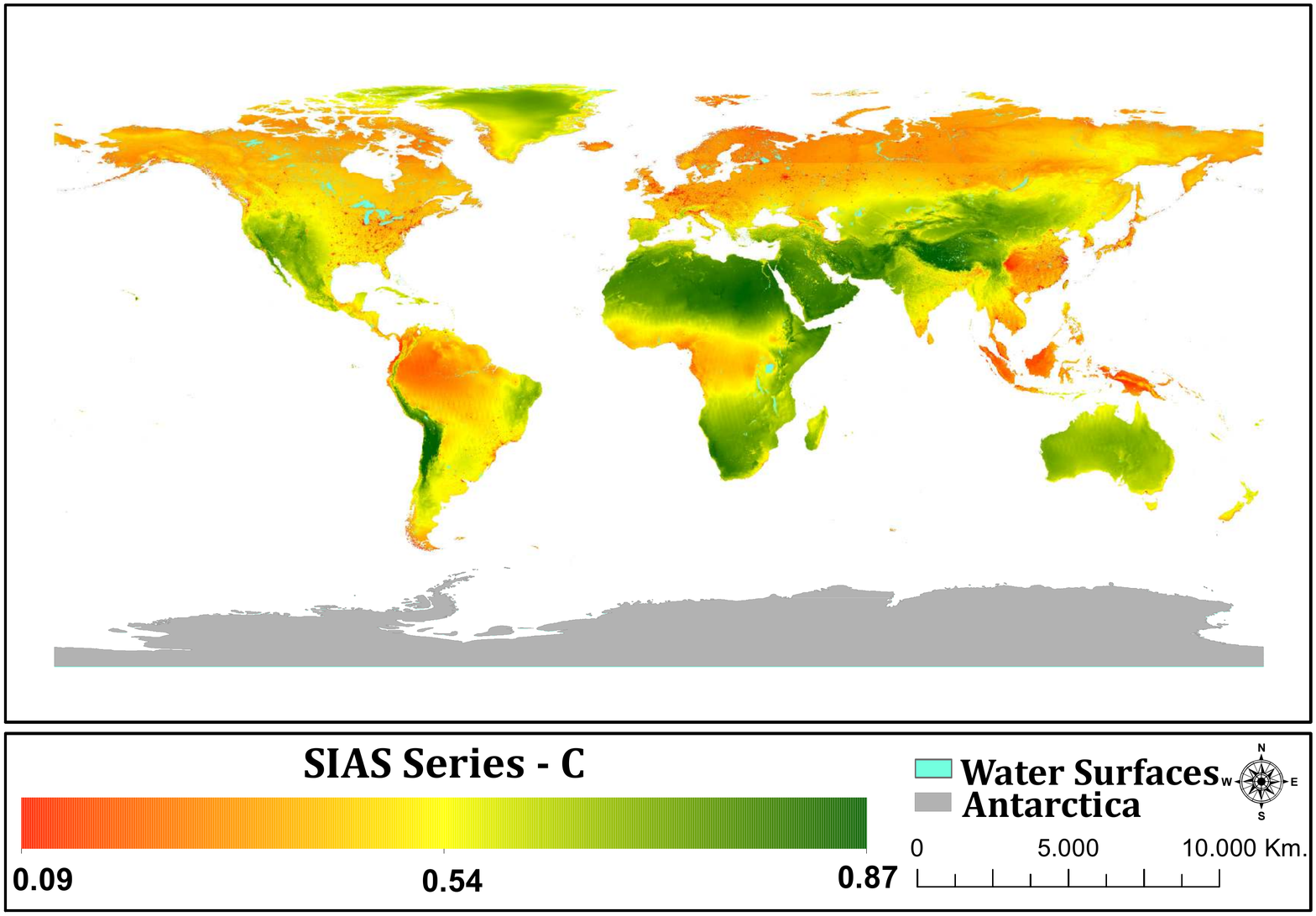}  &
    \includegraphics[height=5.45cm]{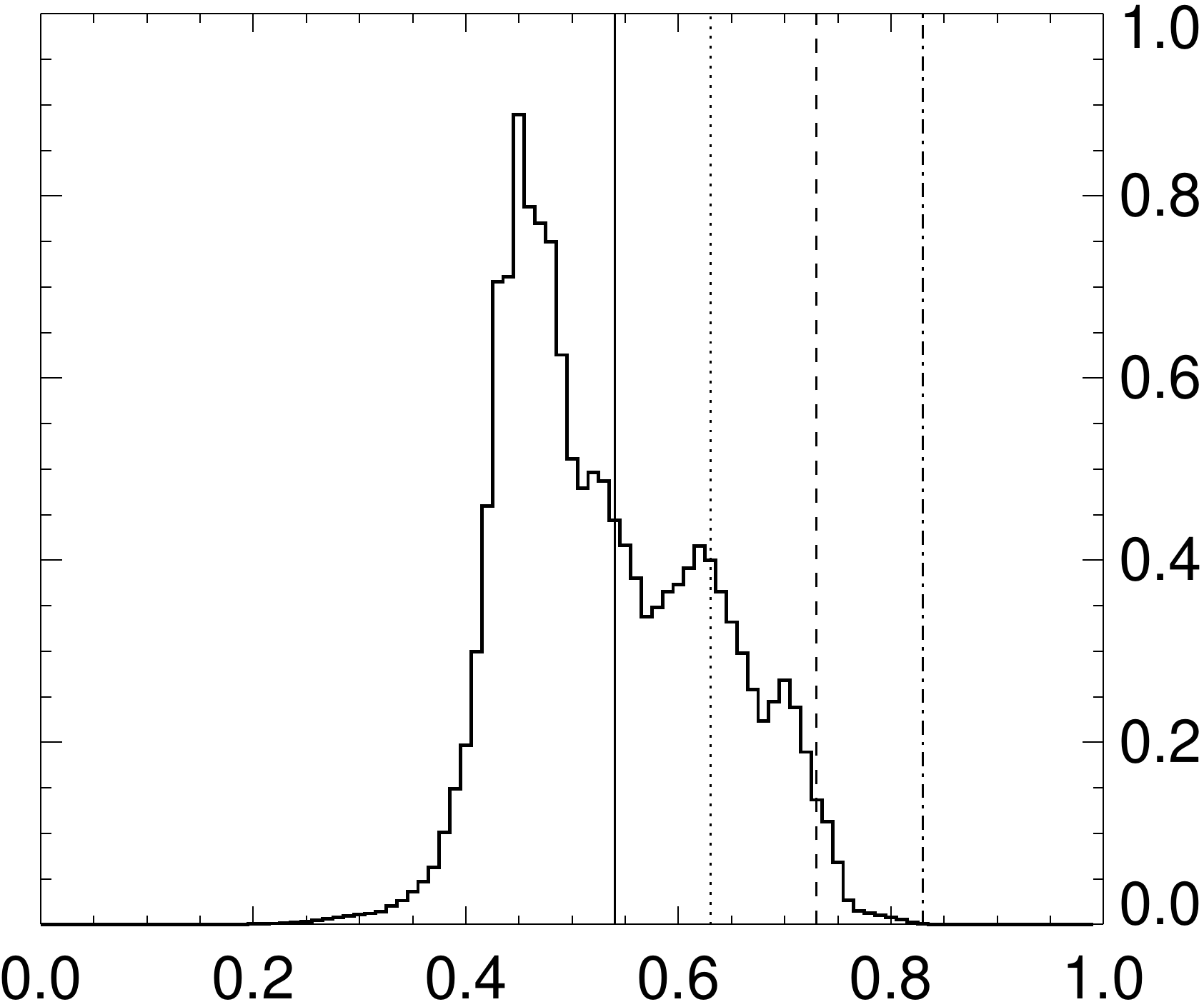}    \\
    \includegraphics[height=5.45cm]{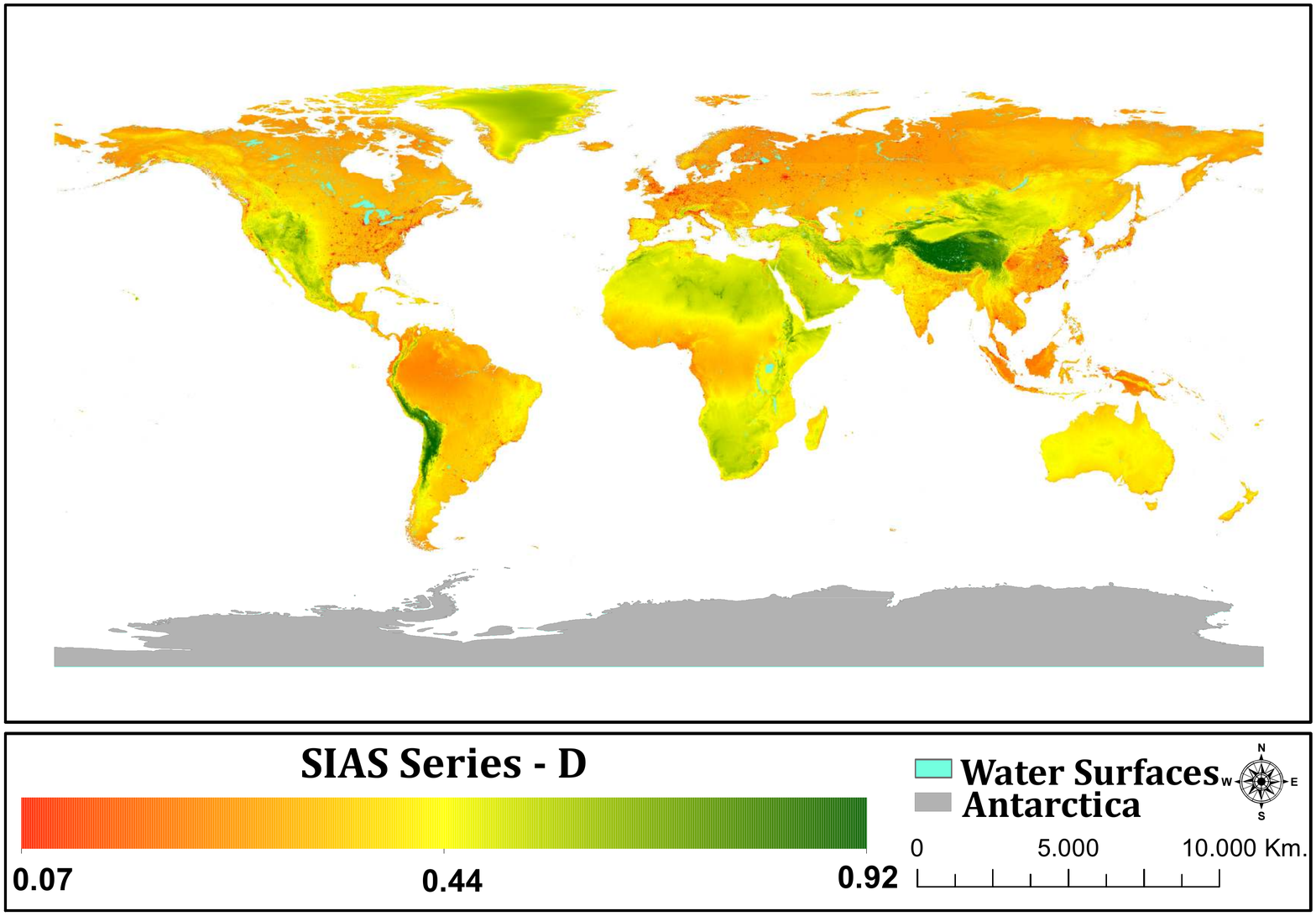}  &
    \includegraphics[height=5.45cm]{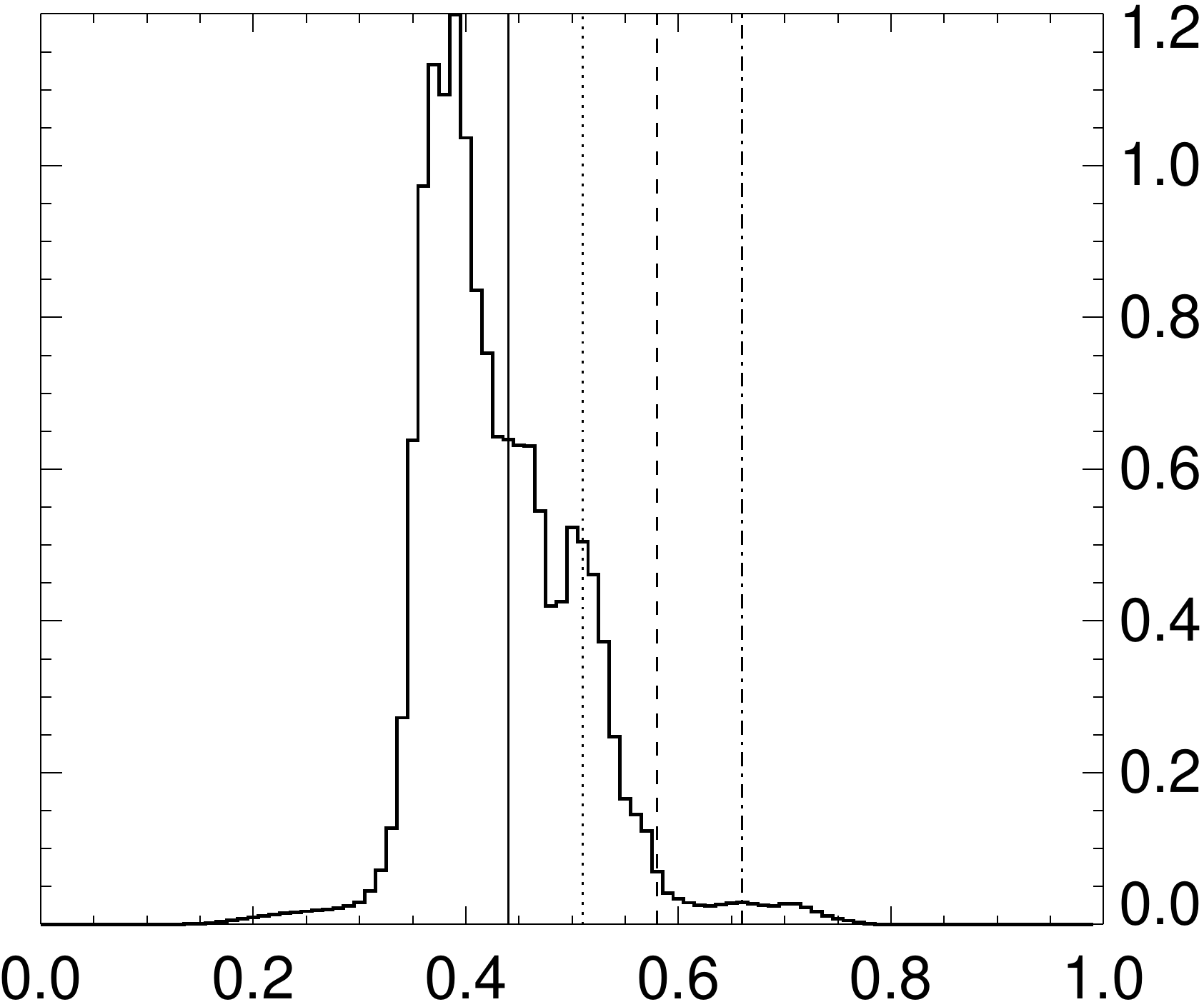}    \\
    \end{tabular}
    \caption{%
    SIAS Series created from GIS/MCDA analysis.
    Layouts (left column) of SIAS Series A (top) to D (bottom) are given respectively, along with their histograms (right column).
    Frequency of all histograms are normalized with $10^7$.
    The histogram also includes the mean, $\mu$ (solid line), $\mu+1\sigma$ (dotted line), $\mu+2\sigma$ (dashed line), $\mu+3\sigma$ (dash-dotted line).
    Statistical values are given in Table \ref{T:results}.%
    }
    \label{F:sias}
\end{figure*}

Important outcomes of four SIAS Series are given below:
\begin{itemize}
\item
First of all, we confirm the common sense of “astronomical site selection criteria”:
Good sites can be searched around the top of mountains (regions having dark green color in Fig. \ref{F:sias}).

\item
If a site location is good enough (see discussion below) then it is found to be good in all four series.
Therefore, the selected site status will be independent of the series (i.e weights of the layers).
Thus, one can select and use output of a suitable series depending on the purpose of the selection.

\item
Among all four series there is no single location on Earth having SIAS equals to 1.00.
This is expected because naturally this value cannot be achieved under Earth's atmosphere.

\item
A quick visual search on the results shown in Fig. \ref{F:sias}, can be summarized as follows.
``The best'' locations having maximum SIAS value in all series, corresponding to regions having dark green color and $\ge 3\sigma$ level are
(a) Mid--western Andes in South America,
(b) Tibetan Plateau in west China.

\item
Similarly, ``the good'' locations will have SIAS values corresponding to regions having light green colors and 2$\sigma$ level are
(a) Greenland, (b) West of North America, (c) Iran and Arabian Peninsula,
(d) West of South America, (e) Northern, Middle Eastern and Southern regions of Africa.

\item
Therefore, the rest SIAS values in all series can be thought as ``the worst'' locations, corresponding to regions having yellow, orange and red colors, and $\le1\sigma$ value.

\item
Even though there is no obvious longitudinal localization on SIAS values, there exist two latitude bands that can be marked as ``the good or the best'':
One in 10--50 North and the other in 10--40 South.
These regions are expected because they fall close to global cloud circulation patterns of Earth.
\end{itemize}

\subsection{Discussion}
\begin{itemize}
\item
This work is only limited to find or to localize ``regions'' all around the terrestrial land surfaces; not to find individual coordinates.

\item
SIAS usage can be simplified as Series B:
Since some of the layers are correlated with CC, DEM and AL layers, these three layers are found to be representing all layers when ``astronomical site selection'' is aimed.

\item
Since the resolution of layer datasets are finalized at 1 km resolution, site selection cannot give definitive results (e.g specific coordinates) unless datasets having finer resolutions were included (e.g in meter order).
Considering all the series, $3\sigma$ level corresponds to $\sim$ 2--3 \% of land surface which is equivalent to 3.5--5.0 million sq km.
In size, this value is too big.
Using a rough but quick calculation, tens of thousands of astronomical observatory site could be found.

\item
Therefore, the best way to use SIAS Series is ``to eliminate'' the worst ($\le1\sigma$), not ``to find the best'' ($\ge3\sigma$).
Moreover, the final site selection has to be done by performing long--term bottom--up, on--site measurements.
\end{itemize}

\subsection{Outcomes}
\begin{figure*}
    \centering
    \begin{tabular}{@{}P{1.5cm}@{}c@{}c@{}c@{}c@{}}
    & SIAS Series A & SIAS Series B & SIAS Series C & SIAS Series D \\
Hawaii 
    & \includegraphics[height=3.3cm,valign=c]{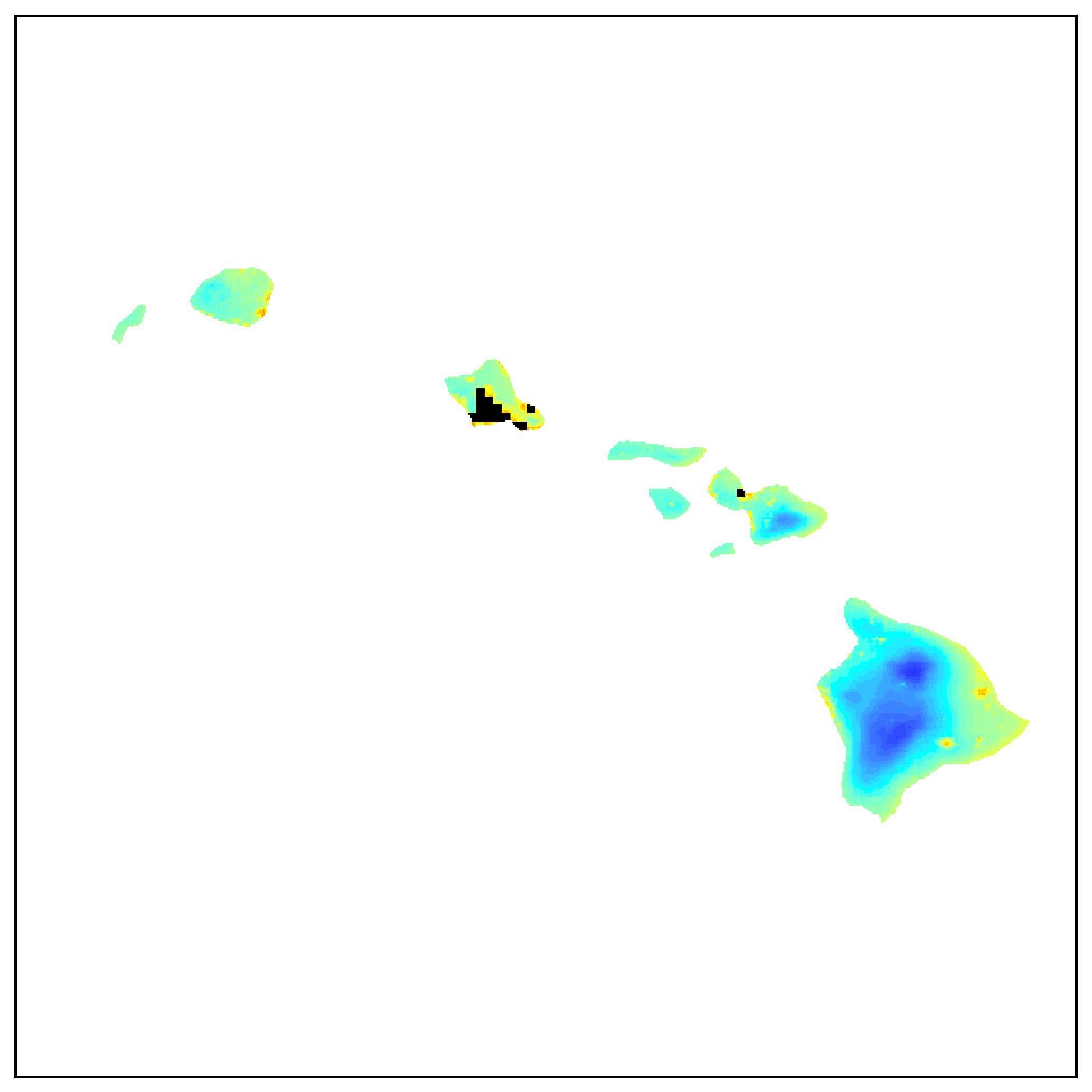}
    & \includegraphics[height=3.3cm,valign=c]{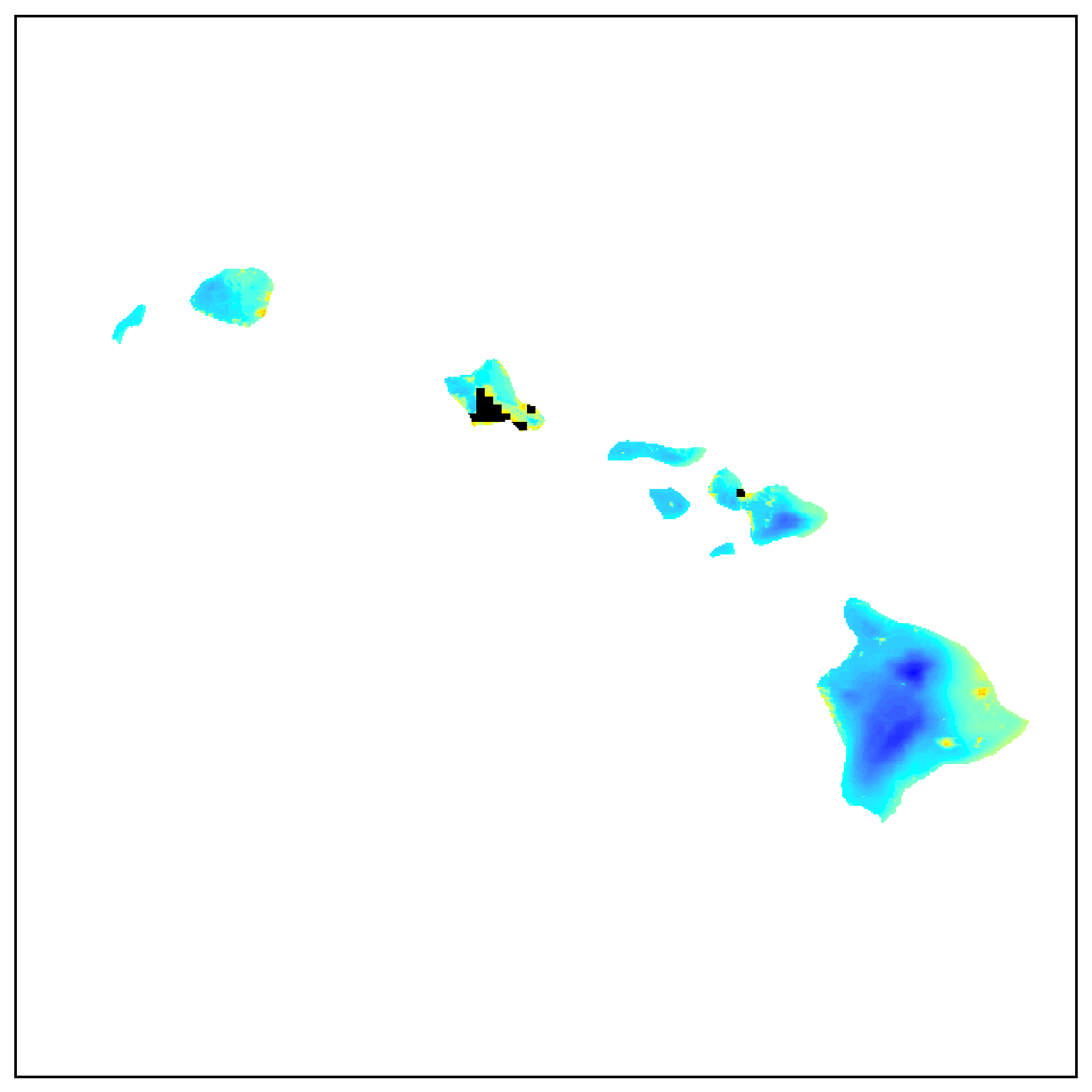}
    & \includegraphics[height=3.3cm,valign=c]{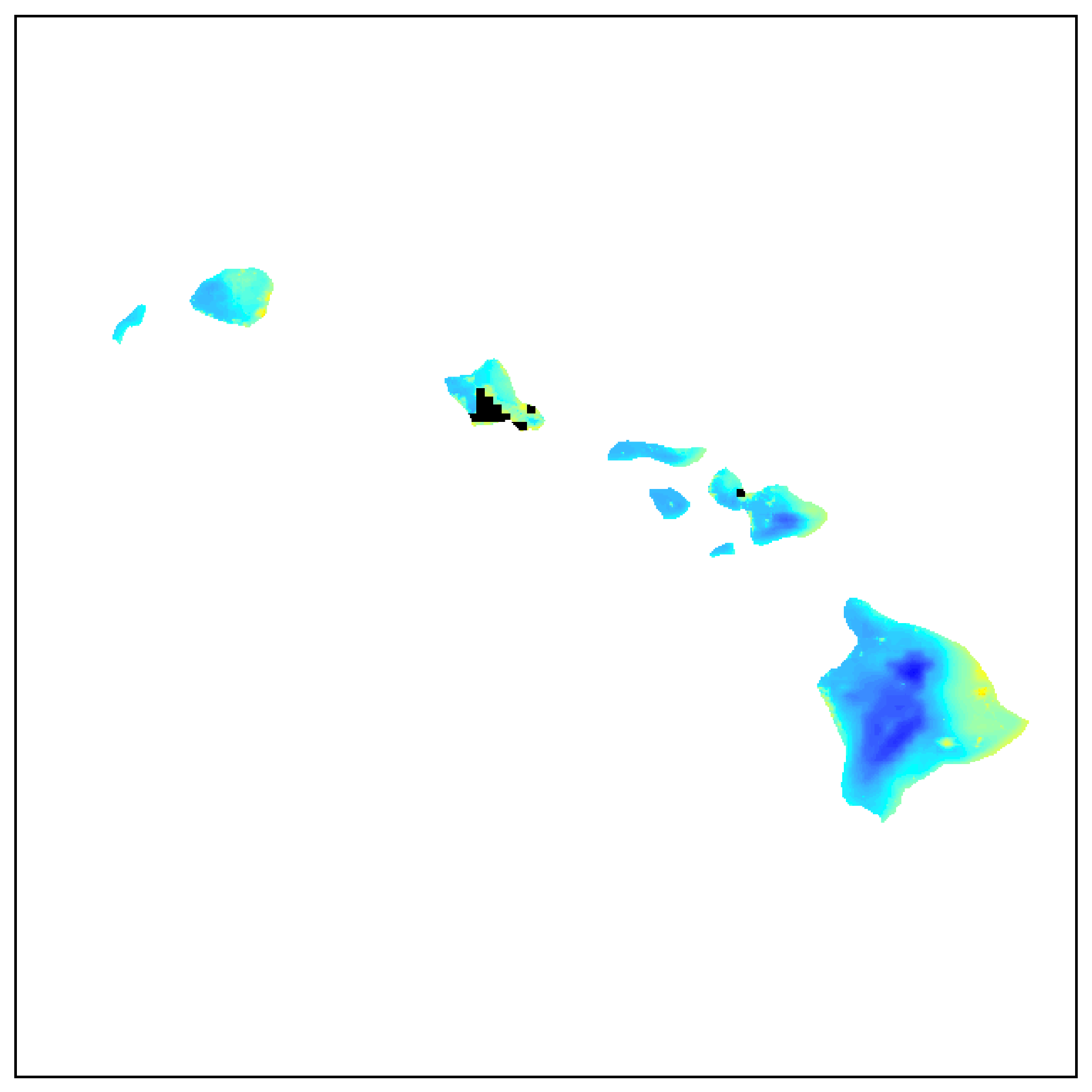}
    & \includegraphics[height=3.3cm,valign=c]{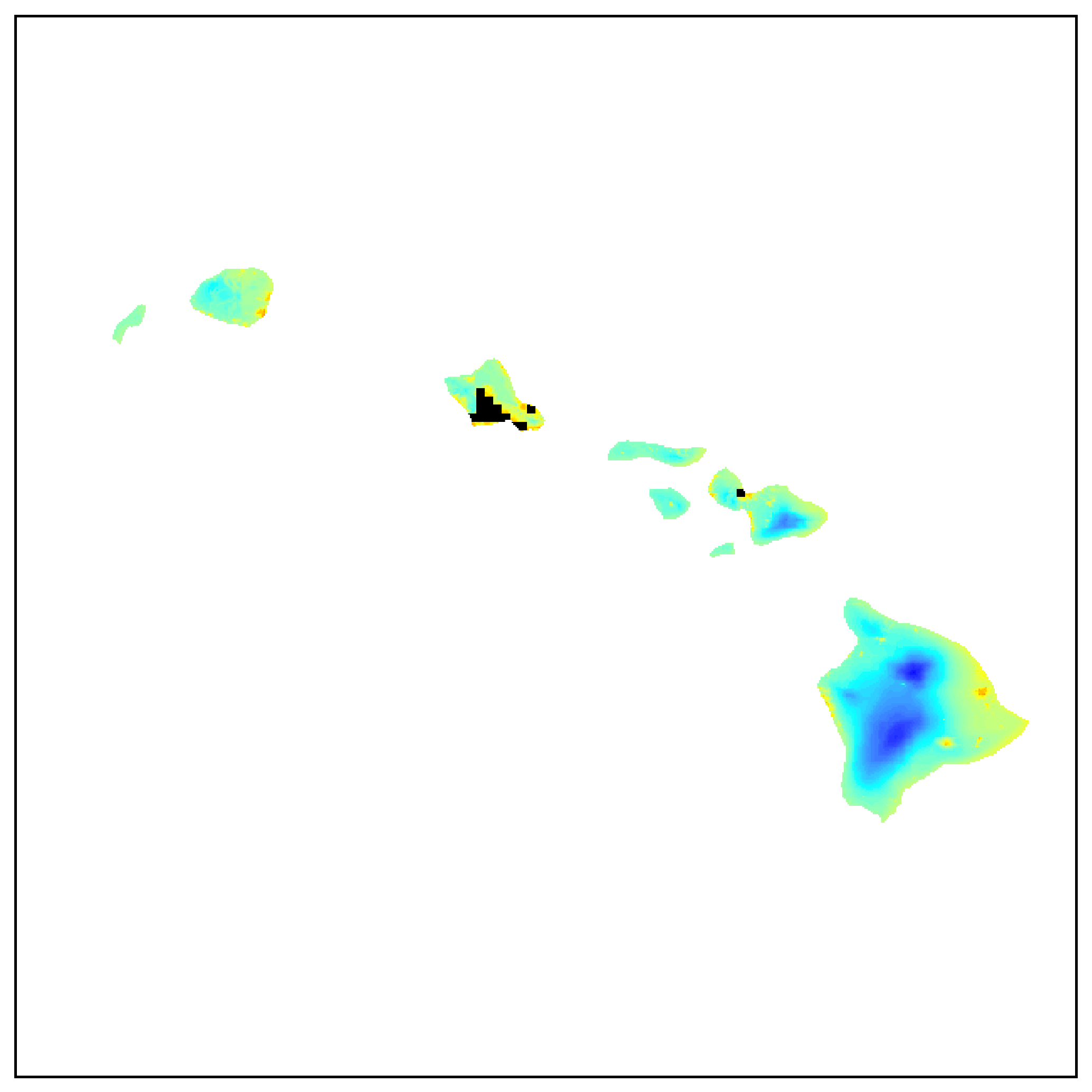}\\
    & ~0.49\hfill 0.85~
    & ~0.15\hfill 0.78~
    & ~0.14\hfill 0.80~
    & ~0.12\hfill 0.71 \\
Chile
    & \includegraphics[height=3.3cm,valign=c]{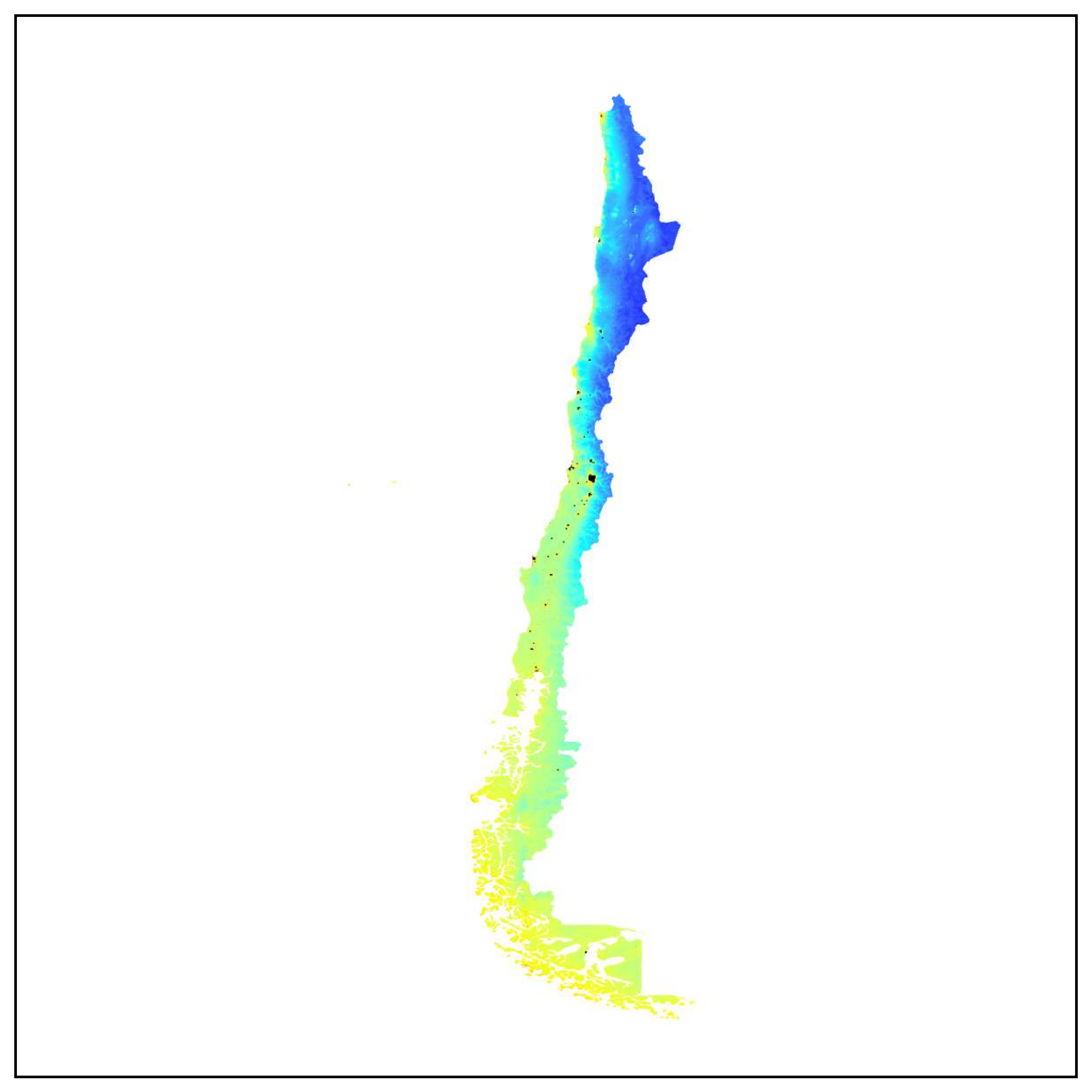}
    & \includegraphics[height=3.3cm,valign=c]{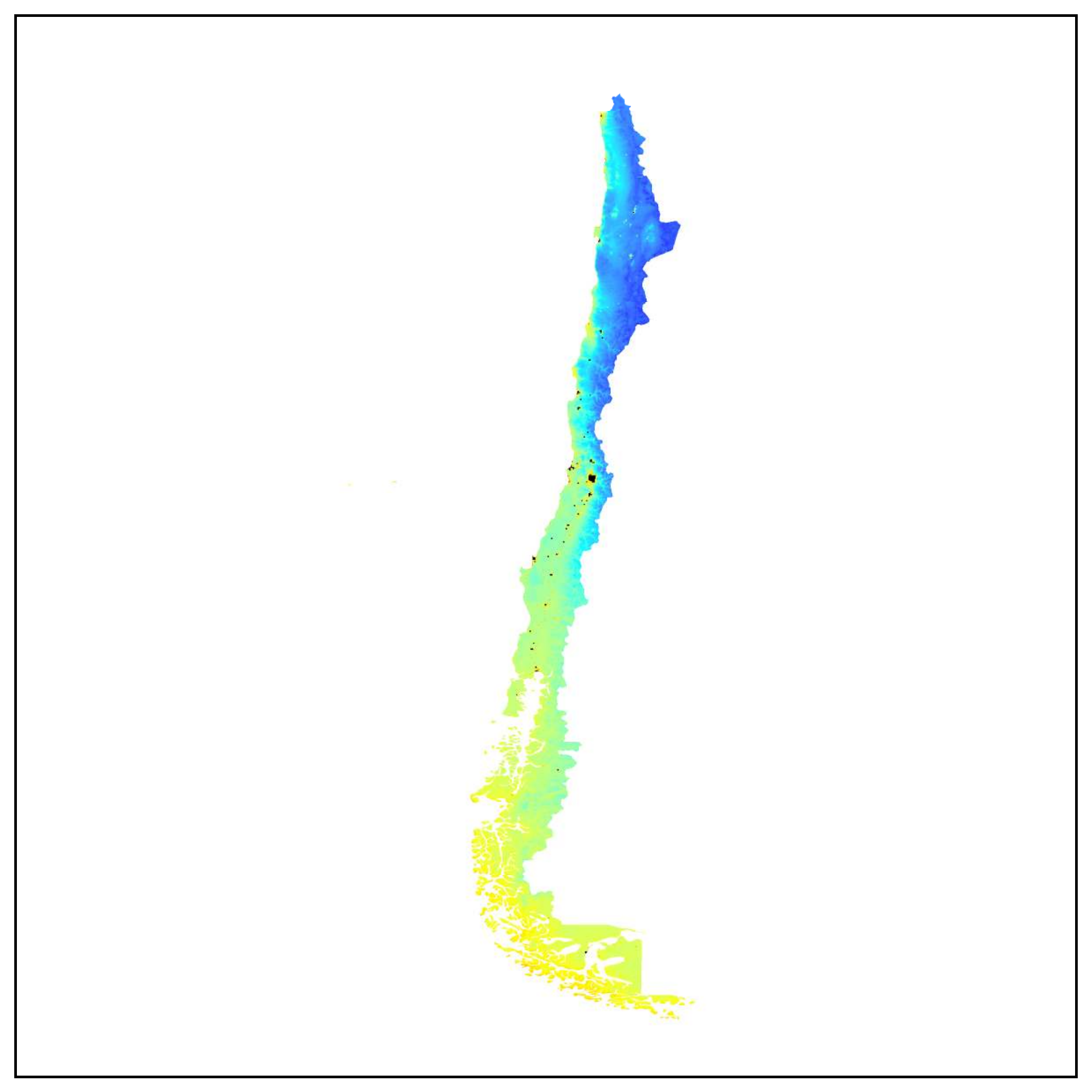}
    & \includegraphics[height=3.3cm,valign=c]{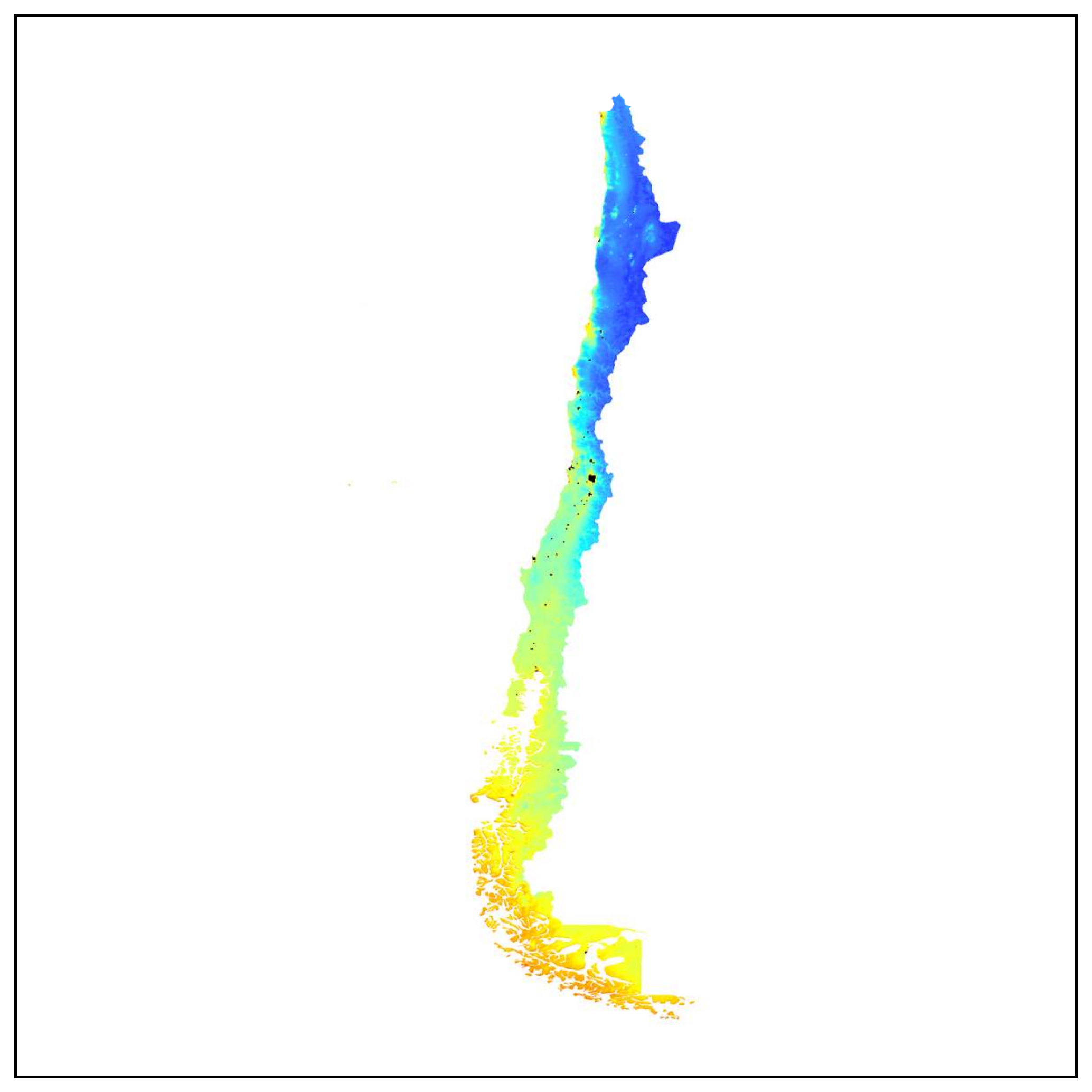}
    & \includegraphics[height=3.3cm,valign=c]{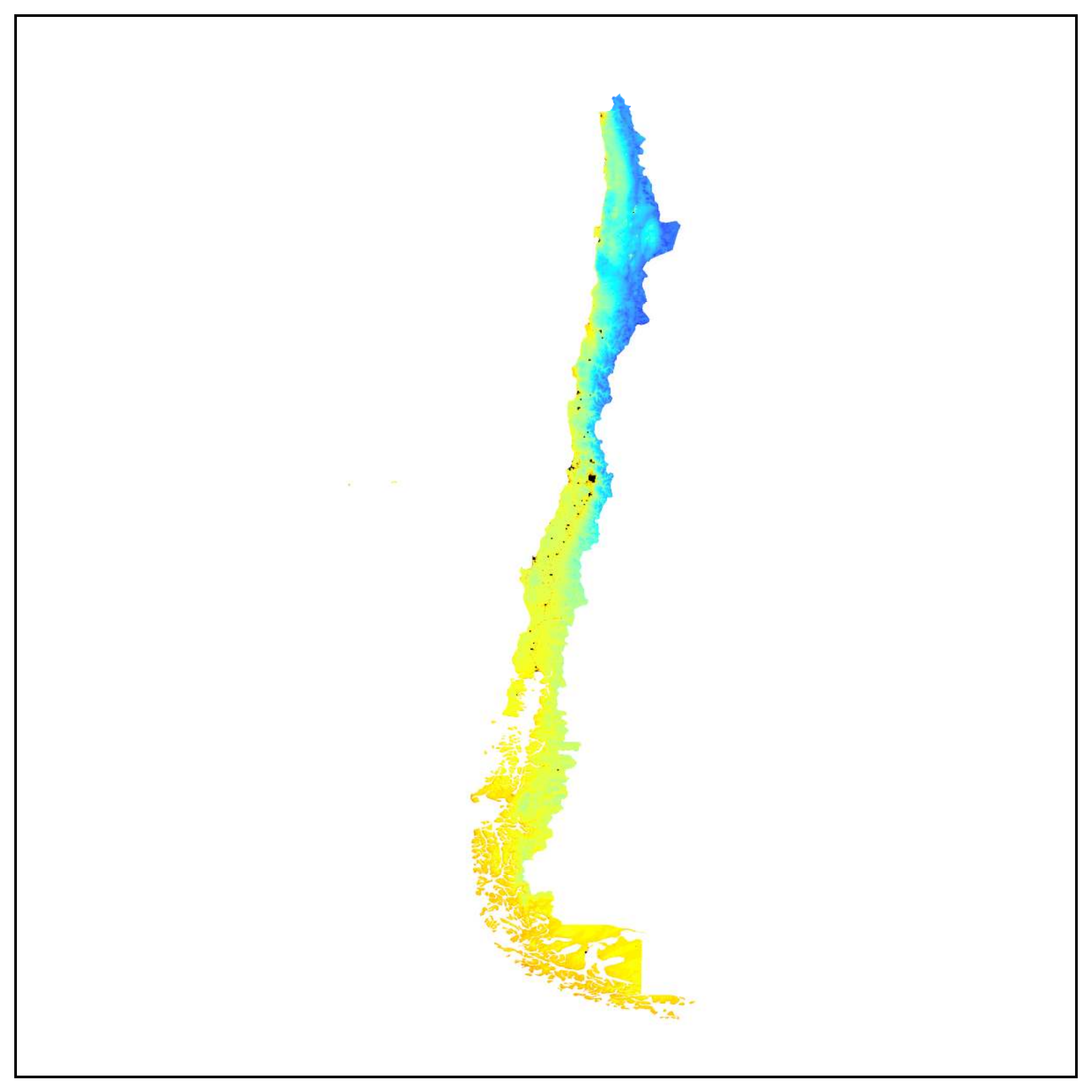}\\
    & ~0.54\hfill 0.91~
    & ~0.19\hfill 0.87~
    & ~0.21\hfill 0.87~
    & ~0.15\hfill 0.84 \\
Canary Islands
    & \includegraphics[height=3.3cm,valign=c]{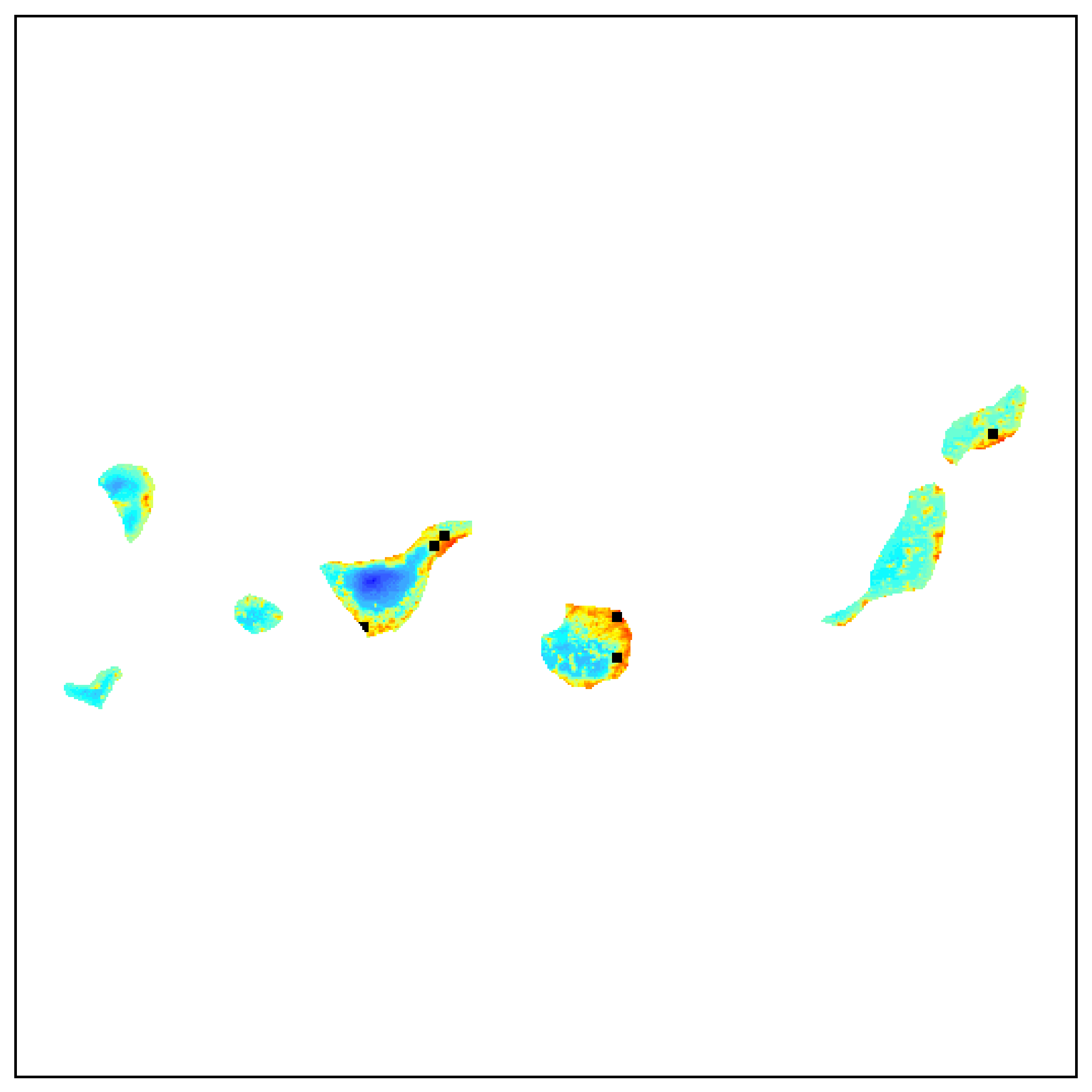}
    & \includegraphics[height=3.3cm,valign=c]{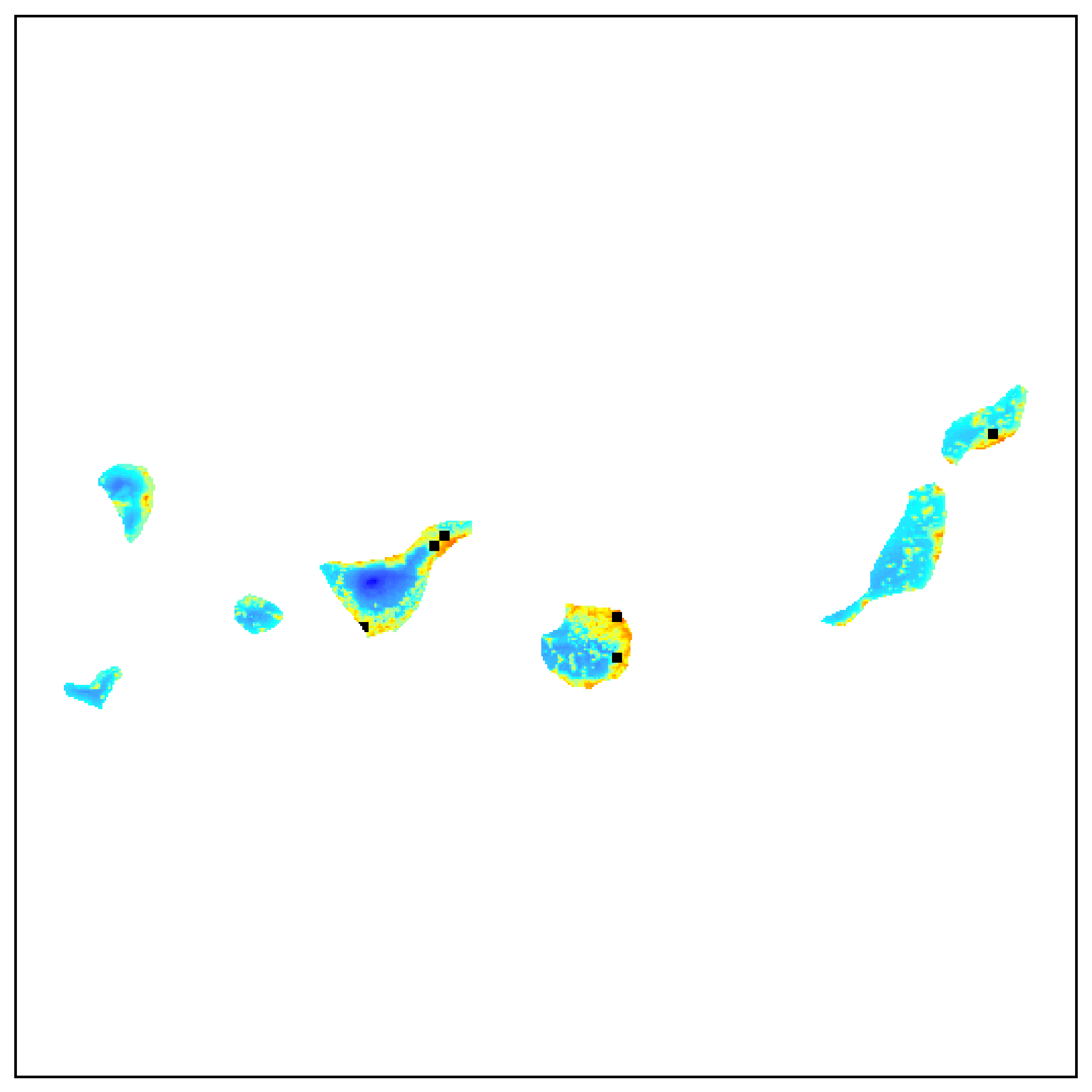}
    & \includegraphics[height=3.3cm,valign=c]{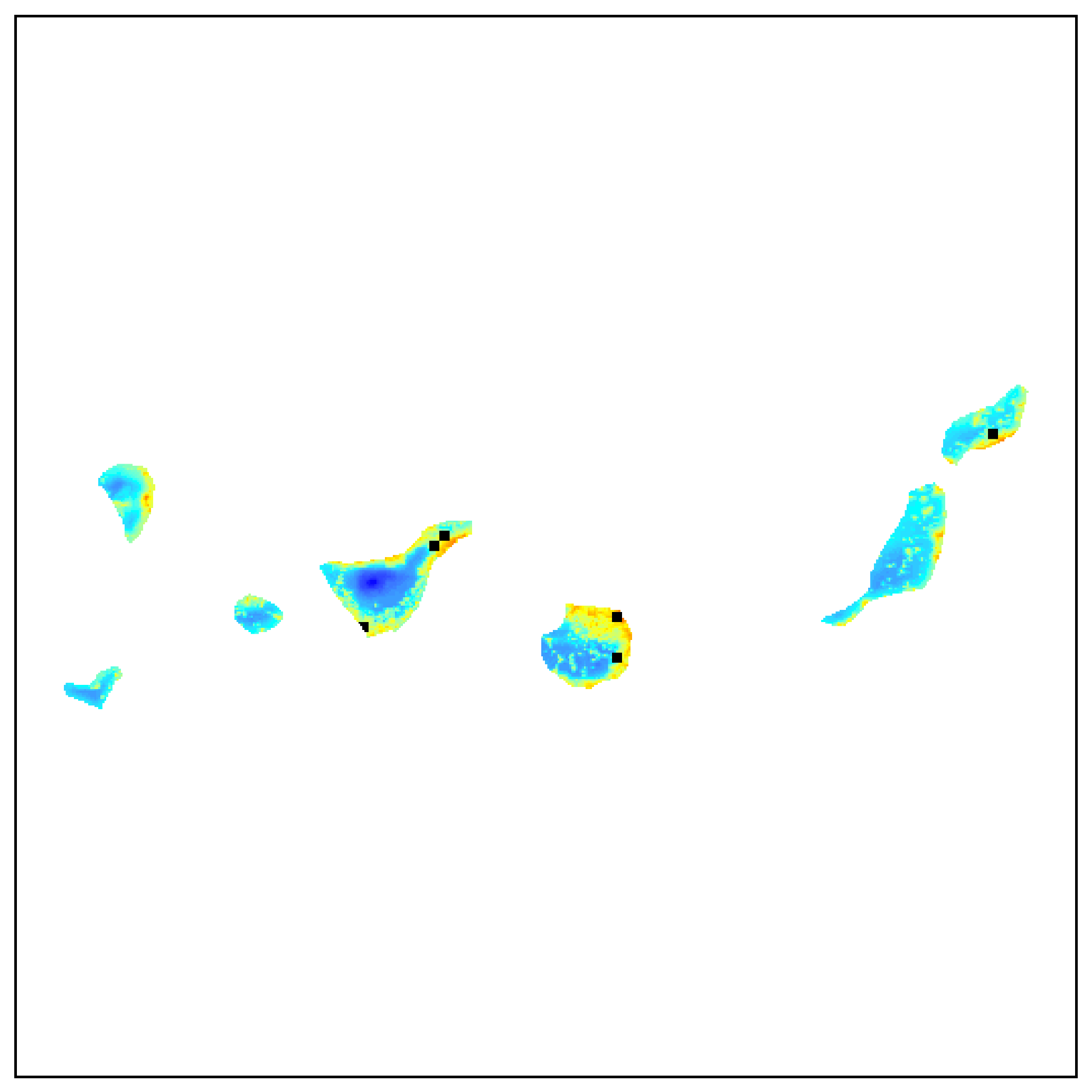}
    & \includegraphics[height=3.3cm,valign=c]{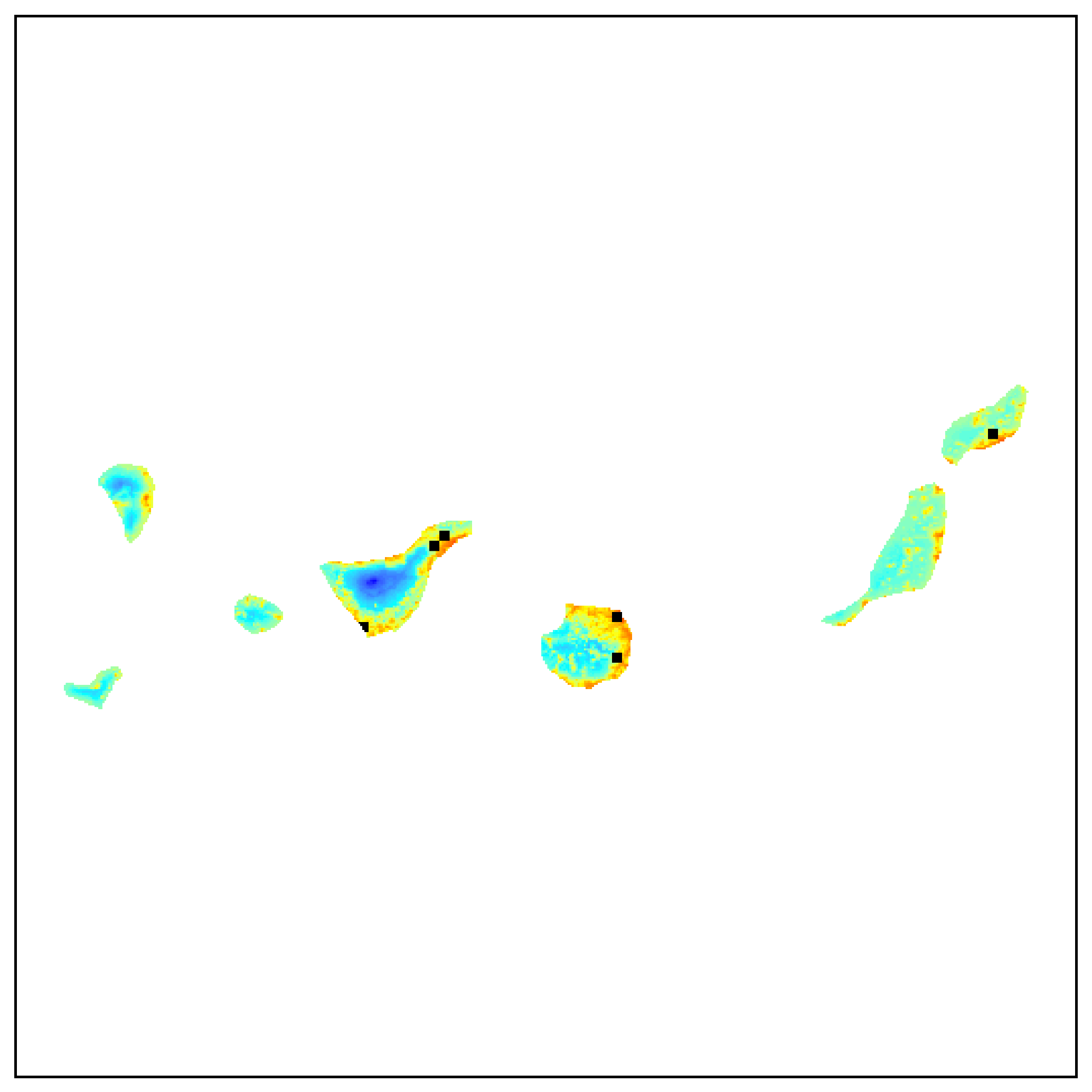}\\
    & ~0.53\hfill 0.81~
    & ~0.19\hfill 0.75~
    & ~0.23\hfill 0.77~
    & ~0.15\hfill 0.67 \\
Iran
    & \includegraphics[height=3.3cm,valign=c]{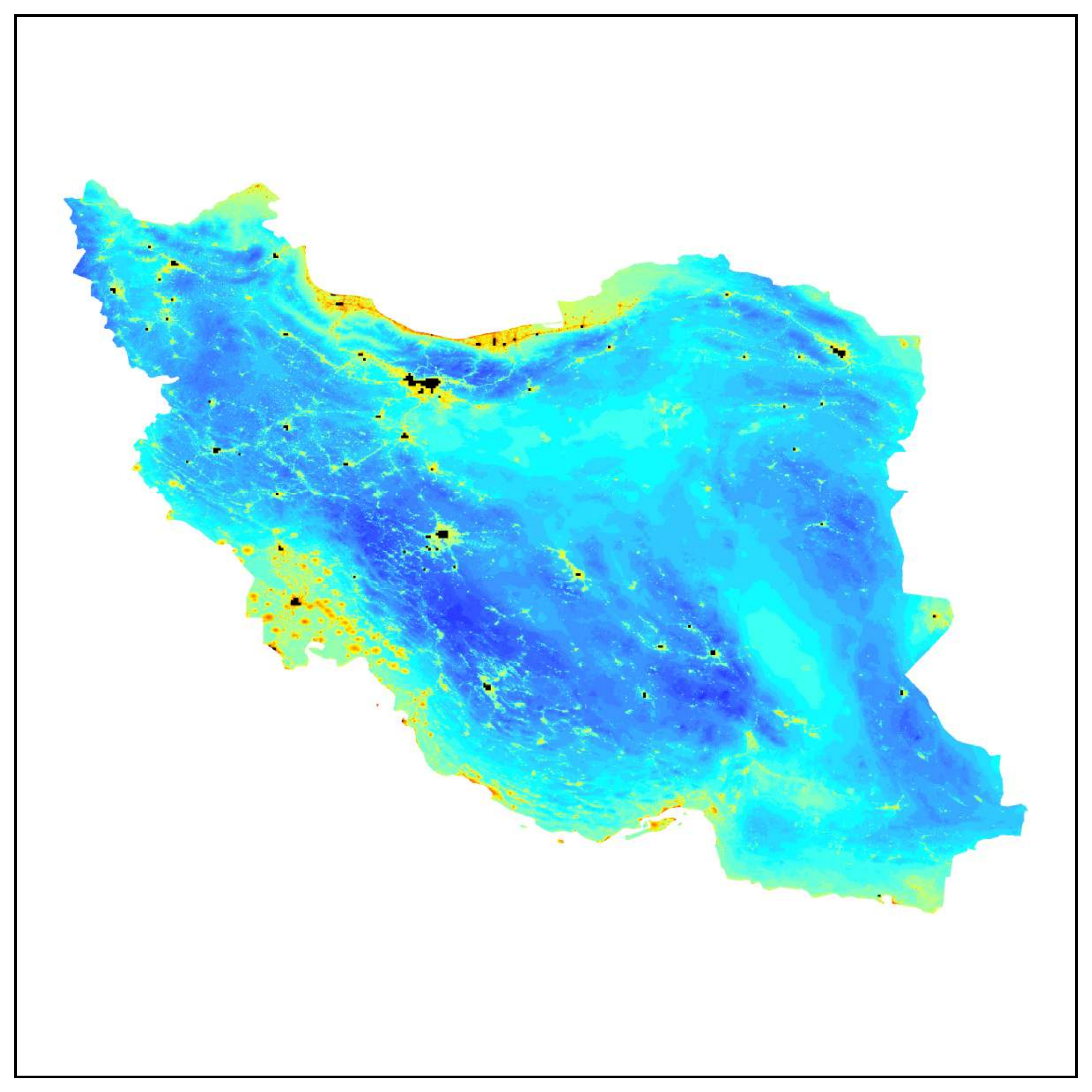}
    & \includegraphics[height=3.3cm,valign=c]{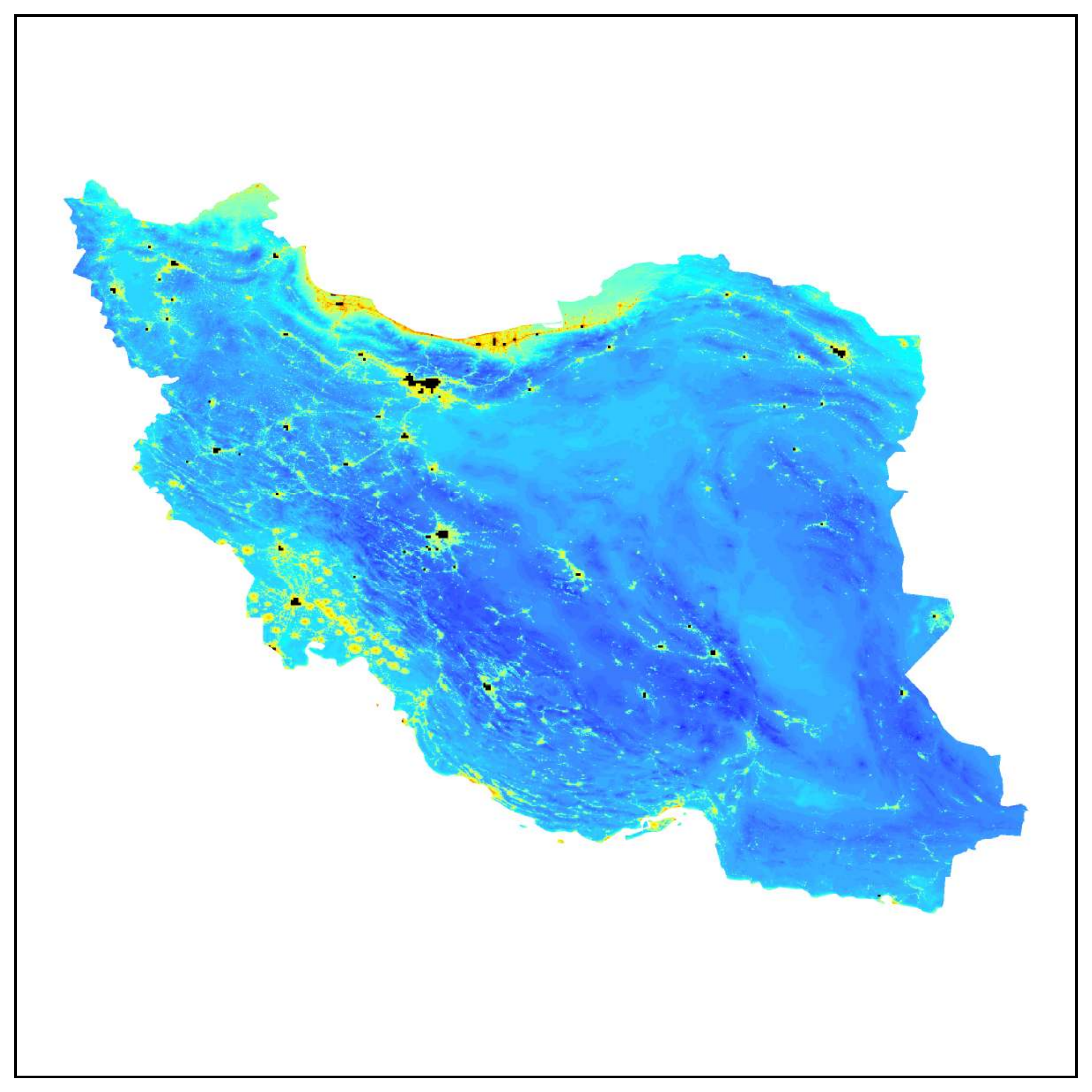}
    & \includegraphics[height=3.3cm,valign=c]{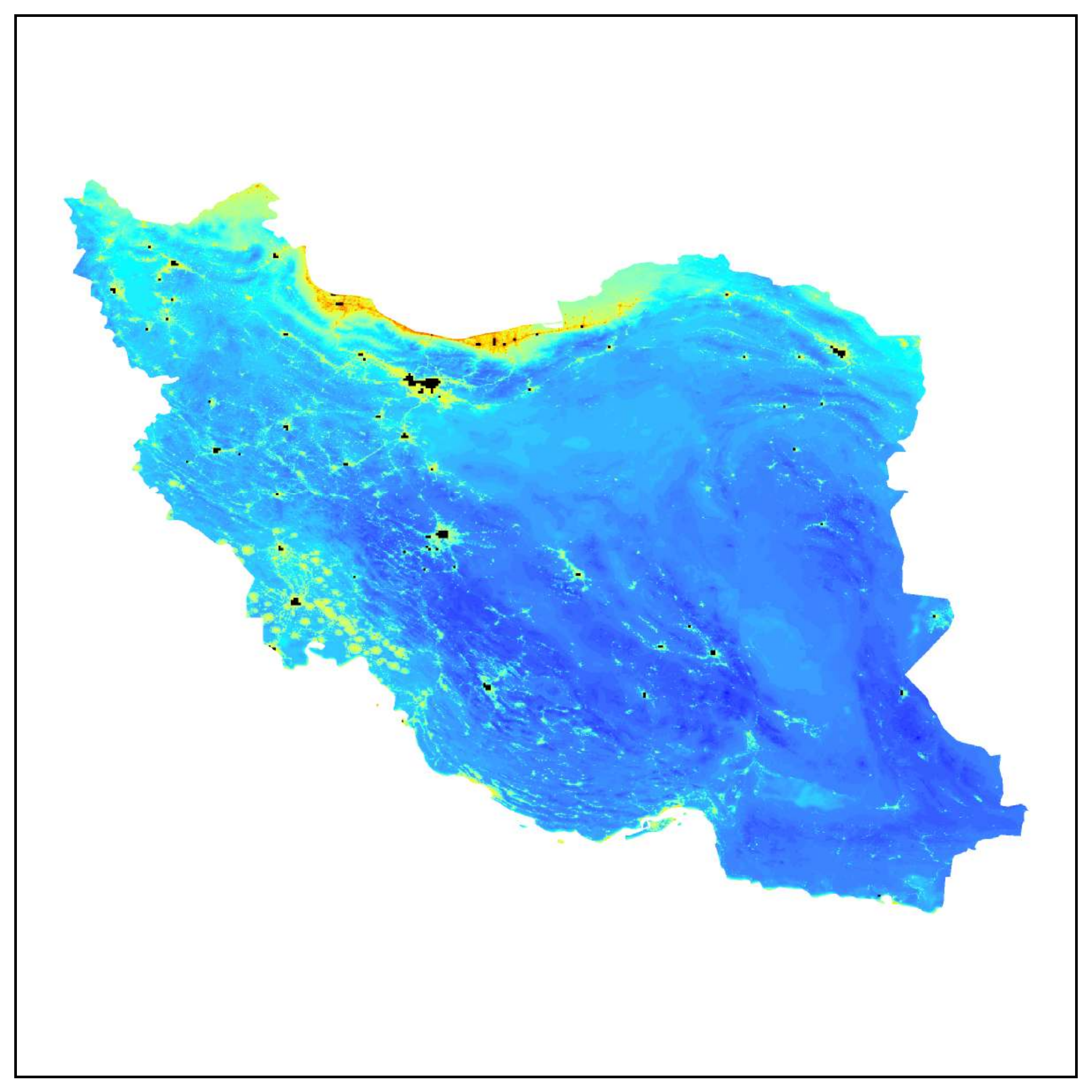}
    & \includegraphics[height=3.3cm,valign=c]{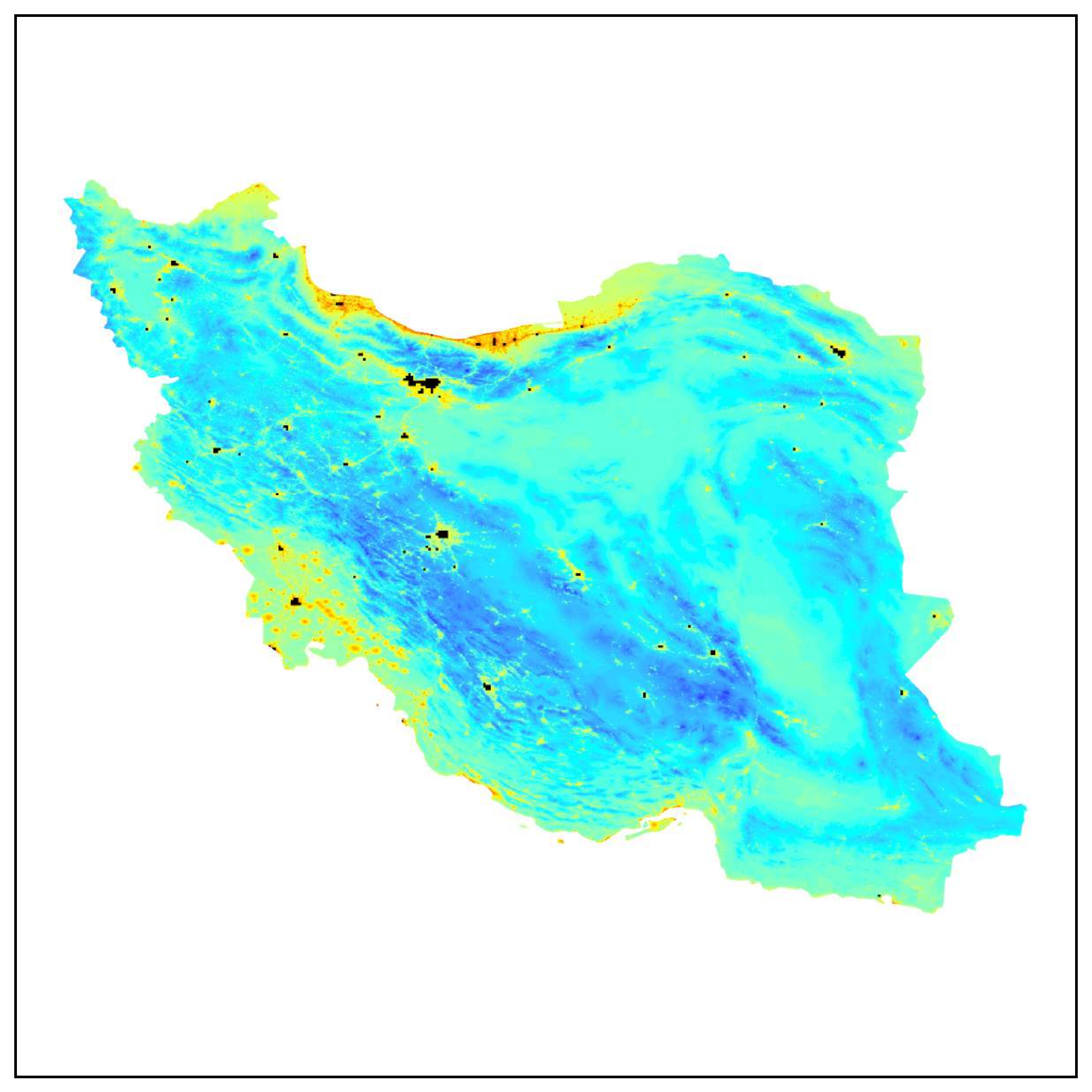}\\
    & ~0.54\hfill 0.85~
    & ~0.20\hfill 0.78~
    & ~0.25\hfill 0.80~
    & ~0.16\hfill 0.73 \\
Pakistan
    & \includegraphics[height=3.3cm,valign=c]{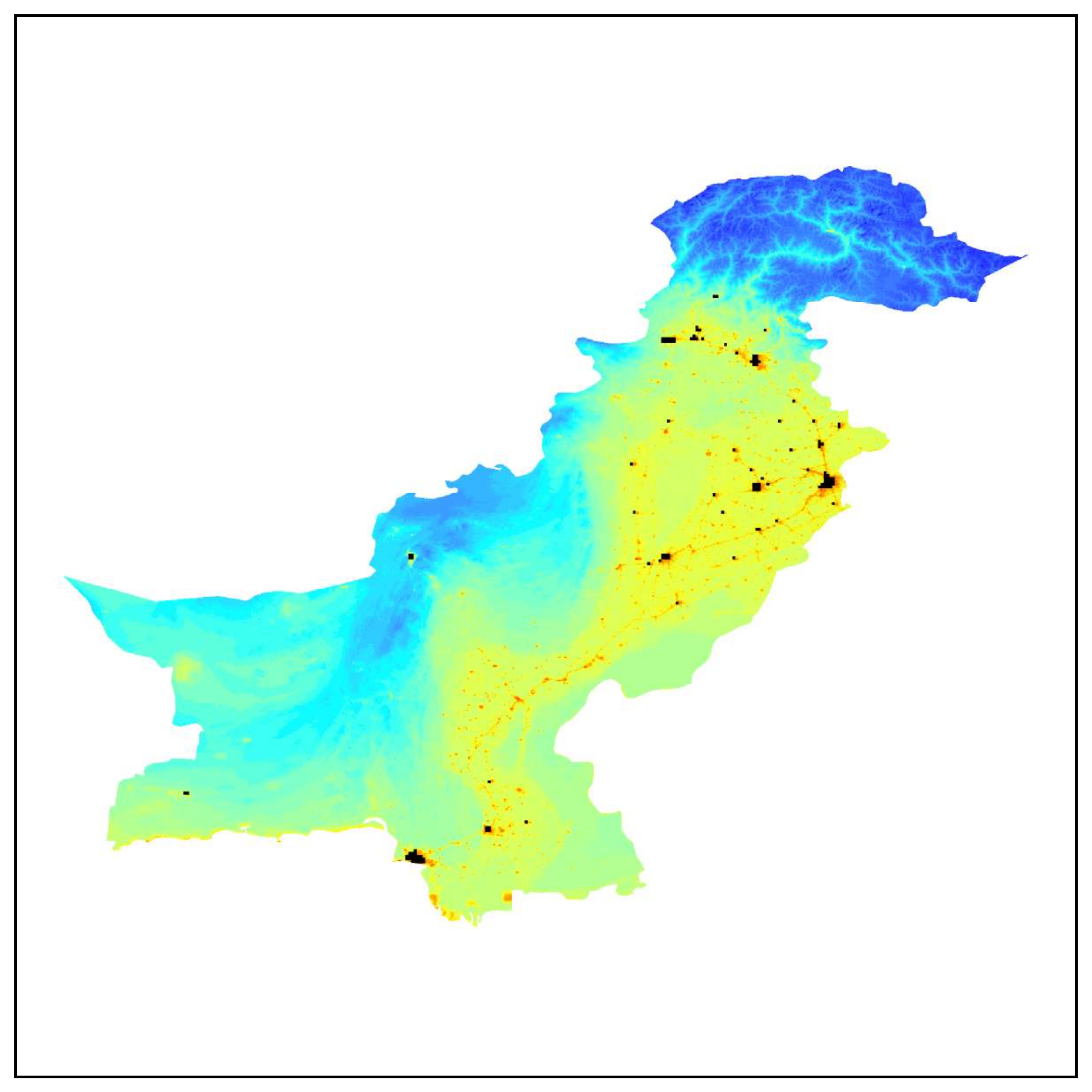}
    & \includegraphics[height=3.3cm,valign=c]{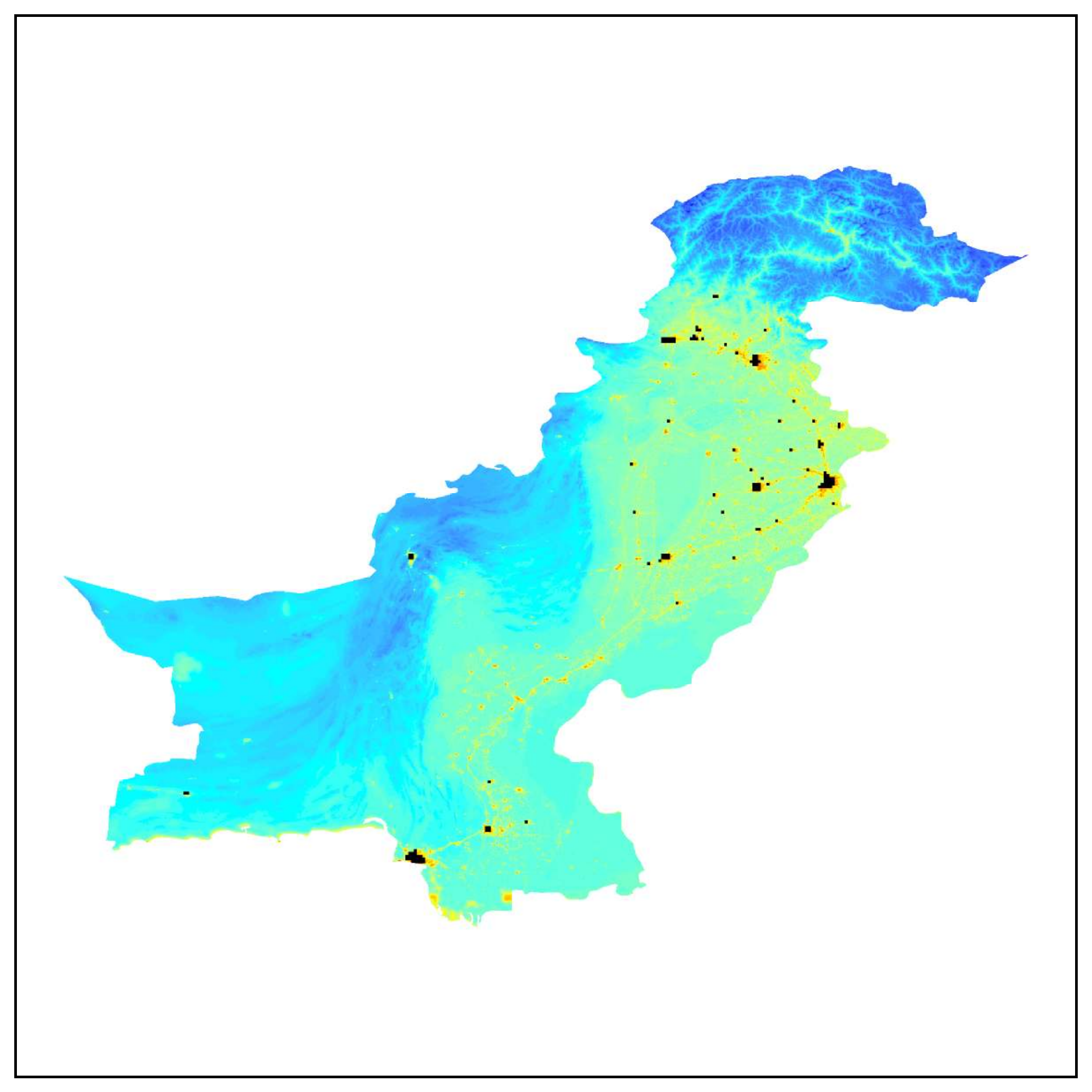}
    & \includegraphics[height=3.3cm,valign=c]{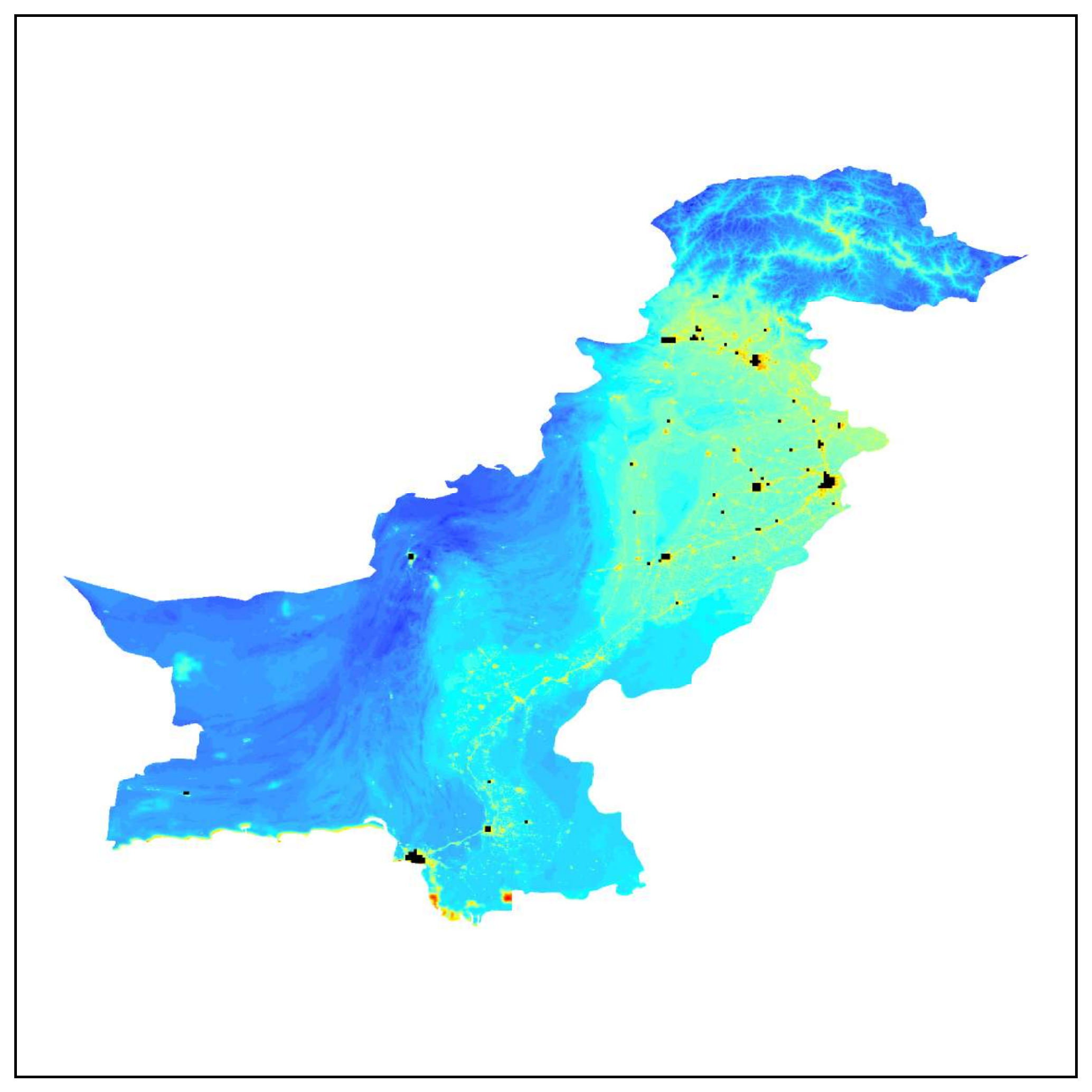}
    & \includegraphics[height=3.3cm,valign=c]{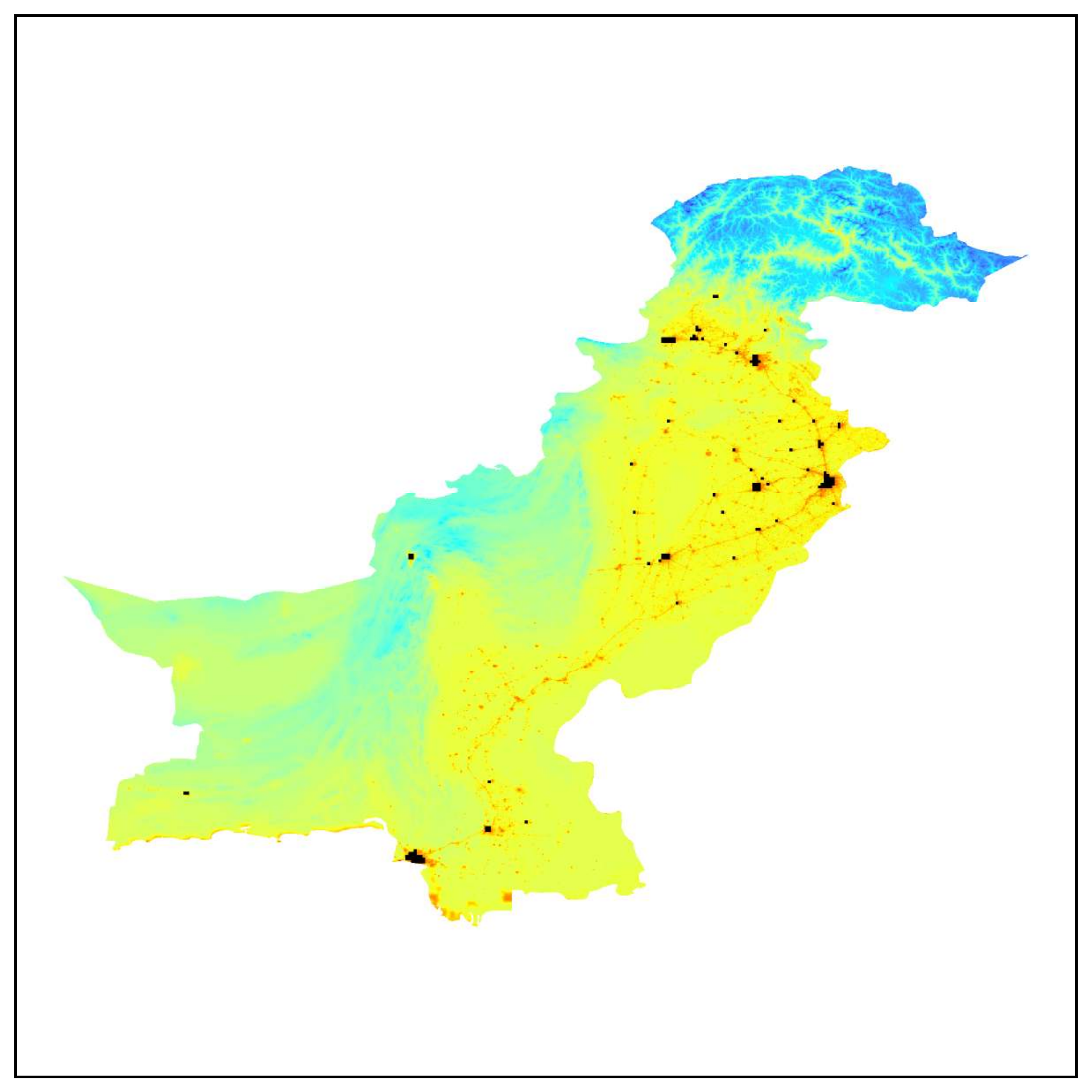}\\
    & ~0.55\hfill 0.89~
    & ~0.27\hfill 0.84~
    & ~0.34\hfill 0.80~
    & ~0.22\hfill 0.86 \\
China
    & \includegraphics[height=3.3cm,valign=c]{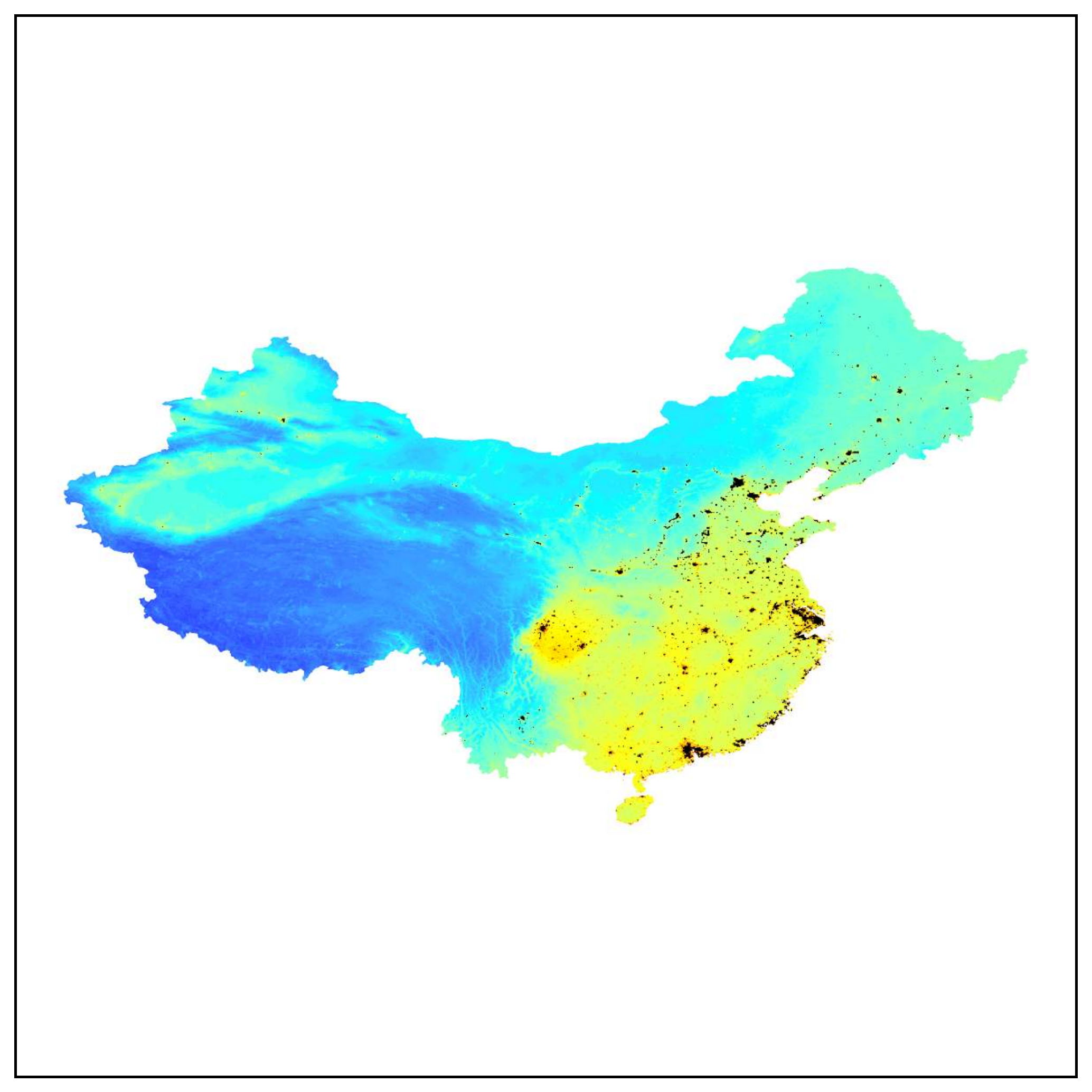}
    & \includegraphics[height=3.3cm,valign=c]{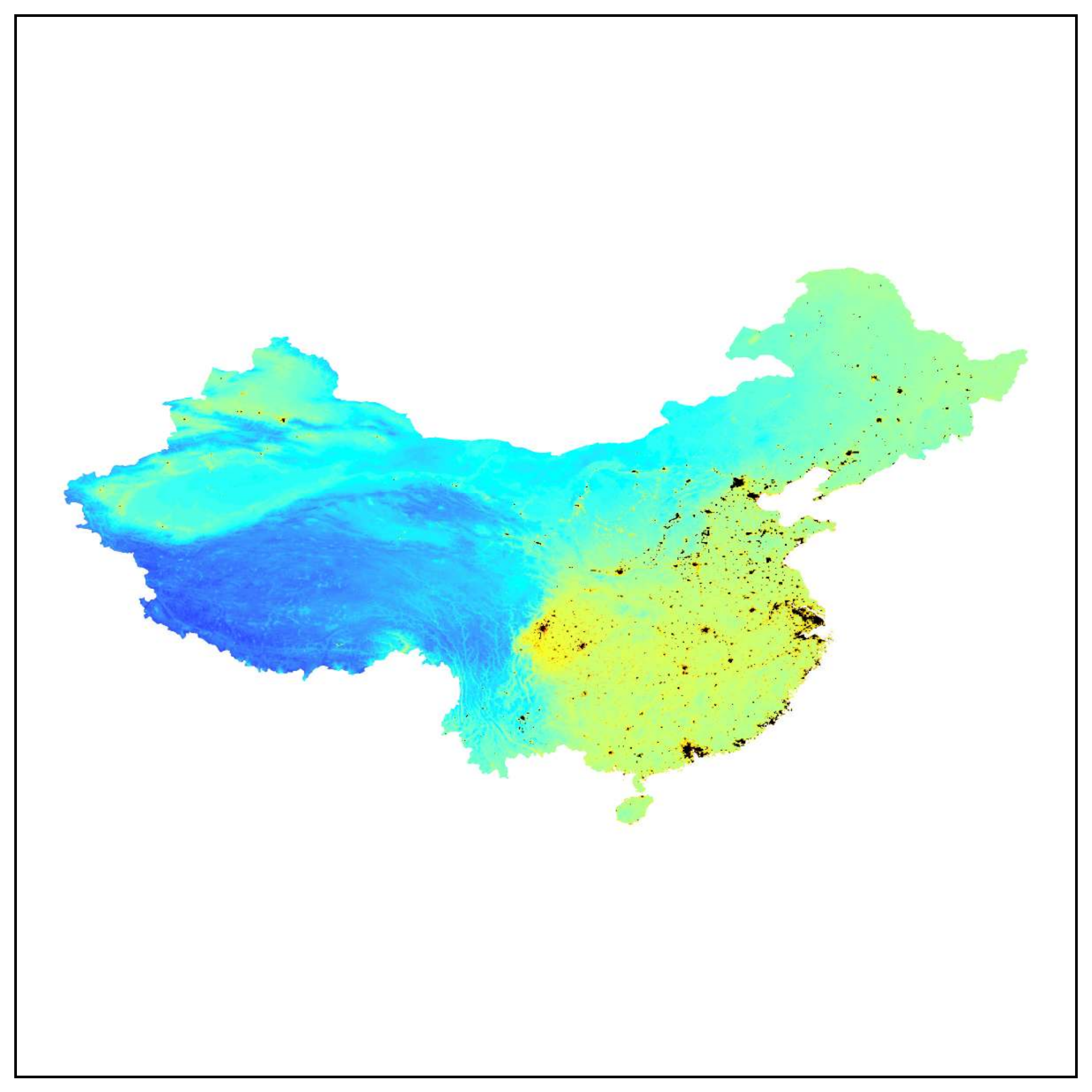}
    & \includegraphics[height=3.3cm,valign=c]{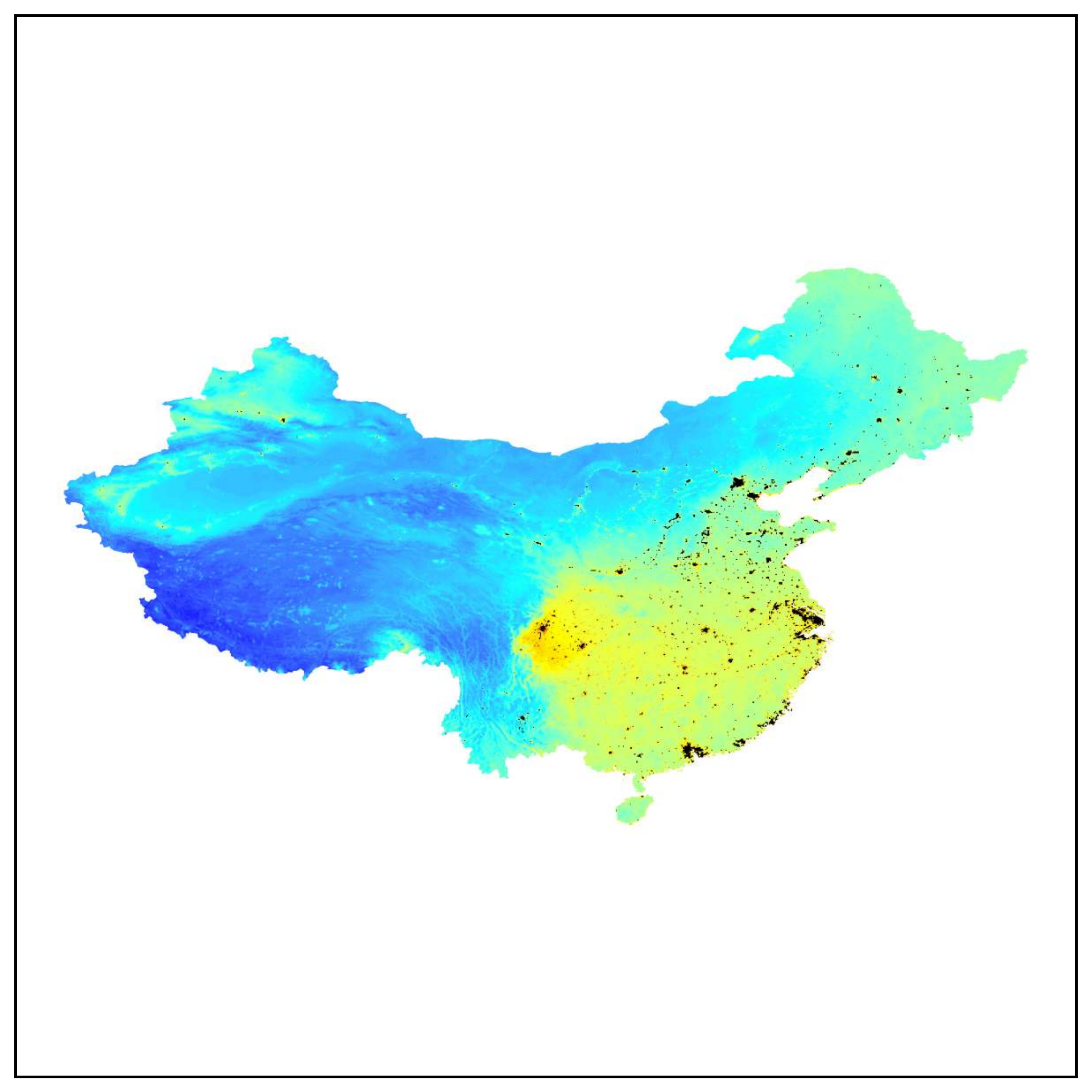}
    & \includegraphics[height=3.3cm,valign=c]{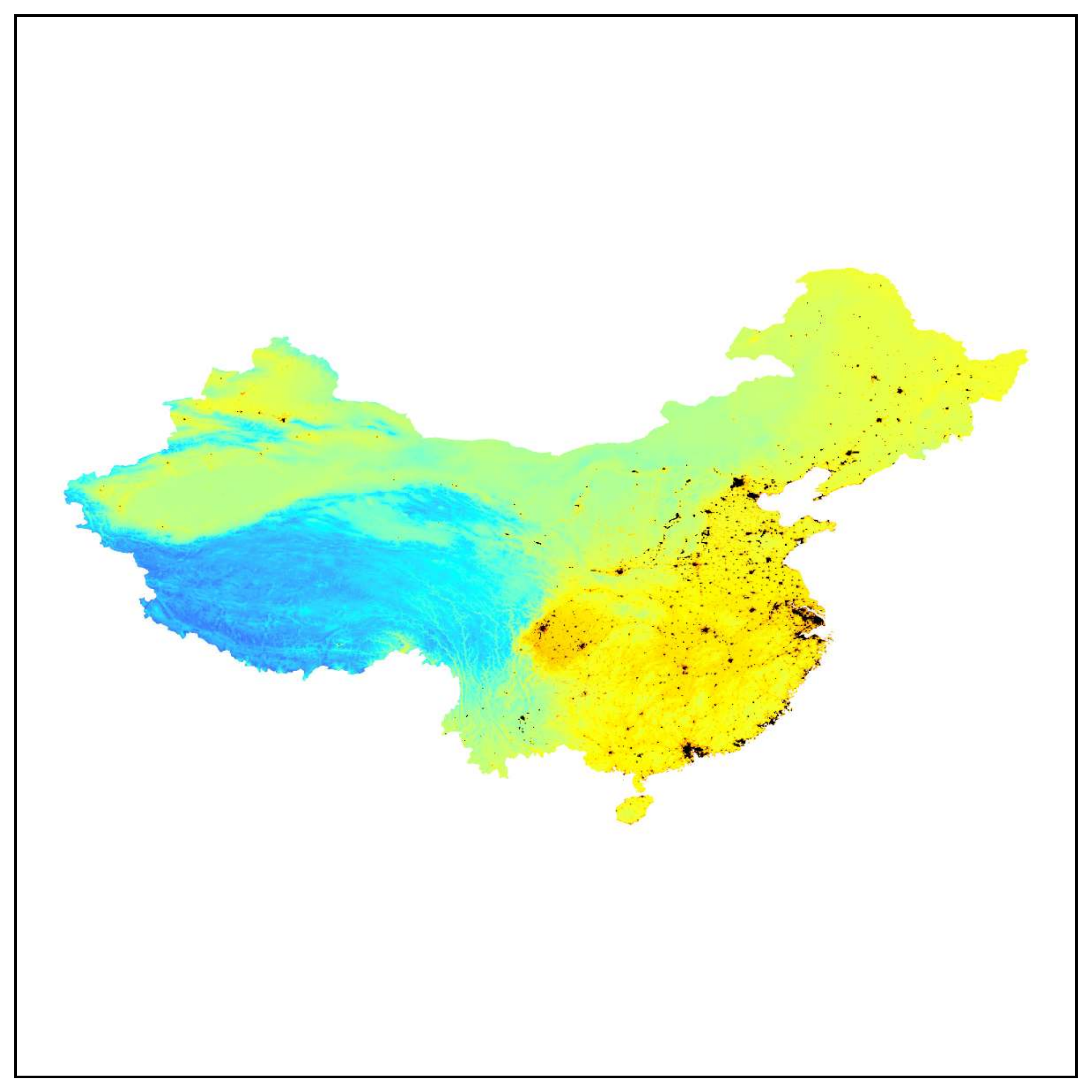}\\
    & ~0.44\hfill 0.93~
    & ~0.13\hfill 0.89~
    & ~0.13\hfill 0.84~
    & ~0.12\hfill 0.91 \\
    \end{tabular}
    \caption{%
    The results of GIS/MCDA analysis in all series for a limited number of countries (top to bottom panels with longitude order):
    Hawaii, Chile, Canary Islands, Iran, Pakistan and China.
    SIAS Series A-D are given from left to right.
    Minimum (left -- red in color grading) and maximum (right -- blue in color grading) values of each series are given below the corresponding map.
    Color black represents built-up area within the country.
    All country layouts will be available online.%
    }
    \label{F:country}
\end{figure*}

Layout of SIAS Series analysis are created for a total of 247 countries.
However, countries with small land surface area ($\le 10,000$ sq km) or countries with many small islands are excluded.
Therefore, country layouts were created only for 168 countries.
Among all the others, six countries were selected to compare the results of this work.
They are either well-known in observational astronomy or a recent site selection study was performed on the site (westward longitude listed first):
Hawaii, Chile, Canary Islands, Iran, Pakistan, China.
Layouts of these regions are given in Fig. \ref{F:country}.
Layouts for the other countries are given as a supplement through publisher's repository.

Due to the finalized resolution in the datasets, it would be wrong to label the whole country as ``best, good, worst'' from just statistics of its SIAS Series;
localized best ``locations'' could be found in the worst ``regions''.
For example, in Hawaii and the Canary Islands:
While shores of the islands fall into the worst category, volcanic mountains and their caldera are at the best category.
Similarly, this is the same for Chile, China, Iran and Pakistan where they have recently performed a site selection study.

SIAS Series analysis are also created specifically for 15 observatory sites having +4 m telescopes (in Table \ref{T:obs}).
Using the 3$\sigma$ rule given in Table \ref{T:results}, the following points are noted for these observatories:
(1) All are around or above 1$\sigma$ in all of the Series;
(2) Most of them are also close to 2$\sigma$ level in all of the Series.

In addition to this subset, the same analysis has been performed for all observatories, in total 2,123.
They will be accessible as a supplement through publisher’s repository.
When all of them included, only approximately 10\%, 2.5\%, 0.3\% falls in 1$\sigma$, 2$\sigma$, 3$\sigma$ levels, respectively.
However, to have a complete statistics for current observatory sites, the database of the list has to be updated so that it only includes professional sites.

The SIAS Series analysis for the observatories is given in the Table \ref{T:earlyresult}.
Using the 3$\sigma$ rule given in Table \ref{T:results}, the following points are noted for the listed regions:
(1) Most are around or above 1$\sigma$ in all of the Series;
(2) Almost half of them are above 2$\sigma$ level in all of the Series.

The most important outcome of this study is to create a database of ``Suitability Index'' for any terrestrial coordinate on Earth.
This tabulated database will also contain all Series studied in this work.
Therefore, amateurs, institutions or countries aiming to construct an observatory could create a short--list of potential site locations using SIAS values created for each country without spending time and budget.

The outcomes and datasets of this study has been made available through a web site, namely ``Astro GIS Database'' on \texttt{\url{www.astrogis.org}}.
\section{\textbf{Supplementary Material}}
\label{sec:Supl}

Additional supporting information include two files in the online version of this article and in \cite{2019arXiv191201911A}.

\textbf{\texttt{GSSA-Table6-Full.txt}}: Complete version of Table \ref{T:obs}, values of 6 Layers (CC, DEM, PWV, AOD, WS, LULC) and values of 4 SIAS Series (A, B, C, D) for all observatories.

\textbf{\texttt{GSSA-Figure4-Full.pdf}}: Complete version of Fig. \ref{F:country}, Layouts of 4 SIAS Series (A, B, C, D) for all countries having surface area > $\sim$ 10 km$^{2}$.

\section*{Acknowledgement}
This research was supported by the Scientific and Technological Research Council of Turkey (TÜBİTAK) through project number 117F309.
We thank to Esragül Atalay, Derya Öztürk Çetni and Hülya Işık for their efforts in downloading datasets.
We also thank supervisors of the project, namely İbrahim Küçük, Süha Berberoğlu and Mustafa Atılan.
We thank to Mehmet E. Özel for his helpful comments.
We thank to Alişan Aktay for design of the https://www.astrogis.org website.

\bibliographystyle{mnras}
\bibliography{GSSA-0}

\begin{thebibliography}{}
\makeatletter
\relax
\def\mn@urlcharsother{\let\do\@makeother \do\$\do\&\do\#\do\^\do\_\do\%\do\~}
\def\mn@doi{\begingroup\mn@urlcharsother \@ifnextchar [ {\mn@doi@}
  {\mn@doi@[]}}
\def\mn@doi@[#1]#2{\def\@tempa{#1}\ifx\@tempa\@empty \href
  {http://dx.doi.org/#2} {doi:#2}\else \href {http://dx.doi.org/#2} {#1}\fi
  \endgroup}
\def\mn@eprint#1#2{\mn@eprint@#1:#2::\@nil}
\def\mn@eprint@arXiv#1{\href {http://arxiv.org/abs/#1} {{\tt arXiv:#1}}}
\def\mn@eprint@dblp#1{\href {http://dblp.uni-trier.de/rec/bibtex/#1.xml}
  {dblp:#1}}
\def\mn@eprint@#1:#2:#3:#4\@nil{\def\@tempa {#1}\def\@tempb {#2}\def\@tempc
  {#3}\ifx \@tempc \@empty \let \@tempc \@tempb \let \@tempb \@tempa \fi \ifx
  \@tempb \@empty \def\@tempb {arXiv}\fi \@ifundefined
  {mn@eprint@\@tempb}{\@tempb:\@tempc}{\expandafter \expandafter \csname
  mn@eprint@\@tempb\endcsname \expandafter{\@tempc}}}

\bibitem[\protect\citeauthoryear{{Abdelaziz}, {Guebsi}, {Guessoum}  \&
  {Flamant}}{{Abdelaziz} et~al.}{2017}]{abdelaziz2017}
{Abdelaziz} G.,  {Guebsi} R.,  {Guessoum} N.,   {Flamant} C.,  2017, in Journal
  of Physics Conference Series. p. 012070,
  \mn@doi{10.1088/1742-6596/869/1/012070}

\bibitem[\protect\citeauthoryear{{Ackerman} \& {Cox}}{{Ackerman} \&
  {Cox}}{1981}]{ackerman1981}
{Ackerman} S.~A.,  {Cox} S.~K.,  1981, \mn@doi [J. Appl. Meteorol.]
  {10.1175/1520-0450(1981)020<0581:COSAAS>2.0.CO;2}, \href
  {https://ui.adsabs.harvard.edu/abs/1981JApMe..20..581A} {20, 581}

\bibitem[\protect\citeauthoryear{{Ackerman}, {Strabala}, {Menzel}, {Frey},
  {Moeller}  \& {Gumley}}{{Ackerman} et~al.}{1998}]{ackerman1998}
{Ackerman} S.~A.,  {Strabala} K.~I.,  {Menzel} W.~P.,  {Frey} R.~A.,  {Moeller}
  C.~C.,   {Gumley} L.~E.,  1998, \mn@doi [\jgr] {10.1029/1998JD200032}, \href
  {https://ui.adsabs.harvard.edu/abs/1998JGR...10332141A} {103, 32,141}

\bibitem[\protect\citeauthoryear{{Aksaker} et~al.,}{{Aksaker}
  et~al.}{2015}]{aksaker2015}
{Aksaker} N.,  et~al., 2015, \mn@doi [Exp. Astron.]
  {10.1007/s10686-015-9458-x}, \href
  {https://ui.adsabs.harvard.edu/abs/2015ExA....39..547A} {39, 547}

\bibitem[\protect\citeauthoryear{{Aksaker}, {Yerli}, {Erdo{\u{g}}an}, {Kurt},
  {Kaba}, {Bayazit}  \& {Yesilyaprak}}{{Aksaker}
  et~al.}{2019}]{2019arXiv191201911A}
{Aksaker} N.,  {Yerli} S.~K.,  {Erdo{\u{g}}an} M.~A.,  {Kurt} Z.,  {Kaba} K.,
  {Bayazit} M.,   {Yesilyaprak} C.,  2019, arXiv e-prints, \href
  {https://ui.adsabs.harvard.edu/abs/2019arXiv191201911A} {p. arXiv:1912.01911}

\bibitem[\protect\citeauthoryear{{Br{\'e}on} \& {Colzy}}{{Br{\'e}on} \&
  {Colzy}}{1999}]{breon1999}
{Br{\'e}on} F.-M.,  {Colzy} S.,  1999, \mn@doi [J. Appl. Meteorol.]
  {10.1175/1520-0450(1999)038<0777:CDFTSP>2.0.CO;2}, \href
  {https://ui.adsabs.harvard.edu/abs/1999JApMe..38..777B} {38, 777}

\bibitem[\protect\citeauthoryear{Churchman \& Ackoff}{Churchman \&
  Ackoff}{1954}]{churchman1954}
Churchman C.~W.,  Ackoff R.~L.,  1954, \mn@doi [Oper. Res.]
  {10.1287/opre.2.2.172}, 2, 172

\bibitem[\protect\citeauthoryear{{Daniyal} \& {Hassan Kazmi}}{{Daniyal} \&
  {Hassan Kazmi}}{2019}]{daniyal2019}
{Daniyal} A.,  {Hassan Kazmi} S.~J.,  2019, \mn@doi [RAA]
  {10.1088/1674-4527/19/9/129}, \href
  {https://ui.adsabs.harvard.edu/abs/2019RAA....19..129D} {19, 129}

\bibitem[\protect\citeauthoryear{{Elvidge}, {Baugh}, {Zhizhin}, {Hsu}  \&
  {Ghosh}}{{Elvidge} et~al.}{2017}]{elvidge2017}
{Elvidge} C.~D.,  {Baugh} K.,  {Zhizhin} M.,  {Hsu} F.~C.,   {Ghosh} T.,  2017,
  \mn@doi [IJRS] {10.1080/01431161.2017.1342050}, \href
  {https://ui.adsabs.harvard.edu/abs/2017IJRS...38.5860E} {38, 5860}

\bibitem[\protect\citeauthoryear{{Falchi} et~al.,}{{Falchi}
  et~al.}{2016}]{falchi2016}
{Falchi} F.,  et~al., 2016, \mn@doi [Sci. Adv.] {10.1126/sciadv.1600377}, \href
  {https://ui.adsabs.harvard.edu/abs/2016SciA....2E0377F} {2, e1600377}

\bibitem[\protect\citeauthoryear{{Feizizadeh}, {Jankowski}  \&
  {Blaschke}}{{Feizizadeh} et~al.}{2014a}]{feizizadeh2014a}
{Feizizadeh} B.,  {Jankowski} P.,   {Blaschke} T.,  2014a, \mn@doi [Comput.
  Geosci.] {10.1016/j.cageo.2013.11.009}, \href
  {https://ui.adsabs.harvard.edu/abs/2014CG.....64...81F} {64, 81}

\bibitem[\protect\citeauthoryear{{Feizizadeh}, {Shadman Roodposhti},
  {Jankowski}  \& {Blaschke}}{{Feizizadeh} et~al.}{2014b}]{feizizadeh2014b}
{Feizizadeh} B.,  {Shadman Roodposhti} M.,  {Jankowski} P.,   {Blaschke} T.,
  2014b, \mn@doi [Comput. Geosci.] {10.1016/j.cageo.2014.08.001}, \href
  {https://ui.adsabs.harvard.edu/abs/2014CG.....73..208F} {73, 208}

\bibitem[\protect\citeauthoryear{{Frey}, {Ackerman}, {Liu}, {Strabala},
  {Zhang}, {Key}  \& {Wang}}{{Frey} et~al.}{2008}]{frey2008}
{Frey} R.~A.,  {Ackerman} S.~A.,  {Liu} Y.,  {Strabala} K.~I.,  {Zhang} H.,
  {Key} J.~R.,   {Wang} X.,  2008, \mn@doi [J. Atmos. Oceanic Tech.]
  {10.1175/2008JTECHA1052.1}, \href
  {https://ui.adsabs.harvard.edu/abs/2008JAtOT..25.1057F} {25, 1057}

\bibitem[\protect\citeauthoryear{{Friedl} et~al.,}{{Friedl}
  et~al.}{2002}]{friedl2002}
{Friedl} M.~A.,  et~al., 2002, \mn@doi [Remote Sens. Environ.]
  {10.1016/S0034-4257(02)00078-0}, \href
  {https://ui.adsabs.harvard.edu/abs/2002RSEnv..83..287F} {83, 287}

\bibitem[\protect\citeauthoryear{{Friedl}, {Sulla-Menashe}, {Tan}, {Schneider},
  {Ramankutty}, {Sibley}  \& {Huang}}{{Friedl} et~al.}{2010}]{friedl2010}
{Friedl} M.~A.,  {Sulla-Menashe} D.,  {Tan} B.,  {Schneider} A.,  {Ramankutty}
  N.,  {Sibley} A.,   {Huang} X.,  2010, \mn@doi [Remote Sens. Environ.]
  {10.1016/j.rse.2009.08.016}, \href
  {https://ui.adsabs.harvard.edu/abs/2010RSEnv.114..168F} {114, 168}

\bibitem[\protect\citeauthoryear{García-Cascales \&
  Sánchez-Lozano}{García-Cascales \& Sánchez-Lozano}{2013}]{garcia2013}
García-Cascales M.~S.,  Sánchez-Lozano J.,  2013, \mn@doi [Renew. Sust.
  Energ. Rev.] {10.1016/j.rser.2013.03.019}, 24, 544

\bibitem[\protect\citeauthoryear{Glickman}{Glickman}{2000}]{glickman2000}
Glickman T.~S.,  2000, Glossary of Meteorology (2nd ed.).
American Meteorological Society, Boston, Mass.

\bibitem[\protect\citeauthoryear{{Gorsevski} \& {Jankowski}}{{Gorsevski} \&
  {Jankowski}}{2010}]{gorsevski2010}
{Gorsevski} P.~V.,  {Jankowski} P.,  2010, \mn@doi [Comput. Geosci.]
  {10.1016/j.cageo.2010.03.001}, \href
  {https://ui.adsabs.harvard.edu/abs/2010CG.....36.1005G} {36, 1005}

\bibitem[\protect\citeauthoryear{{Graham}, {Sarazin}, {Beniston}, {Collet},
  {Hayoz}, {Neun}  \& {Casals}}{{Graham} et~al.}{2005}]{graham2005}
{Graham} E.,  {Sarazin} M.,  {Beniston} M.,  {Collet} C.,  {Hayoz} M.,  {Neun}
  M.,   {Casals} P.,  2005, \mn@doi [Meteorol. Appl.]
  {10.1017/S1350482705001520}, \href
  {https://ui.adsabs.harvard.edu/abs/2005MeApp..12...77G} {12, 77}

\bibitem[\protect\citeauthoryear{{Graham}, {Sarazin}, {Kurlandczyk}, {Neun}  \&
  {Matzler}}{{Graham} et~al.}{2008}]{graham2008}
{Graham} E.,  {Sarazin} M.,  {Kurlandczyk} H.,  {Neun} M.,   {Matzler} C.,
  2008, {Site selection for extremely large telescopes using the FriOWL
  software and global re-analysis climate data}.
Proceedings of SPIE, p. 70121Y, \mn@doi{10.1117/12.787847}

\bibitem[\protect\citeauthoryear{{Hellemeier}, {Yang}, {Sarazin}  \&
  {Hickson}}{{Hellemeier} et~al.}{2019}]{hellemeier2019}
{Hellemeier} J.~A.,  {Yang} R.,  {Sarazin} M.,   {Hickson} P.,  2019, \mn@doi
  [\mnras] {10.1093/mnras/sty2982}, \href
  {https://ui.adsabs.harvard.edu/abs/2019MNRAS.482.4941H} {482, 4941}

\bibitem[\protect\citeauthoryear{{Hotan}, {Tingay}  \& {Glazebrook}}{{Hotan}
  et~al.}{2013}]{hotan2012}
{Hotan} C.~E.,  {Tingay} S.~J.,   {Glazebrook} K.,  2013, \mn@doi [\pasa]
  {10.1017/pasa.2012.002}, \href
  {https://ui.adsabs.harvard.edu/abs/2013PASA...30....2H} {30, e002}

\bibitem[\protect\citeauthoryear{{Intrieri}, {Shupe}, {Uttal}  \&
  {McCarty}}{{Intrieri} et~al.}{2002}]{Intrieri2002}
{Intrieri} J.~M.,  {Shupe} M.~D.,  {Uttal} T.,   {McCarty} B.~J.,  2002,
  \mn@doi [JGRAS Oceans] {10.1029/2000JC000423}, \href
  {https://agupubs.onlinelibrary.wiley.com/doi/abs/10.1029/2000JC000423} {107,
  SHE 5}

\bibitem[\protect\citeauthoryear{Jiang \& Eastman}{Jiang \&
  Eastman}{2000}]{jiang2000}
Jiang H.,  Eastman J.~R.,  2000, \mn@doi [Int J Geogr Inf Sci]
  {10.1080/136588100240903}, 14, 173

\bibitem[\protect\citeauthoryear{{K{\"u}{\c{c}}{\"u}k}
  et~al.,}{{K{\"u}{\c{c}}{\"u}k} et~al.}{2012}]{kucuk2012}
{K{\"u}{\c{c}}{\"u}k} I.,  et~al., 2012, \mn@doi [Exp. Astron.]
  {10.1007/s10686-011-9264-z}, \href
  {https://ui.adsabs.harvard.edu/abs/2012ExA....33....1K} {33, 1}

\bibitem[\protect\citeauthoryear{{Lau} \& {Crane}}{{Lau} \&
  {Crane}}{1997}]{lau1997}
{Lau} N.-C.,  {Crane} M.~W.,  1997, \mn@doi [Mon. Weather Rev.]
  {10.1175/1520-0493(1997)125<3172:CSASOO>2.0.CO;2}, \href
  {https://ui.adsabs.harvard.edu/abs/1997MWRv..125.3172L} {125, 3172}

\bibitem[\protect\citeauthoryear{{Liu}, {Yao}, {Wang}, {Ma}, {He}  \&
  {Wang}}{{Liu} et~al.}{2010}]{Liu2010}
{Liu} L.-Y.,  {Yao} Y.-Q.,  {Wang} Y.-P.,  {Ma} J.-L.,  {He} B.-L.,   {Wang}
  H.-S.,  2010, \mn@doi [Research in Astronomy and Astrophysics]
  {10.1088/1674-4527/10/10/009}, \href
  {https://ui.adsabs.harvard.edu/abs/2010RAA....10.1061L} {10, 1061}

\bibitem[\protect\citeauthoryear{{Long}, {Sabburg}, {Calb{\'o}}  \&
  {Pag{\`e}s}}{{Long} et~al.}{2006}]{long2006}
{Long} C.~N.,  {Sabburg} J.~M.,  {Calb{\'o}} J.,   {Pag{\`e}s} D.,  2006,
  \mn@doi [J. Atmos. Oceanic Tech.] {10.1175/JTECH1875.1}, \href
  {https://ui.adsabs.harvard.edu/abs/2006JAtOT..23..633L} {23, 633}

\bibitem[\protect\citeauthoryear{Nas, Cay, Iscan  \& Berktay}{Nas
  et~al.}{2009}]{nas2009}
Nas B.,  Cay T.,  Iscan F.,   Berktay A.,  2009, \mn@doi [Environ. Monit.
  Assess.] {10.1007/s10661-008-0713-8}, 160, 491

\bibitem[\protect\citeauthoryear{{Nasiri}, {Darudi}, {Khosroshahi}  \&
  {Sarazin}}{{Nasiri} et~al.}{2019}]{nasiri2019}
{Nasiri} S.,  {Darudi} A.,  {Khosroshahi} H.~G.,   {Sarazin} M.,  2019, \mn@doi
  [\mnras] {10.1093/mnras/stz726}, \href
  {https://ui.adsabs.harvard.edu/abs/2019MNRAS.486.4226N} {486, 4226}

\bibitem[\protect\citeauthoryear{Noorollahi, Yousefi  \& Mohammadi}{Noorollahi
  et~al.}{2016}]{noorollahi2016}
Noorollahi Y.,  Yousefi H.,   Mohammadi M.,  2016, \mn@doi [Sustain. Energy
  Technol. Assess.] {10.1016/j.seta.2015.11.007}, 13, 38

\bibitem[\protect\citeauthoryear{{Norris}}{{Norris}}{1999}]{norris1999}
{Norris} J.~R.,  1999, \mn@doi [J. Clim.]
  {10.1175/1520-0442(1999)012<1864:OTAPAI>2.0.CO;2}, \href
  {https://ui.adsabs.harvard.edu/abs/1999JCli...12.1864N} {12, 1864}

\bibitem[\protect\citeauthoryear{{Nurbandi}, {Ramadhani Yusuf}, {Prasetya}  \&
  {Dimas Afrizal}}{{Nurbandi} et~al.}{2016}]{nurbandi2016}
{Nurbandi} W.,  {Ramadhani Yusuf} F.,  {Prasetya} R.~a.,   {Dimas Afrizal} M.,
  2016, in IOP Conference Series: Earth and Environmental Science. p. 012040,
  \mn@doi{10.1088/1755-1315/47/1/012040}

\bibitem[\protect\citeauthoryear{{Osborn} \& {Sarazin}}{{Osborn} \&
  {Sarazin}}{2018}]{2018MNRAS.480.1278O}
{Osborn} J.,  {Sarazin} M.,  2018, \mn@doi [\mnras] {10.1093/mnras/sty1898},
  \href {https://ui.adsabs.harvard.edu/abs/2018MNRAS.480.1278O} {480, 1278}

\bibitem[\protect\citeauthoryear{{Otarola} et~al.,}{{Otarola}
  et~al.}{2019}]{2019PASP..131d5001O}
{Otarola} A.,  et~al., 2019, \mn@doi [\pasp] {10.1088/1538-3873/aafb78}, \href
  {https://ui.adsabs.harvard.edu/abs/2019PASP..131d5001O} {131, 045001}

\bibitem[\protect\citeauthoryear{{Pfister}, {McKenzie}, {Liley}, {Thomas},
  {Forgan}  \& {Long}}{{Pfister} et~al.}{2003}]{pfister2003}
{Pfister} G.,  {McKenzie} R.~L.,  {Liley} J.~B.,  {Thomas} A.,  {Forgan} B.~W.,
    {Long} C.~N.,  2003, \mn@doi [J. Appl. Meteorol.]
  {10.1175/1520-0450(2003)042<1421:CCBOAI>2.0.CO;2}, \href
  {https://ui.adsabs.harvard.edu/abs/2003JApMe..42.1421P} {42, 1421}

\bibitem[\protect\citeauthoryear{{Platnick}, {King}, {Ackerman}, {Menzel},
  {Baum}, {Riedi}  \& {Frey}}{{Platnick} et~al.}{2003}]{platnick2003}
{Platnick} S.,  {King} M.~D.,  {Ackerman} S.~A.,  {Menzel} W.~P.,  {Baum}
  B.~A.,  {Riedi} J.~C.,   {Frey} R.~A.,  2003, \mn@doi [IEEE Trans. Geosci.
  Remote Sens.] {10.1109/TGRS.2002.808301}, \href
  {https://ui.adsabs.harvard.edu/abs/2003ITGRS..41..459P} {41, 459}

\bibitem[\protect\citeauthoryear{{Racine}}{{Racine}}{2005}]{Racine2005}
{Racine} R.,  2005, \mn@doi [\pasp] {10.1086/429307}, \href
  {https://ui.adsabs.harvard.edu/abs/2005PASP..117..401R} {117, 401}

\bibitem[\protect\citeauthoryear{Remer, Tanré  \& Kaufman}{Remer
  et~al.}{2009}]{remer2009}
Remer L.~A.,  Tanré D.,   Kaufman Y.~J.,  2009, Algorithm for remote sensing
  of tropospheric aerosol from MODIS: Collection 5

\bibitem[\protect\citeauthoryear{Rikalovic, Cosic  \& Lazarevic}{Rikalovic
  et~al.}{2014}]{rikalovic2014}
Rikalovic A.,  Cosic I.,   Lazarevic D.,  2014, \mn@doi [Procedia Eng.]
  {10.1016/j.proeng.2014.03.090}, 69

\bibitem[\protect\citeauthoryear{{Sarazin}, {Graham}  \&
  {Kurlandczyk}}{{Sarazin} et~al.}{2006}]{sarazin2006}
{Sarazin} M.,  {Graham} E.,   {Kurlandczyk} H.,  2006, The Messenger, \href
  {https://ui.adsabs.harvard.edu/abs/2006Msngr.125...44S} {125, 44}

\bibitem[\protect\citeauthoryear{{Saunders} et~al.,}{{Saunders}
  et~al.}{2009}]{saunders2009}
{Saunders} W.,  et~al., 2009, \mn@doi [\pasp] {10.1086/605780}, \href
  {https://ui.adsabs.harvard.edu/abs/2009PASP..121..976S} {121, 976}

\bibitem[\protect\citeauthoryear{{Sayer}, {Munchak}, {Hsu}, {Levy},
  {Bettenhausen}  \& {Jeong}}{{Sayer} et~al.}{2014}]{sayer2014}
{Sayer} A.~M.,  {Munchak} L.~A.,  {Hsu} N.~C.,  {Levy} R.~C.,  {Bettenhausen}
  C.,   {Jeong} M.~J.,  2014, \mn@doi [JGR] {10.1002/2014JD022453}, \href
  {https://ui.adsabs.harvard.edu/abs/2014JGRD..11913965S} {119, 13,965}

\bibitem[\protect\citeauthoryear{{Sch{\"o}ck} et~al.,}{{Sch{\"o}ck}
  et~al.}{2009}]{TMT2009}
{Sch{\"o}ck} M.,  et~al., 2009, \mn@doi [\pasp] {10.1086/599287}, \href
  {https://ui.adsabs.harvard.edu/abs/2009PASP..121..384S} {121, 384}

\bibitem[\protect\citeauthoryear{Soltani \& Marandi}{Soltani \&
  Marandi}{2011}]{soltani2011}
Soltani A.,  Marandi I.,  2011, \mn@doi [JUEE] {10.4090/juee.2011.v5n1.032043},
  5, 32

\bibitem[\protect\citeauthoryear{Sulla~Menashe \& Friedl}{Sulla~Menashe \&
  Friedl}{2018}]{sullamenashe}
Sulla~Menashe D.,  Friedl M.~A.,  2018, Technical report, User Guide to
  Collection 6 MODIS Land Cover (MCD12Q1 and MCD12C1) Product, \url
  {https://modis.ornl.gov/documentation/guides/MCD12_User_Guide_V6.pdf}.
NASA, GSFC, \url
  {https://modis.ornl.gov/documentation/guides/MCD12_User_Guide_V6.pdf}

\bibitem[\protect\citeauthoryear{Tzeng \& Huang}{Tzeng \&
  Huang}{2013}]{tzeng2011}
Tzeng G.-H.,  Huang J.-J.,  2013, Multiple attribute decision making methods
  and applications, Taylor and Francıs, \mn@doi{10.1007/978-3-642-48318-9}

\bibitem[\protect\citeauthoryear{{Umar}, {Zainal Abidin}  \& {Abidin
  Ibrahim}}{{Umar} et~al.}{2014}]{umar2014}
{Umar} R.,  {Zainal Abidin} Z.,   {Abidin Ibrahim} Z.,  2014, in Journal of
  Physics Conference Series. p. 012009, \mn@doi{10.1088/1742-6596/539/1/012009}

\bibitem[\protect\citeauthoryear{Uyan}{Uyan}{2013}]{uyan2013}
Uyan M.,  2013, \mn@doi [Renew. Sust. Energ. Rev.]
  {https://doi.org/10.1016/j.rser.2013.07.042}, 28, 11

\bibitem[\protect\citeauthoryear{{Varela}, {Bertolin},
  {Mu{\~n}oz-Tu{\~n}{\'o}n}, {Ortolani}  \& {Fuensalida}}{{Varela}
  et~al.}{2008}]{varela2008}
{Varela} A.~M.,  {Bertolin} C.,  {Mu{\~n}oz-Tu{\~n}{\'o}n} C.,  {Ortolani} S.,
   {Fuensalida} J.~J.,  2008, \mn@doi [\mnras]
  {10.1111/j.1365-2966.2008.13803.x}, \href
  {https://ui.adsabs.harvard.edu/abs/2008MNRAS.391..507V} {391, 507}

\bibitem[\protect\citeauthoryear{{Varela} et~al.,}{{Varela}
  et~al.}{2014}]{varela2014}
{Varela} A.~M.,  et~al., 2014, \mn@doi [\pasp] {10.1086/676135}, \href
  {https://ui.adsabs.harvard.edu/abs/2014PASP..126..412V} {126, 412}

\bibitem[\protect\citeauthoryear{{Venkatappa}, {Sasaki}, {Shrestha}, {Tripathi}
   \& {Ma}}{{Venkatappa} et~al.}{2019}]{ventkappa2019}
{Venkatappa} {Sasaki} {Shrestha} {Tripathi}  {Ma} 2019, \mn@doi [Remote
  Sensing] {10.3390/rs11131514}, \href
  {https://ui.adsabs.harvard.edu/abs/2019RemS...11.1514V} {11, 1514}

\bibitem[\protect\citeauthoryear{{Vernin} et~al.,}{{Vernin}
  et~al.}{2011}]{vernin2011}
{Vernin} J.,  et~al., 2011, \mn@doi [\pasp] {10.1086/662995}, \href
  {https://ui.adsabs.harvard.edu/abs/2011PASP..123.1334V} {123, 1334}

\bibitem[\protect\citeauthoryear{{Yao}}{{Yao}}{2005}]{yao2005}
{Yao} Y.,  2005, \mn@doi [J. Korean Astron. Soc.] {10.5303/JKAS.2005.38.2.113},
  \href {https://ui.adsabs.harvard.edu/abs/2005JKAS...38..113Y} {38, 113}

\bibitem[\protect\citeauthoryear{{van Belle}, {Meinel}  \& {Meinel}}{{van
  Belle} et~al.}{2004}]{vanbelle2004}
{van Belle} G.~T.,  {Meinel} A.~B.,   {Meinel} M.~P.,  2004, {The scaling
  relationship between telescope cost and aperture size for very large
  telescopes}.
SPIE Proceeding, pp 563--570, \mn@doi{10.1117/12.552181}

\makeatother
\end{thebibliography}

\begin{table*}
	\caption{%
	A short summary of earlier works on astronomical site selection.
	Abbreviations for Layers are divided into two: Main and Others.
	\underline{Main Layers} (marked with `X' when the layer is used in the reference):
	DEM: Digital Elevation Model,
	CC: Cloud Coverage,
	AL: Artificial Light.
	\underline{Other Layers} (in alphabetical order):
	A: Accessibility,
	AOD: Aerosol Optical Depth,
	AT: Air Temperature,
	AU: Aurora,
	AI: Airglow, 
	ATE: Atmospheric Termal Emission,
	BL: Boundary Layer,
	BP: Barometric Pressure,
	D: Dewpoint,
	FS: Free Seeing,
	GH: Geopotential Height,
	IH: Inversion Height,
	IF: Inversion Frequency,
	LU: Land Use
	LC: Land Cover,
	LS: Local Specifications,
	O: Orography,
	OLR: Outgoing Longwave Radiation,
	OT: Optical Turbulence,
	PWV: Precipitable Water Vapor,
	RH: Relative Humidity,
	S: Seismicity,
	SB: Sky Brightness,
	SD: Sunshine Duration,
	SW: Surface Wind,
	T: Temperature,
	TS: Terrain Slope
	V: Vegetation,
	VP: Vapor Pressure,
	VV: Vertical Velocities,
	WD: Wind Direction,
	WS: Wind Speed.%
}
	\small\addtolength{\tabcolsep}{-2pt}
	\begin{tabular}{@{}cccccl@{}}
	\hline\hline
Region
    &   \multicolumn{3}{@{}c@{}}{Main Layers}
    &   Other Layers
    &   References \\
    &	DEM &	CC &	AL &&	\\
    \hline
West China
    &	X
    &	X
    &	-
    &	A, PWV, SD, SW, VP
    &	\cite{yao2005}	\\
Global
    &	X
    &	X
    &	-
    &	AOD, AT, D, GH, O, PWV, V, VV, OLR, WS
    &	\cite{graham2008} \\
South Pole
    &	X
    &	X
    &	-
    &	AI, AU, BL, FS, SB, ST, PWV
    &	\cite{saunders2009}	\\
Chile
    &	-
    &	X
    &	-
    &	AT, BL, FS, OT, PWV, RH, WS
    &	\cite{vernin2011} \\
Australia
    &	X
    &	X
    &	-
    &	-	
    &	\cite{hotan2012} \\
Spain, Chile, Argentina
    &	-
    &	-
    &	-
    &	AT, BP, RH, WD, WS
    &	\cite{varela2014} \\
Turkey
    &	X
    &	X
    &	X
    &	AOD, PWV, WS
    &	\cite{aksaker2015} \\
MENA
    &	X
    &	X
    &	X
    &	AOD, AT, PWV, RH, WS
    &	\cite{abdelaziz2017} \\
Iran
    &	X
    &	X
    &	X
    &	AOD, AL, IH, IF, LS, PWV, RH, S, SB, WS
    &	\cite{nasiri2019} \\
Pakistan
    &	X
    &	X
    &	X
    &	A, AOD, LU, LC, S, TS, WS
    &	\cite{daniyal2019} \\
    \hline\hline
	\end{tabular}
	\label{T:EarlyWork}
\end{table*}

\begin{table*}
    \centering
    \caption{%
    Source, location and characteristics of datasets.
    Some datasets use more than one product.
    Resolution given here represent original dataset; the final resolution is fixed to 1 km.
    Time coverage is usually taken from the satellite repositories.
    Two products of Aqua/Terra satellites have different time coverage: Feb 2000 to date and 2002 to date for MOD and MYD products, respectively.
    Time coverage of CC was started from July 2003 due to irregularities in earlier records from MYD35\_L2 product.%
    }
    \begin{tabular}{@{}cccccc@{}}
    \hline \hline
\multirow{2}{*}{Layer}
    & Satellite/Instrument
    & Product
    & Resolution
    & Coverage
    & Size\\
    \cline{2-6}
    & \multicolumn{5}{l}{Resource Location}\\
    \hline
\multirow{3}{*}{CC}
    & Aqua/Terra-MODIS
    & MOD35\_L2
    & 1 km 
    & 2000-2019
    & 24 TB\\
    & (NASA/LAADS DAAC)
    & MYD35\_L2
    & 
    & 2003-2019
    & \\\cline{2-6}
    & \multicolumn{5}{l}{\url{ladsweb.modaps.eosdis.nasa.gov/archive/allData/61/}}\\
    \hline
\multirow{2}{*}{DEM}
    & USGS - EROS Data Center (EDC)
    & GTOPO30 
    & 1 km 
    & 1996
    & 1.7 GB\\\cline{2-6}
    & \multicolumn{5}{l}{\url{earthexplorer.usgs.gov}}\\
    \hline
\multirow{2}{*}{AL}
    & NOAA/SUOMI-NPP
    & VIIRS/DNB 
    & 750 m 
    & 2016
    & 25 GB\\\cline{2-6}
    & \multicolumn{5}{l}{\url{https://eogdata.mines.edu/download_dnb_composites.html}}\\
    \hline
\multirow{3}{*}{PWV}
    & Aqua/Terra-MODIS
    & MOD07\_L2
    & 5 km 
    & 2000-2019
    & 10 TB\\
    & (NASA/LAADS DAAC)
    & MYD07\_L2 
    & 
    & 2002-2019
    & \\\cline{2-6}
    & \multicolumn{5}{l}{\url{ladsweb.modaps.eosdis.nasa.gov/archive/allData/61/}}\\
    \hline
\multirow{4}{*}{AOD}
    & Aqua/Terra-MODIS
    & MOD04\_3K
    & 3 / 10 km
    & 2000-2019
    & 24 TB \\
    & (NASA/LAADS DAAC)
    & MYD04\_3K
    & 
    & 2002-2019
    & \\
    & 
    & MOD04\_L2 
    & 
    & 
    & \\\cline{2-6}
    & \multicolumn{5}{l}{\url{ladsweb.modaps.eosdis.nasa.gov/archive/allData/61/}}\\
    \hline
\multirow{2}{*}{WIND}
    & Model Measurement
    & Global Wind Atlas 
    & 225 m
    & 2019 
    & 14 GB\\\cline{2-6}
    & \multicolumn{5}{l}{\url{globalwindatlas.info}}\\
    \hline
\multirow{2}{*}{LULC}
    & Aqua/Terra - MODIS 
    & MCD12Q1  
    & 10 km
    & 2018 
    & 200 MB\\\cline{2-6}
    & \multicolumn{5}{l}{\url{ladsweb.modaps.eosdis.nasa.gov/opendap/allData/6/MCD12Q1/contents.html}}\\
    \hline\hline
    \end{tabular}
    \label{T:source}
\end{table*}
\begin{table*}
    \centering
    \caption{%
    Summary of input and output of MCDA analysis for all the layers.
    AOD is a dimensionless number giving the total amount of aerosol within the vertical column over the site location.
    Since final layer resolution is fixed to 1 km, ``Original Resolution'' is given here.
    Definition used in Process column:
    NN: Nearest Neighbor Algorithm;
    LFZ: Linear Fuzzy Logic;
    L: Logarithm (base 10);
    SFL: S-shaped Inverse Sigmoidal Fuzzy Logic.%
    }
    \begin{tabular}{c c cc ccc cc}
    \hline \hline
Layer
    & \multicolumn{1}{C{3cm}}{Original Resolution}
    & \multicolumn{2}{c}{Process}
    & \multicolumn{2}{c}{Layer Value} &
    & \multicolumn{2}{C{3cm}}{Scale} \\
    & (km)
    & \multicolumn{1}{c}{IN}
    & \multicolumn{1}{c}{OUT}
    & \multicolumn{1}{c}{Min}
    & \multicolumn{1}{c}{Max}
    & \multicolumn{1}{c}{Unit}
    & \multicolumn{1}{c}{0 (Worst)}
    & \multicolumn{1}{c}{1 (Best)} \\
    \hline
CC
    & 3
    & NN & LFZ
    & 0.3 & 0.95 & \%
    & Cloudy & Clear\\
DEM
    & 1
    & -	& LFZ
    & -408 & 8,685 & m
    & Lowest & Highest\\
AL
    & 0.375
    & NN & L+SFL
    & 0	& 1682.71 & W cm$^{-2}$ sr$^{-1}$
    & Bright & Dark\\
PWV
    & 5
    & NN & LFZ
    & 0.68 & 53.54 & mm
    & High & Low\\
AOD
    & 3, 10
    & NN & LFZ
    & 0 & 4.98 & -
    & High & Low\\
WS
    & 0.225
    & NN & LFZ
    & 0.01 & 84.67 & m s$^{-1}$
    & Fast & Slow\\
LULC
    & 5
    & NN & -
    & - & - & CLASS
    & - & -\\
    \hline\hline
\end{tabular}
\label{T:summary}
\end{table*}
\begin{table}
    \caption{%
    Series created in MCDA analysis for the layers discussed above.
    In each series a weight is given to each targeted layer.
    In all four series created, layer list of earlier works were modified in this work.%
    }
    \begin{tabular}{@{}ccccccc@{}}
    \hline \hline
SIAS Series
    & CC    & DEM   & AL    & PWV   & AOD   & WS \\
    \hline
A   & 1     & 1     & 1     & 1     & 1     & 1 \\ 
B   & 1     & 1     & 1     & -     & -     & - \\ 
C   & 2     & 1     & 1     & -     & -     & - \\ 
D   & 1     & 2     & 1     & -     & -     & - \\ 
    \hline \hline
    \end{tabular}
\label{T:layers}
\end{table}
\begin{table*}
    \caption{%
    The results for a selected list of previous works on astronomical site selection.
    The table is in increasing publication year order.
    The AL layer is excluded because all of the selected observatory sites are located in dark regions (having AL values as 0.00).
    LULC identification numbers are taken from Table 3 of \citet{sullamenashe}.
    \underline{Regions from earlier studies:}
    KALSU (West China); Shiquanhe (West China); OMA (West China); Chajnantor (Chile);
    Cerros Tolar (Chile); Cerro Armazones (Chile); Cerro Tolonchar (Chile); San Pedro M\'{a}rtir (Mexico); Mauna Kea (Hawaii); Mt Meharry (Australia); Mereenie (Australia); ORM (Spain); Aklim (Morocco); Ventarrones (Chile); Macon (Argentina); Dinava (Iran); Gargash (Iran); Nok Kundi (Pakistan); Kalat (Pakistan); Kalam (Pakistan).
    \underline{References:} [1] \citep{yao2005}, [2] \citep{sarazin2006}, [3] \citep{TMT2009}, [4] \citep{hotan2012}, [5] \citep{varela2014}, [6] \citep{nasiri2019}, [7] \citep{daniyal2019}.%
}
    \centering
    \begin{tabular}{@{}l@{}ccc cccccc cccc@{}}
    \hline \hline
\multicolumn{4}{@{}l}{Region}
    & \multicolumn{6}{|c|}{Layers}
    & \multicolumn{4}{c}{SIAS Series} \\
    \hline 
Name
    & Ref.  & $\lambda$ & $\Phi$
    & CC    & DEM   & PWV   & AOD   & WS    & LULC
    & A	    & B     & C     & D \\
    \hline
KALSU 
    & [1] &	74.80	&	38.15
    &	0.31	&	4360	&	2.13	&	0.39	&	9.74	&	16
    &	0.83	&	0.73	&	0.71	&	0.67	\\
Shiquanhe
    & [1] &	80.02	&	32.32
    &	0.20	&	4788	&	1.9     &	0.25	&	6.93	&	16
    &	0.87	&	0.78	&	0.78	&	0.73	\\
OMA
    & [1] &	83.07	&	32.55
    &	0.24	&	4976	&	2.08	&	0.29	&	7.03	&	16
    &	0.86	&	0.78	&	0.77	&	0.73	\\
Chajnantor
    & [2] &	-67.75	&	23.02
    &	0.13	&	5032	&	2.43	&	0.05	&	8.33	&	16
    &	0.88	&	0.81	&	0.82	&	0.76	\\
Cerro Tolar
    & [3] &	-70.09	&	-21.96
    &	0.07	&	2150	&	11.28	&	0.08	&	6.04	&	16
    &	0.82	&	0.73	&	0.77	&	0.62	\\
Cerro Armazones
    & [3] &	-70.18	&	-24.58
    &	0.05	&	2885	&	7.35	&	0.05	&	4.91	&	16
    &	0.85	&	0.76	&	0.81	&	0.66	\\
Cerro Tolonchar 
    & [3] & -67.97	&	-23.93
    &	0.13	&	4255	&	2.65	&	0.04	&	4.38	&	16
    &	0.87	&	0.79	&	0.80	&	0.72	\\
San Pedro {M\'{a}rtir}
    & [3] &	-115.46 &	 31.04
    &	0.17	&	2657	&	15.15	&	0.12	&	4.78	&	10
    &	0.79	&	0.71	&	0.74	&	0.62	\\
Mauna Kea
    & [4] &	-155.48	&	19.83
    &	0.14	&	3865	&	13.75	&	0.07	&	5.42	&	16
    &	0.83	&	0.77	&	0.79	&	0.69	\\
Mt. Meharry
    & [4] &	118.58	&	22.98
    &	0.28	&	1100	&	18.93	&	0.01	&	6.49	&	7
    &	0.73	&	0.62	&	0.63	&	0.50	\\
Mereenie
    & [4] &	132.20	&	23.88
    &	0.25	&	790     &	15.75	&	0.00	&	6.48	&	7
    &	0.75	&	0.61	&	0.64	&	0.49	\\
ORM
    & [5] &	-17.89	&	28.76
    &	0.28	&	2139	&	21.47	&	0.25	&	N/A     &	9
    &	0.74	&	0.64	&	0.65	&	0.55	\\
Aklim
    & [5] &	-2.43	&	34.91
    &	0.38	&	146     &	17.34	&	0.11	&	4.28	&	12
    &	0.71	&	0.55	&	0.56	&	0.43	\\
Ventarrones
    & [5] &	-70.40	&	24.62
    &	0.05	&	2528	&	9.50	&	10      &	5.69	&	16
    &	0.83	&	0.75	&	0.80	&	0.64	\\
Macon
    & [6] &	-67.78	&	22.93
    &	0.14	&	5130	&	2.55	&	0.10	&	9.09	&	16
    &	0.88	&	0.81	&	0.82	&	0.76	\\
Dinava
    & [6] &	50.54	&	34.09
    &	0.24	&	1542	&	12.07	&	0.27	&	5.12	&	7
    &	0.77	&	0.65	&	0.67	&	0.54	\\
Gargash
    & [6] &	51.31	&	33.67
    &	0.24	&	3359	&	10.41	&	0.19	&	9.42	&	10
    &	0.80	&	0.71	&	0.72	&	0.64	\\
Nok Kundi
    & [6] &	62.75	&	28.82
    &	0.12	&	675     &	16.04	&	0.35	&	6.59	&	16
    &	0.74	&	0.63	&	0.69	&	0.50	\\
Kalat
    & [6] &	27.70	&	27.70
    &	0.14	&	1430	&	18.31	&	0.30	&	6.35	&	16
    &	0.77	&	0.68	&	0.72	&	0.56	\\
Kalam
    & [6] &	34.00	&	34.00
    &	0.27	&	610     &	18.37	&	0.59	&	4.83	&	10
    &	0.72	&	0.61	&	0.63	&	0.49	\\
    \hline\hline
    \end{tabular}
    \label{T:earlyresult}
\end{table*}
\begin{table*}
    \caption{%
    The results for astronomical sites hosting a +4 m class telescope.
    The table is in decreasing latitude order from South to North.
    The AL layer is excluded because all of the selected observatories are located in dark regions (having AL values as 0.00) except MDO with AL value of 49.5.
    LULC identification numbers are taken from Table 3 of \citet{sullamenashe}.
    \underline{Observatory abbreviations} (given geographic coordinates usually represents a single telescope among others):
    SAAO: South African Astronomical Obs.;
    CPO: Cerro Pachón Obs.;
    CTIO: Cerro Tololo Inter-American Obs.;
    LCO: Las Campanas Obs.;
    CA: Cerro Armazones.;
    PO: Paranal Obs.;
    MKO: Mauna Kea Obs.;
    ORM: Roque de los Muchachos Obs.;
    MDO: McDonald Obs.;
    FLWO: Fred Lawrence Whipple Obs.;
    KPNO: Kitt Peak National Obs.;
    MGIO: Mount Graham International Obs.;
    PalO: Palomar Obs.;
    LO: Lowell Obs.;
    DAG: Eastern Anatolia Obs.;
    BAO: Beijing Astronomical Obs.;
    SAO: Special Astrophysical Obs.;
    MRO: Maple Ridge Obs.
    Full observatory list will be available online.%
}
    \centering
    \begin{tabular}{@{}l@{}ccc cccccc cccc@{}}
    \hline \hline
\multicolumn{4}{@{}l}{Observatory}
    & \multicolumn{6}{|c|}{Layers}
    & \multicolumn{4}{c}{SIAS Series} \\
    \hline 
\multicolumn{2}{@{}l}{Abbreviation}
    & $\lambda$ & $\Phi$
    & CC    & DEM   & PWV   & AOD   & WS    & LULC
    & A	    & B     & C     & D \\
    \hline
\multicolumn{2}{@{}l}{SAAO}
    & 20.81 & -32.37
    & 0.15  & 1640  & 9.71 & N/A   & 7.57 & 2
    & 0.80  & 0.68  & 0.72 & 0.57\\
\multicolumn{2}{@{}l}{CPO}
    & -70.73& -30.24
    & 0.19  & 2600  & 7.07 & 0.26  & 4.98 & 2
    & 0.81  & 0.70  & 0.72 & 0.61\\
\multicolumn{2}{@{}l}{CTIO}
    & -70.80& -30.16
    & 0.18  & 1751  & 8.21 & 0.30  & 3.74 & 2
    & 0.80  & 0.68  & 0.71 & 0.57\\
\multicolumn{2}{@{}l}{LCO}
    & -70.70& -29.00
    & 0.12  & 2144  & 7.48 & 0.26  & 3.48 & 2
    & 0.82  & 0.71  & 0.75 & 0.60\\
\multicolumn{2}{@{}l}{CA}
    & -70.20& -24.60
    & 0.05  & 2789  & 7.35 & 0.06  & 4.74 & 16
    & 0.85  & 0.77  & 0.81 & 0.66\\
\multicolumn{2}{@{}l}{PO}
    & -70.40& -24.62
    & 0.05  & 2374  & 9.50 & 0.10  & 5.99 & 7
    & 0.83  & 0.74  & 0.79 & 0.63\\
\multicolumn{2}{@{}l}{MKO}
    & -155.47& 19.82
    & 0.14  & 4120  & 13.13 & 0.08 & 5.33 & 7
    & 0.83  & 0.78  & 0.79  & 0.71\\
\multicolumn{2}{@{}l}{ORM}
    & -17.89& 28.75
    & 0.28  & 2214  & 21.47 & 0.25 & 5.57 & 4
    & 0.74  & 0.65  & 0.66  & 0.56\\
\multicolumn{2}{@{}l}{MDO}
    & 104.02& 30.67
    & 0.30  & 1953  & 12.81 & 0.09 & 5.71 & 2
    & 0.77  & 0.64  & 0.65  & 0.54\\
\multicolumn{2}{@{}l}{FLWO}
    & -110.88& 31.68
    & 0.30  & 2369  & 12.25 & 0.11 & 6.31 & 5
    & 0.77  & 0.66  & 0.66  & 0.57\\
\multicolumn{2}{@{}l}{KPNO}
    & -111.59& 31.95
    & 0.27  & 1810  & 13.80 & 0.13 & 6.13 & 2
    & 0.76  & 0.65  & 0.66  & 0.54\\
\multicolumn{2}{@{}l}{MGIO}
    & -109.89& 32.70
    & 0.29  & 3081  & 11.20 & 0.10 & 7.01 & 5
    & 0.80  & 0.68  & 0.68  & 0.61\\
\multicolumn{2}{@{}l}{PalO}
    & -116.86& 33.35
    & 0.22  & 1676  & 14.20 & 0.09 & 2.77 & 4
    & 0.77  & 0.66  & 0.68  & 0.55\\
\multicolumn{2}{@{}l}{LO}
    & -111.42& 34.74
    & 0.29  & 2330  &  9.43 & 0.06 & 3.23 & 9
    & 0.79  & 0.66  & 0.67 & 0.57\\
\multicolumn{2}{@{}l}{DAG}
    & 41.23& 39.78
    & 0.36  & 2989  &  5.48 & 0.17 & 6.66 & 5
    & 0.79  & 0.66  & 0.65 & 0.59\\
\multicolumn{2}{@{}l}{BAO}
    & 117.57& 40.39
    & 0.40  & 825   & 11.52 & 0.22 & 2.02 & 1
    & 0.73  & 0.57  & 0.57  & 0.46\\
\multicolumn{2}{@{}l}{SAO}
    & 41.44& 43.64
    & 0.41 & 1995 & 8.48 & 0.09 & 4.77 & 8
    & 0.77  & 0.61  & 0.59  & 0.52\\
\multicolumn{2}{@{}l}{MRO}
    & -122.57& 49.28
    & 0.60  & 356   & 11.89 & 0.17 & 1.47 & 1
    & 0.69  & 0.47  & 0.45  & 0.37\\
    \hline \hline
    \end{tabular}
    \label{T:obs}
\end{table*}
\begin{table*}
    \caption{%
    The statistics of SIAS Series.
    See Section \ref{S:conc} for explanation and details.%
    }
    \begin{tabular}{@{}c cccc @{}C{10mm}C{10mm}C{10mm}C{10mm}C{10mm}C{10mm}@{}}
    \cline{2-11}
    &&&&&\multicolumn{6}{c@{}}{Sigma Level}\\
    &    \multicolumn{4}{c@{}}{Stat. of the series}
    &    \multicolumn{6}{c@{}}{Observatories (N)- Ratio (\%)}\\
    &&&&&\multicolumn{6}{c@{}}{Surface Area ($\times 10^{6}$ sq.km) - Ratio to Total Land (\%)}\\
    \hline
Series
    & Min
    & $\mu$
    & $\sigma$
    & Max
    & \multicolumn{2}{C{2cm}}{$\mu + 1\sigma$}
    & \multicolumn{2}{C{2cm}}{$\mu + 2\sigma$}
    & \multicolumn{2}{C{2cm}@{}}{$\mu + 3\sigma$}\\
    \hline
\multirow{3}{*}{A}
    & \multirow{3}{*}{0.38}
    & \multirow{3}{*}{0.71}
    & \multirow{3}{*}{0.05}
    & \multirow{3}{*}{0.93}
    & \multicolumn{2}{c}{0.77}
    & \multicolumn{2}{c}{0.82}
    & \multicolumn{2}{c@{}}{0.87} \\
    \cline{6-11}
    &&&&
    & 205
    & 9.66
    & 37
    & 1.74
    & 3
    & 0.14\\
    \cline{6-11}
    &&&&
    & 63.80
    & 37.38
    & 19.59
    & 11.47
    & 3.79
    & 2.22 \\
    \hline
    \multirow{3}{*}{B}
    & \multirow{3}{*}{0.09}
    & \multirow{3}{*}{0.54}
    & \multirow{3}{*}{0.08}
    & \multirow{3}{*}{0.89}
    & \multicolumn{2}{c}{0.62}
    & \multicolumn{2}{c}{0.70}
    & \multicolumn{2}{c@{}}{0.77} \\
    \cline{6-11}
    &&&&
    & 219
    & 10.32
    & 39
    & 1.84
    & 10
    & 0.47\\
    \cline{6-11}
    &&&&
    & 43.26
    & 25.35
    & 27.87
    & 16.33
    & 3.67
    & 2.15 \\
    \hline
    \multirow{3}{*}{C}
    & \multirow{3}{*}{0.09}
    & \multirow{3}{*}{0.54}
    & \multirow{3}{*}{0.10}
    & \multirow{3}{*}{0.87}
    & \multicolumn{2}{c}{0.63}
    & \multicolumn{2}{c}{0.73}
    & \multicolumn{2}{c@{}}{0.83} \\
    \cline{6-11}
    &&&&
    & 235
    & 11.07
    & 43
    & 2.03
    & 1
    & 0.05\\
    \cline{6-11}
    &&&&
    & 41.67
    & 24.42
    & 29.36
    & 17.20
    & 4.08
    & 2.39 \\
    \hline
    \multirow{3}{*}{D}
    & \multirow{3}{*}{0.07}
    & \multirow{3}{*}{0.44}
    & \multirow{3}{*}{0.07}
    & \multirow{3}{*}{0.92}
    & \multicolumn{2}{c}{0.51}
    & \multicolumn{2}{c}{0.58}
    & \multicolumn{2}{c@{}}{0.66} \\
    \cline{6-11}
    &&&&
    & 239
    & 11.26
    & 89
    & 4.19
    & 10
    & 0.47\\
    \cline{6-11}
    &&&&
    & 45.18
    & 26.47
    & 22.53
    & 13.20
    & 5.13
    & 3.01 \\
    \hline\hline
    \end{tabular}
    \label{T:results}
\end{table*}
%

\bsp	
\label{lastpage}
\end{document}